\documentclass[3p,10pt]{elsarticle}
 
\journal{Journal of Computational Physics}
 
\biboptions{sort&compress}

\usepackage{amsmath}
\usepackage{amssymb}
\usepackage{amsthm}
\usepackage{mathtools}
\usepackage{mathrsfs}
\usepackage{esvect}
\usepackage{nicefrac}
\usepackage{cancel}
\usepackage{wasysym}
\usepackage{siunitx}

\usepackage{graphicx}
\usepackage{float}
\usepackage{wrapfig}
\usepackage{subcaption}
\usepackage{xcolor}
\usepackage{multirow}
\usepackage{tabularx}
\usepackage{longtable}

\usepackage{multicol}
\usepackage[strict]{changepage}
\usepackage{relsize}
\usepackage{comment}
 
\usepackage{xurl}
\usepackage{hyperref}
\usepackage{algorithm}
\usepackage{algpseudocode}
 
\let\oldsqrt\sqrt
\def\sqrt{\mathpalette\DHLhksqrt}
\def\DHLhksqrt#1#2{%
\setbox0=\hbox{$#1\oldsqrt{#2}$}\dimen0=\ht0
\advance\dimen0-0.2\ht0
\setbox2=\hbox{\vrule height\ht0 depth -\dimen0}%
{\box0\lower0.4pt\box2}}

\begin{document}

\begin{frontmatter}
 
\title{von Neumann Stability Analysis of Braginskii MHD:\\
Thermal Conduction, Viscosity, Magnetic Diffusion, and Thermoelectric Effects;\\
and an Implementation in Athena++}
 
\author[leeds]{Alexandra F. Albery\corref{cor1}}
\ead{pybla@leeds.ac.uk}
\author[leeds]{Julian M. Pittard}
\cortext[cor1]{Corresponding author}
 
\affiliation[leeds]{organization={School of Physics and Astronomy, University of Leeds},
             addressline={Woodhouse},
             city={Leeds},
             postcode={LS2 9JT},
             country={United Kingdom}}
 
\begin{abstract}
Athena++ is a finite-volume astrophysical magnetohydrodynamics (MHD) code. As of writing, isotropic thermal conduction and kinematic viscosity, Ohmic resistivity, and ambipolar diffusion are present in the main branch. Here we present a fork of Athena++, ``Athena-Aniso''. It implements fully anisotropic thermal conduction (parallel, perpendicular, and the Righi--Leduc effect); anisotropic viscosity (all five Braginskii viscosity tensors); the Hall effect; and several thermoelectric effects (Peltier, Seebeck, Nernst, and Ettingshausen). These effects can be weighted by ionisation fraction in a partially ionised gas, with the neutral part diffusing isotropically. We have also added isotropic bulk viscosity. The Biermann battery is not implemented here.

The dissipative terms have super-time-stepping support, while the non-dissipative terms are combined with the ideal MHD flux. A CFL condition for all terms is derived and implemented using von Neumann stability analysis. The dissipative terms are found to be stable under first-order time integration. The non-dissipative terms are found to be stable under third-order time integration.

We additionally derive new signal speeds for a modified HLLE Riemann solver. These suppress grid-scale non-linear instabilities still present in the non-dissipative terms.

The Nernst/Ettingshausen system is found to be ill-posed unless it is stabilised by both perpendicular thermal conduction and perpendicular magnetic diffusion. We derive a well-posedness criterion for this.

We provide numerical tests of each implemented effect against derived analytical solutions.
\end{abstract}
 
 
\end{frontmatter}
\newpage
\section{Introduction}
Isotropic diffusion (thermal conduction, viscosity, Ohmic resistivity, and ambipolar diffusion) is present in many MHD code bases, including Athena++ \cite{Stone2020}, which we will be using henceforth. The GPU version of Athena++, AthenaK, has support for parallel anisotropic thermal conduction \cite{stone_2026}. Anisotropic thermal conduction and viscosity split into parallel and perpendicular components are also present in some codes such as the finite volume astrophysical codes PLUTO \cite{Mignone_2012} and AREPO \cite{Springel_2010, Kannan_2016, Berlok_2020}; and the multi-physics code Gizmo \cite{Hopkins_2016}.

The FLASH code \cite{Fryxell_2000} supports the Hall effect, anisotropic resistivity, the Righi--Leduc effect, and thermoelectric effects (including the Biermann battery). In its explicit extended-MHD timestep, the non-ideal magnetic terms are recast as advection by characteristic drift velocities. These are summed with the fluid velocity into an effective electron velocity, and the CFL condition uses the larger of the fluid and effective electron velocities plus the fast-magnetosonic speed.
For stability, the thermoelectric drifts are each passed through a flux limiter that restricts them relative to a user-set fraction of the electron thermal speed (some of these terms are disabled at shocks). The Hall and battery terms enter through a wave speed: the fast-magnetosonic speed is replaced by the maximum of itself, the whistler phase speed, a Hall drift-wave speed, and a thermal--magnetic (battery) wave speed \footnote{FLASH 4.8: hy\_uhd\_computeDtExtMHD.F90}.

The Hall effect on its own is well known to be unstable to first-order (and second-order) naive time integration \cite{Falle_2003}; however, it can be stabilised by the presence of ambipolar diffusion \cite{osullivan_2006}. It has also been shown that the Hall effect is conditionally stable at RK3 using spectral methods \cite{kunz_2013}. With finite-volume methods the Hall effect still suffers from non-linear instabilities using RK3, despite being shown to be linearly stable via von Neumann stability analysis \cite{iwasaki_2025}. Several modifications to the MHD solver have been proposed to stabilise the Hall effect at the grid scale including artificial hyper-resistivity \cite{Birn_2001} \cite{VIGANO_20211} and modifying the Riemann solver to use the whistler wave phase speed (Hall-HLL) \cite{Toth_2008, Lesur_2014, marchand_2019}.

Explicit integration of dissipative terms is restricted by the timestep $\Delta t_\mathrm{par} \propto \Delta x^2/D$, where $D$ is the diffusion coefficient. This becomes prohibitively restrictive at high resolution and/or when diffusion is strong. Super-time-stepping (STS) relaxes this constraint while remaining explicit. STS schemes construct stability polynomials that extend stability further along the negative real axis, so that a super-step of $s$ stages can be $\propto s^{2}$ times the explicit limit, whereas $s$ ordinary explicit sub-steps advance only $s$ times it. The Chebyshev-based STS of \cite{Alexiades_1996} is first-order accurate and requires a tunable damping parameter, whereas the Runge--Kutta--Legendre (RKL) schemes \cite{Meyer_2012, Meyer_2014} need no such parameter. RKL1 and RKL2 are first- and second-order accurate in time respectively. Athena++ implements both as an operator-split update of the dissipative terms: one super-step equal to the MHD timestep for RKL1, or two of half that length for RKL2 - one before, and one after the MHD update, giving a speed-up roughly proportional to $s$ \cite{Stone2020}. Because the RKL stability regions are built along the negative real axis, they contain no part of the imaginary axis: like RK1 and RK2, RKL1 and RKL2 amplify every purely imaginary mode. The non-dissipative terms, whose eigenvalues lie on the imaginary axis, therefore gain nothing from STS and must be integrated with the main MHD update.

In this work we present a new implementation of anisotropic dissipative and non-dissipative effects in Athena++ (``Athena-Aniso''). Effects acting on the momentum and energy have their fluxes calculated directly at cell faces and are integrated with either the main MHD solver or the STS solver. Effects acting on the magnetic field have their EMFs calculated directly at cell edges and passed to the upwind constrained transport (UCT) solver \cite{Stone_2009, Felker_2018} (also with STS support). In Section~2 we present the non-ideal MHD equations that are solved. In Section~3 we perform von Neumann stability analysis on the non-ideal terms to derive their CFL conditions, and determine which time integrator is required for stability. The dissipative and non-dissipative parts are first separated to provide the CFL conditions required for super-time-stepping the dissipative parts; then the unsplit system is analysed for non-STS use. In Section~4 we derive the modified HLLE signal speeds. In Section~5 we present our implementation in Athena++. In Section~6 we test the new physics against analytical results. In Section~7 we conclude and summarise our results. We provide the expanded form of the viscosity terms in Appendix~A. Intermediate steps of the derivations of Sections~3 and~4 are given in Appendices~B and~C, and the set-up of the noise tests of Section~6 in Appendix~D. We provide numerical validation of our derived CFL conditions with physical diffusion coefficients in Appendix~E.
\section{Governing Equations}
Neglecting all operator-split source terms (such as radiative heating/cooling, self-gravity, and orbital advection) and relativistic effects, Athena-Aniso solves the following equations of non-ideal MHD \cite{Braginskii_1965, Huba_2008, Morales_2018, Stone2020} \cite[pp.~405--406]{Kulsrud_2020}
\begin{subequations}
    \label{non-ideal-mhd-eq}
    \begin{equation}
        \frac{\partial \rho}{\partial t} + \vec{\nabla}\cdot[\rho \vec{v}]=0,
    \end{equation}
    \begin{align}
         \frac{\partial \rho \vec{v}}{\partial t} + \vec{\nabla}\cdot\left[ 
\rho \vec{v}\otimes \vec{v} -\vec{B}\otimes\vec{B}+\left(P+\frac{|\vec{B}|^{2}}{2}\right)\boldsymbol{I} -\boldsymbol{\Pi}_{\text{aniso}}\right] = \vec{0}, 
    \end{align}
    \begin{align}
        \frac{\partial \varepsilon}{\partial t} &+ \vec{\nabla} \cdot \left[ \left( 
 \varepsilon + P + \frac{|\vec{B}|^{2}}{2} \right)\vec{v} - \vec{B}(\vec{B}\cdot\vec{v}) \right. \left. +  \vec{q}_{\text{aniso}} + \vec{q}_{J} - \boldsymbol{\Pi}_{\text{aniso}}\cdot\vec{v} \right.  \\ 
 & \left. + \left( \eta\vec{J} + \eta_{AD}\hat{b}\times(\vec{J}\times\hat{b}) +\eta_{H}\vec{J}\times\hat{b} + \vec{F}_{T}\right)\times \vec{B} \right]  = 0, \nonumber
    \end{align}
    \begin{equation}
        \frac{\partial \vec{B}}{\partial t} - \vec{\nabla}\times\left[\vec{v}\times\vec{B} - \eta\vec{J} - \eta_{AD}\hat{b}\times(\vec{J}\times\hat{b}) - \eta_{H}\vec{J}\times\hat{b} - \vec{F}_{T}\right] = \vec{0}.
    \end{equation}
\end{subequations}
Here, $\rho$ is the mass density, $\vec{v}$ is the velocity vector, $P$ is the thermal pressure, $\vec{B}$ is the magnetic field, $\hat{b}=\vec{B}/|\vec{B}|$ is the magnetic field unit vector, and $\vec{J}=\vec{\nabla}\times\vec{B}$ is the current density (in Athena++ code units). We take a single temperature model with $T=T_{e}=T_{i}=P/\rho$. $\eta$, $\eta_{AD}$, and $\eta_{H}$ are the Ohmic, ambipolar and Hall coefficients respectively.

The total energy per unit volume is comprised of the internal, kinematic and magnetic energy
\begin{equation}
    \varepsilon = \varepsilon_{\text{int}} + \varepsilon_{\text{kin}} + \varepsilon_{\text{mag}}, \qquad \varepsilon_{\text{int}} = \frac{P}{\gamma -1}, \quad \varepsilon_{\text{kin}} = \frac{1}{2}\rho|\vec{v}|^{2}, \quad \varepsilon_{\text{mag}} = \frac{1}{2}|\vec{B}|^{2}.
\end{equation}
$\vec{q}_{\text{aniso}}$ is the heat flux vector; it contains isotropic neutral thermal conduction and ionised Braginskii electron thermal conduction weighted by level of ionisation \cite{Braginskii_1965}
\begin{subequations}
\label{eqn:conduction}
\begin{equation}
    \label{kappa-aniso}
    \vec{q}_{\text{aniso}} = (1-x_{\text{ion}})\vec{q}_{\text{neu}}+x_{\text{ion}}\vec{q}_{e}, \qquad \vec{q}_{\text{neu}} = -\kappa_{n}\rho\vec{\nabla}[P/\rho], \qquad \vec{q}_{e} = \vec{q}_{\parallel} + \vec{q}_{\perp} + \vec{q}_{\wedge},
\end{equation}
\begin{equation}
\vec{q}_{\parallel}=-\kappa_{\parallel}\rho\left(\hat{b}\cdot\vec{\nabla} [P/\rho]\right)\hat{b}, \quad
  \vec{q}_{\perp} = -\kappa_{\perp}\rho\left(\vec{\nabla}[P/\rho] - \left(\hat{b}\cdot\vec{\nabla} [P/\rho]\right)\hat{b} \right), \quad
\vec{q}_{\wedge} = -\kappa_{\wedge}\rho\hat{b}\times\vec{\nabla}[P/\rho].
\end{equation}
\end{subequations}
Where $x_{\text{ion}}$ is the mass fraction of ionised gas and $\kappa_{*}$ are the thermal diffusivity coefficients. The parallel and perpendicular parts are the electron thermal conduction along and across the magnetic field respectively, while the wedge term is the Righi--Leduc effect (also referred to as the thermal Hall effect) \cite{Callen_1948}.

$\boldsymbol{\Pi}_{\text{aniso}}$ is the viscous stress, comprising molecular bulk viscosity, isotropic neutral viscosity, and Braginskii ion viscosity, with the latter two weighted by the ionisation fraction \cite{Braginskii_1965}
\begin{subequations}
\label{visc-aniso}
\begin{equation}
\boldsymbol{\Pi}_{\text{aniso}}= \boldsymbol{\Pi}_{\text{bulk}}+ (1- x_{\text{ion}})\boldsymbol{\Pi}_{\text{neu}} + x_{\text{ion}}\boldsymbol{\Pi}_{\text{ion}}, \qquad
    \boldsymbol{\Pi}_{\text{bulk}} = \zeta(\vec{\nabla}\cdot\vec{v})\boldsymbol{I}, \qquad
\boldsymbol{\Pi}_{\text{neu}} = \nu\rho\boldsymbol{W}, \qquad
\boldsymbol{\Pi}_{\text{ion}} = \sum_{m=0}^{4}\boldsymbol{\Pi}^{(m)},
\end{equation}
\begin{equation}
        \boldsymbol{\Pi}^{(0)} = \frac{3}{2}\rho\nu_{\parallel}\left(\hat{b}\otimes\hat{b}-\frac{1}{3}\boldsymbol{I}\right)\left(\hat{b}\cdot\boldsymbol{W}\cdot\hat{b}\right),
    \end{equation}
    \begin{equation}
    \boldsymbol{\Pi}^{(1)} =  \rho\nu_{\perp1}\left[\boldsymbol{W}_{\perp}
+\frac{1}{2}\boldsymbol{I}_{\perp}(\hat{b}\cdot\boldsymbol{W}\cdot\hat{b}) \right],
    \end{equation}
    \begin{equation}
    \boldsymbol{\Pi}^{(2)} =  4\rho\nu_{\perp2}\left[\boldsymbol{I}_{\perp} \cdot \boldsymbol{W} \cdot \left(\hat{b}\otimes\hat{b}\right)
+\left(\hat{b}\otimes\hat{b}\right) \cdot \boldsymbol W \cdot \boldsymbol{I}_{\perp}
\right],
    \end{equation}
    \begin{equation}
        \boldsymbol{\Pi}^{(3)} = -\frac{1}{2}\rho\nu_{\wedge1}\left[\hat{b}\times\boldsymbol{W}\cdot\boldsymbol{I}_{\perp} -\boldsymbol{I}_{\perp}\cdot \boldsymbol{W} \times \hat{b} \right],
    \end{equation}
    \begin{equation}
        \boldsymbol{\Pi}^{(4)} = -2\rho\nu_{\wedge2}\left[\hat{b}\times\boldsymbol{W}\cdot(\hat{b}\otimes\hat{b})
-(\hat{b}\otimes\hat{b})\cdot\boldsymbol{W}\times\hat{b}
\right].
    \end{equation}
\end{subequations}
Where $\zeta$ is the bulk dynamic viscosity and $\nu_{*}$ are the kinematic neutral and ionised viscosities. $\boldsymbol{I}_{\perp}$ is the magnetic perpendicular identity tensor, $\boldsymbol{W}$ the trace-free symmetric rate-of-strain tensor, and $\boldsymbol{W}_{\perp}$ its perpendicular components \cite{fitzpatrick_2016}
\begin{equation}
    \label{p_aniso-perp}
    \boldsymbol{I}_{\perp} = \boldsymbol{I} - \hat{b}\otimes\hat{b}, \qquad
    \boldsymbol{W} = \vec{\nabla}[\vec{v}] +  \vec{\nabla}[\vec{v}]^{T} - \frac{2}{3}(\vec{\nabla}\cdot\vec{v})\boldsymbol{I}, \qquad
    \boldsymbol{W}_{\perp} = \boldsymbol{I}_{\perp}\cdot\boldsymbol{W}\cdot\boldsymbol{I}_{\perp}.
\end{equation}
The parallel ($\boldsymbol{\Pi}^{(0)}$) and perpendicular ($\boldsymbol{\Pi}^{(1)} + \boldsymbol{\Pi}^{(2)}$) components are the ion kinematic viscosity along and across the magnetic field respectively, while the wedge terms ($\boldsymbol{\Pi}^{(3)} + \boldsymbol{\Pi}^{(4)}$) are referred to as gyroviscosity \cite{Braginskii_1965} or the viscous Hall effect \cite{Lingam_2015}. The factors of 4 and 2 are retained so that, in the strongly magnetised limit, $\nu_{\perp2}=\nu_{\perp1}$ and $\nu_{\wedge2}=\nu_{\wedge1}$ \cite{Braginskii_1965}.

$\vec{q}_{J}$ is what we term the thermoelectric conduction. This is the heat flux due to a current and is given as equation 4.32 in \cite{Braginskii_1965}. The parallel and perpendicular parts are the Peltier effect along and across the magnetic field respectively, while the wedge term is the Ettingshausen effect \cite{Callen_1948}
\begin{subequations}
\label{therm-elec-cond}
\begin{equation}
    \vec{q}_{J} = x_{\text{ion}}(\vec{q}_{J,\parallel} + \vec{q}_{J,\perp} + \vec{q}_{J,\wedge}), \qquad
\vec{q}_{J,\parallel} = -\kappa_{J,\parallel}(\hat{b}\cdot\vec{J})\hat{b}, \qquad
\vec{q}_{J,\perp} = -\kappa_{J,\perp}(\vec{J}-\hat{b}(\hat{b}\cdot\vec{J})), \qquad
\vec{q}_{J,\wedge} = -\kappa_{J,\wedge}\hat{b}\times\vec{J},
\end{equation}
\end{subequations}
where the $\kappa_{J,*}$ are $\frac{T_{e}}{e}\beta^{Tu}$, in the Braginskii description.

$\vec{F}_{T}$ is what we term the thermoelectric force. This is the electric field induced by a temperature gradient. It is given as equation 4.31 in \cite{Braginskii_1965}
\begin{subequations}
    \label{therm-elec-force}
\begin{equation}
\vec{F}_{T} =  \frac{x_{\text{ion}}}{P/\rho}(\vec{F}_{T,\parallel} + \vec{F}_{T,\perp} + \vec{F}_{T,\wedge}),
\end{equation}
\begin{equation}
      \vec{F}_{T,\parallel} = - \kappa_{J,\parallel}\hat{b}\left(\hat{b}\cdot\vec{\nabla}[P/\rho]\right),\qquad  \vec{F}_{T,\perp} = - \kappa_{J,\perp}\left(\vec{\nabla}[P/\rho] - \hat{b}\left(\hat{b}\cdot\vec{\nabla}[P/\rho]\right)\right), \qquad
    \vec{F}_{T,\wedge} = -\kappa_{J,\wedge}\hat{b}\times\vec{\nabla}[P/\rho].
\end{equation}
\end{subequations}
The factors $\frac{1}{P/\rho} \kappa_{J,*}$ are $\frac{1}{e}\beta^{uT}$ in the  Braginskii description. They form an Onsager--Casimir  pair with the $\vec{q}_{J}$ coefficients \cite{Onsager_1931, Casimir_1945}. The $\vec{F}_{T}$ terms are formulated this way to allow for $\kappa_{J,*}$ to be passed to both $\vec{F}_{T}$ and $\vec{q}_{J}$ with one parameter while remaining dimensionally consistent. The parallel and perpendicular parts are the Seebeck effect along and across the magnetic field respectively, while the wedge term is the Nernst effect \cite{Callen_1948}.

The $\kappa_{J,*}$ coefficients are physically functions of temperature and, through the electron Hall parameter $x_e=\omega_e\tau_e\propto BT_e^{3/2}/n_e$, of density and field strength \cite{Braginskii_1965}. The Thomson effect of \cite{Callen_1948} therefore need not be added separately: it arises automatically from $\vec{\nabla}\cdot\vec{q}_{J}$ and $\vec{J}\cdot\vec{F}_{T}$ once $\kappa_{J,*}$ is evaluated locally.

The thermal conduction (neutral and anisotropic),  kinematic (neutral and Braginskii) viscosity, and Hall effect coefficients are in dimensions of diffusivity L$^{2}$T$^{-1}$, the same as Athena++'s isotropic conduction and viscosity, and Ohmic resistivity (their ambipolar diffusion coefficient is in terms of $[B]$L$^{2}$$T^{-1}$, which we have opted not to modify for continuity). The bulk viscosity $\zeta$ is supplied as a dynamic viscosity with dimensions of ML$^{-1}$T$^{-1}$, since the solver does not multiply it by the density. The thermoelectric coefficients are in dimensions of electric current L$^{3/2}$M$^{1/2}$T$^{-2}$, for example $\kappa_{J,\parallel} = 0.71\frac{k_{\mathrm{B}}c}{e\sqrt{4\pi}}T = 0.71\frac{\tilde{m}c}{e\sqrt{4\pi}}\frac{P}{\rho}$ in code units; where $\bar{m}=\mu m$, where $\bar{m}$ is the average particle mass, $\mu$ is the mean molecular weight, and $m$ is the mass of a particle. When no neutral component is enabled the gas is treated as fully ionised, $x_{\text{ion}}=1$, and the ionised terms are not weighted. Either thermoelectric half can be disabled for testing, but alone it breaks the Onsager--Casimir symmetry.
\newpage
\section{CFL Conditions for Non-Ideal Terms}
Here we perform stability analysis on the new physics implemented. Throughout we write the imaginary unit as $\sqrt{-1}$ to avoid confusion with cell indexing.

The timestep limits for the non-ideal terms all follow from a von Neumann analysis of the semi-discrete equations, linearised about a background state with coefficients frozen at their local values \cite{Von-Neumann_1950}. Each term is analysed in one dimension with wavevector $\vec{k}=(k_{x},0,0)$, and the magnetic unit vector is parametrised as
\begin{equation}
    \hat{b}=(\cos\theta,\sin\theta\cos\phi, \sin\theta\sin\phi), \qquad u\equiv\cos^{2}\theta\in[0,1],
\end{equation}
%so that $\theta$ is the angle between $\hat{b}$ and $\vec{k}$. % With $\vec{k}$ along $x$, the linearised operators are invariant under rotations about $\vec{k}$, so their eigenvalues are independent of $\phi$ and we set $\phi=0$ without loss of generality. 

Substituting the spatial part of the Fourier mode ansatz $\vec{U}_{i}^{n}=\tilde{\vec{U}}\exp(n\sigma\Delta t)\exp(\sqrt{-1}k_{x}i\Delta x)$ \cite{Von-Neumann_1950} into the semi-discrete equations replaces the spatial operator by its symbol $\hat{L}(k,u)$: $d\vec{U}_{i}/dt=\hat{L}(k,u)\vec{U}_{i}$. The symbol $\hat{L}$ is assembled from those of the central differences
\begin{equation}
    \frac{U_{i+1}-2U_{i}+U_{i-1}}{\Delta x^{2}}\rightarrow \hat{L}=-\frac{4}{\Delta x^{2}}\sin^{2}\left(\frac{k_{x}\Delta x}{2}\right), \qquad \frac{U_{i+1}-U_{i-1}}{2\Delta x}\rightarrow \hat{L}=\frac{\sqrt{-1}}{\Delta x}\sin(k_{x}\Delta x),
\end{equation}
whose magnitudes peak at $4/\Delta x^{2}$ for $k_{x}\Delta x=\pi$ and at $1/\Delta x$ for $k_{x}\Delta x=\pi/2$ respectively.

An explicit integrator advances each mode as $\vec{U}^{n+1}_{i}=R(\Delta t\hat{L})\vec{U}^{n}_{i}$, where $R(Z)$ is its stability function. For an $s$-stage Runge--Kutta method of order $s\leq4$, $R(Z)=\sum_{m=0}^{s}Z^{m}/m!$ \cite{Reddy_1992}. For an eigenvector of $\hat{L}$ with eigenvalue $\lambda(k,u)$, the temporal part of the ansatz then gives the amplification factor
\begin{equation}\label{eqn:amplification}
    \exp(\sigma\Delta t)=R(Z), \qquad Z\equiv\lambda\Delta t,
\end{equation}
so stability requires $|R(Z)|\leq1$ for every eigenvalue of $\hat{L}$.
The operators can be grouped into two main categories: those with negative real eigenvalues, which are dissipative; and those with purely imaginary eigenvalues, which are non-dissipative. The non-dissipative terms split into two sub-categories: those which are advective, scaling $1/\Delta x$ at a fixed coefficient; or dispersive, scaling $1/\Delta x^{2}$. These require
\begin{equation}
    \Delta t\leq\frac{r_{\text{R}}}{\max(|\lambda(k,u)|)}\quad\text{(dissipative)}, \qquad \Delta t\leq\frac{r_{\text{I}}}{\max(|\lambda(k,u)|)}\quad\text{(non-dissipative)},
\end{equation}
where $r_{\text{R}}$ and $r_{\text{I}}$ are the extents of the stability region along the negative real and imaginary axes respectively. 
\begin{table}[H]
    \centering
    \begin{tabular}{lccc}
        \hline
        Integrator & $R(Z)$ & $r_{\text{R}}$ & $r_{\text{I}}$ \\
        \hline
        RK1 & $1+Z$ & $2$ & $0$ \\
        RK2, VL2 & $1+Z+Z^{2}/2$ & $2$ & $0$ \\
        RK3 & $1+Z+Z^{2}/2+Z^{3}/6$ & $\approx2.51$ & $\sqrt{3}$ \\
        RK4 & $1+Z+Z^{2}/2+Z^{3}/6 + Z^{4}/24$ & $\approx 2.79$ & $2\sqrt{2}$ \\
        RKL1 & Legendre & $s^{2}+s$ & -- \\
        RKL2 & Legendre & $(s^{2}+s-2)/2$ & -- \\
        \hline
    \end{tabular}
    \caption{Stability functions and the extents of their stability regions along the negative real ($r_{\text{R}}$) and imaginary ($r_{\text{I}}$) axes. RKL1 and RKL2 are the $s$-stage super-time-stepping schemes of \cite{Meyer_2014}.}
    \label{tab:stability}
\end{table}
\noindent We take $r_{\text{R}}=2$, which is valid for every integrator in Table~\ref{tab:stability}. RK1 and RK2 have $r_{\text{I}}=0$, therefore non-dissipative terms require at least a third-order integrator. We take the RK3 value of $r_{\text{I}}=\sqrt{3}$. Over orientation, the maximum of $|\lambda|$ lies at $u=0$, at $u=1$, or at an interior critical point $u^{*}$ where $\partial_{u}|\lambda|=0$, which is included only if $u^{*}\in(0,1)$.

In $N_{\text{Dim}}$ dimensions the symbols of the central differences in each direction add \cite[pp.~63--64]{Morton_2005}. For a Laplacian on a uniform grid this gives $\max|\lambda_{ND}| = N_{\text{Dim}}\max|\lambda_{1D}|$ exactly.

We do not include numerical dissipation from the Riemann solver in this analysis as numerical dissipation shifts imaginary eigenvalues into the left half-plane (negative real) and therefore omitting numerical dissipation is conservative for the timestep bound.
\subsection{Thermal Conduction}
For constant $\rho$ and $\hat{b}$, with $P=(\gamma-1)\varepsilon_{\text{int}}$, the conduction term in equation~(\ref{eqn:conduction}) becomes
\begin{equation}
    \partial_{t}\varepsilon_{\text{int}} = (\gamma-1)\vec{\nabla}\cdot\left[\kappa_{0}\vec{\nabla}[\varepsilon_{\text{int}}] + \Delta\kappa\left(\hat{b}\cdot\vec{\nabla}[\varepsilon_{\text{int}}]\right)\hat{b} + \kappa_{\text{w}}\hat{b}\times\vec{\nabla}[\varepsilon_{\text{int}}]\right],
\end{equation}
with $\kappa_{0} = (1-x_{\text{ion}})\kappa_{\text{n}} + x_{\text{ion}}\kappa_{\perp}$, $\Delta\kappa = x_{\text{ion}}(\kappa_{\parallel}-\kappa_{\perp})$, and $\kappa_{\text{w}} = x_{\text{ion}}\kappa_{\wedge}$. Since $\vec{\nabla}\cdot(\hat{b}\times\vec{\nabla}[\varepsilon_{\text{int}}]) = (\vec{\nabla}\times\hat{b})\cdot\vec{\nabla}[\varepsilon_{\text{int}}]$, the $\kappa_{\text{w}}$ term is purely advective and vanishes for uniform $\hat{b}$; it is treated in Section~\ref{sec:cond_wedge}.

In 1D this reduces to $\partial_{t} \varepsilon_{\text{int}} = (\gamma-1)(\kappa_{0}+\Delta\kappa u)\partial_{xx}\varepsilon_{\text{int}}$, whose eigenvalue at the binding mode ($k_{x}\Delta x = \pi$) is $\lambda=-4(\gamma-1)(\kappa_{0}+\Delta\kappa u)/\Delta x^{2}$. As $\kappa_{0}+\Delta\kappa u$ is linear in $u$, its maximum over orientations lies at $u=0$ or $u=1$, and $|R(Z)_{\text{RK1}}| \leq 1$ gives the timestep as
\begin{equation}
    \Delta t_{\kappa,\parallel\perp} = \text{CFL}\frac{\min(\Delta x)^{2}}{2N_{\text{Dim}}(\gamma-1)((1-x_{\text{ion}})\kappa_{n} + x_{\text{ion}}\max(\kappa_{\parallel},\kappa_{\perp}))}.
\end{equation}
This reduces to the standard isotropic result when $\Delta\kappa=0$.
\subsection{Righi--Leduc (Thermal Hall) Effect}\label{sec:cond_wedge}
Returning to Eq.~(\ref{eqn:conduction}), considering the wedge term of the thermal conduction on its own, and allowing the magnetic field unit vector to vary freely in three dimensions, the one-dimensional operator is
\begin{equation}
    \partial_{t}\varepsilon_{\text{int}} = \kappa_{\text{w}}(\gamma-1)(\vec{\nabla} \times \hat{b})_{x}\partial_{x}\varepsilon_{\text{int}}.
\end{equation}
Unlike the parallel and perpendicular conduction, which contain second order derivatives, this is a first-order advective operator, with an effective phase speed set by the curl of the magnetic unit vector. In our flux implementation in Athena++ the transverse temperature gradients are formed with second-order central differences, so the effective discretisation of $\partial_{x}\varepsilon_{\text{int}}$ is centred, and the $x$-component of the curl of the magnetic unit vector enters as a coefficient. Taking the worst case, in which the transverse components of $\hat{b}$ reverse sign across a single cell so that each one-sided difference is at most 2, this coefficient is bounded by
\begin{equation}
  |(\vec{\nabla} \times \hat{b})_{x}|\leq \frac{2}{\Delta y} + \frac{2}{\Delta z} = \frac{2}{\Delta_{\perp}},
  \qquad
  \Delta_{\perp} = \left(\frac{1}{\Delta y} + \frac{1}{\Delta z}\right)^{-1}.
\end{equation}
The stability parameter, $Z_{\kappa,\wedge} = 2\sqrt{-1}\kappa_{\text{w}}(\gamma-1)\sin(k_{x}\Delta x)\Delta t/(\Delta x \Delta_{\perp})$, is purely imaginary, so it requires an integrator with $r_{\text{I}}>0$, such as RK3 or RK4. Requiring $|R(Z_{\kappa,\wedge})|^{2}\leq1$ gives the RK3 condition $\Delta t_{\kappa,\wedge} = \text{CFL}\sqrt{3}\Delta x \Delta_{\perp}/(2(\gamma-1)x_{\text{ion}}\kappa_{\wedge})$ in 1D. In higher dimensions, each active direction $d$ contributes with its own transverse width $\Delta_{\perp,d}=(\sum_{i\neq d}\Delta x_{i}^{-1})^{-1}$, where $i$ runs over all three directions, giving
\begin{equation}
\Delta t_{\kappa,\wedge} = \text{CFL}\frac{\sqrt{3}}{2(\gamma-1)x_{\text{ion}}\kappa_{\wedge}\sum_{d}\left(\Delta x_{d}\Delta_{\perp,d}\right)^{-1}}.
\end{equation}
On a uniform Cartesian grid, $\Delta_{\perp,d}=\Delta x/2$ in every direction and this becomes
\begin{equation}
\Delta t_{\kappa,\wedge,xyz} = \text{CFL}\frac{\min(\Delta x)^{2}}{\frac{4}{\sqrt{3}}(\gamma-1)N_{\text{Dim}}x_{\text{ion}}\kappa_{\wedge}}.
\end{equation}
If STS is disabled (unsplit thermal conduction), then we simply take the inverse sum of both timesteps as a worst-case bound
\begin{equation}
   \Delta t_{\kappa} = (1/\Delta t_{\kappa,\parallel,\perp} + 1/\Delta t_{\kappa,\wedge})^{-1}. 
\end{equation}
\subsection{Magnetic Diffusion}
Considering just the dissipative parts of the non-ideal magnetic terms (Ohmic and ambipolar)
\begin{equation}
    \partial_{t}\vec{B} = \eta\vec{\nabla}^{2}[\vec{B}] + \eta_{AD}\left(\vec{\nabla}^{2}[\vec{B}] + \vec{\nabla}[\hat{b}\cdot(\vec{\nabla}\times\vec{B})]\times\hat{b}\right).
\end{equation}
Assuming $\hat{b}$ is frozen gives
\begin{equation}
\partial_{t}\begin{pmatrix}
    B_{y} \\
    B_{z}
\end{pmatrix} = \begin{pmatrix}
   (\eta + (1-b_{z}^{2})\eta_{AD}) & b_{y}b_{z}\eta_{AD} \\
    b_{y}b_{z}\eta_{AD} & (\eta + (1-b_{y}^{2})\eta_{AD})
\end{pmatrix} \cdot  \partial_{xx}\begin{pmatrix}
    B_{y} \\
    B_{z}
\end{pmatrix}.
\end{equation}
Taking the Fourier space ansatz at the binding mode $k_{x}\Delta x = \pi$, each eigenvector of the diffusivity matrix, with eigenvalue $\lambda_{\eta,\parallel,\perp}$, has the stability parameter $Z_{\eta,\parallel\perp}=-4\Delta t\lambda_{\eta,\parallel,\perp}/\Delta x^{2}$. The eigenvalues of the diffusivity matrix are
\begin{equation}
        \lambda_{\eta,\parallel,\perp,0} = \eta + \eta_{AD}, \qquad
        \lambda_{\eta,\parallel,\perp1} = \eta + \eta_{AD}(1-b_{y}^{2} - b_{z}^{2}).
\end{equation}
As $1-b_{y}^{2} - b_{z}^{2} = b_{x}^{2} = u$, and the maximum value for $u$ is 1; $\lambda_{\eta,\parallel,\perp,\max} = \lambda_{\eta,\parallel,\perp,0} = \lambda_{\eta,\parallel,\perp1}(1)$. We can therefore use $\lambda_{\eta,\parallel,\perp,\max}=\eta+\eta_{AD}$. Forward Euler stability, $|R(Z_{\eta,\parallel\perp})_{\text{RK1}}|^{2}\leq1$, then gives the CFL condition
\begin{equation}
       \Delta t_{\eta,\parallel\perp} = \text{CFL}\frac{\min(\Delta x)^{2}}{2N_{\text{Dim}}(\eta+\eta_{AD})}.
\end{equation}
\subsection{Hall Effect}
Considering the Hall term on its own, $\partial_{t}\vec{B} = -\vec{\nabla}\times\eta_{H}(\vec{J}\times\hat{b})$, with a constant magnetic field unit vector, and $\vec{B}$ only changing in the $x$-direction, this reduces to
\begin{equation}
    \partial_{t}\begin{pmatrix}
        B_{y},B_{z}
    \end{pmatrix}^{T}
    = \eta_{H}b_{x}\partial_{xx}\begin{pmatrix}
        B_{z},-B_{y}
    \end{pmatrix}^{T}.
\end{equation}
In Fourier space the eigenvalues of the coefficient matrix are $\lambda_{\eta,\wedge,\pm} = \pm \sqrt{-1} \eta_{H}b_{x}$, so the stability parameter is $Z_{\eta,\wedge} = \pm\sqrt{-1}\eta_{H}b_{x}4\Delta t\sin^{2}(k_{x}\Delta x/2)/\Delta x^{2}$, maximised at the binding mode $k_{x}\Delta x = \pi$, where the two branches are complex conjugates and give the same $|R(Z_{\eta,\wedge})|$. Solving with the RK3 stability polynomial the CFL condition becomes
\begin{equation}
  \Delta t_{\eta,\wedge} = \text{CFL} \frac{\min(\Delta x)^{2}}{\frac{4}{\sqrt{3}}N_{\text{Dim}} \eta_{H}}.
\end{equation}
In 1D this is in agreement with \cite{kunz_2013}. This is conservative by a factor $\sqrt{N_{\text{Dim}}}$ relative to the bound of \cite{iwasaki_2025} for the edge-centred Hall EMF in Athena++.
\subsection{Unsplit Magnetic Diffusion}
Summing the effects of Ohmic resistivity, ambipolar diffusion, and the Hall effect, the matrix equation becomes
\begin{equation}
    \frac{d}{dt}\begin{pmatrix}
        \tilde{B}_{y} \\
        \tilde{B}_{z}
    \end{pmatrix} = 
\begin{pmatrix}- b_{z}^{2} \eta_{AD} + \eta + \eta_{AD} & b_{x} \eta_{H} + b_{y} b_{z} \eta_{AD} \\
-b_{x} \eta_{H} + b_{y} b_{z} \eta_{AD} & - b_{y}^{2} \eta_{AD} + \eta + \eta_{AD}\end{pmatrix}\cdot \frac{-4}{\Delta x^{2}}
    \begin{pmatrix}
         \tilde{B}_{y} \\
         \tilde{B}_{z}
    \end{pmatrix}.
\end{equation}
The eigenvalues of this are
\begin{equation}
 \lambda_{\eta,\pm}  = \eta + \frac{1}{2}\eta_{AD}(u+1) \pm \frac{1}{2}\sqrt{\eta_{AD}^{2}(u-1)^{2}-4\eta_{H}^{2}u}.
\end{equation}
This maximises at $u=1$, $\lambda_{\eta,\max} = \eta + \eta_{AD} \pm \sqrt{-1}\eta_{H}$. The positive real part makes the system conditionally stable under forward Euler, and solving $|R(Z_{\eta})|^{2} \leq 1$ for RK1 gives
\begin{equation}
    \Delta t_{\eta,\text{RK1}} = \text{CFL}\frac{\min(\Delta x)^{2}(\eta+\eta_{AD})}{2 N_{\text{Dim}}((\eta+\eta_{AD})^{2}+\eta_{H}^{2})}.
\end{equation}
In Appendix~E we show that this limit is too restrictive to be usable when the Hall term is significant, so an integrator with $r_{\text{I}}>0$ is functionally required. For complex $Z_{\eta}$, the RK3 condition $|R(Z_{\eta})|^{2}\leq1$ reduces to a quintic in $\Delta t$, so we instead take the conservative bound
\begin{equation}
    \Delta t_{\eta,\text{RK3}} = \text{CFL}\frac{\min(\Delta x)^{2}}{\frac{4}{\sqrt{3}}\sqrt{(\eta + \eta_{AD})^{2} + \eta_{H}^{2}}}.
\end{equation}
Since $\text{Re}(\lambda_{\eta})\geq0$ and $|\lambda_{\eta}|\leq\sqrt{(\eta+\eta_{AD})^{2}+\eta_{H}^{2}}$, this keeps every $Z_{\eta}$ in the left half-disk $|Z_{\eta}|\leq\sqrt{3}$. Because $R(Z)$ is a polynomial, by the maximum modulus principle it suffices that $|R|\leq1$ on the boundary of this half-disk, which holds for RK3. The bound is therefore sufficient, though not exact.

This is the limit applied when $\eta_{H}>0$; without the Hall term the unsplit limit reduces to $\Delta t_{\eta,\parallel,\perp}$. The forward-Euler limit $\Delta t_{\eta,\text{RK1}}$ is shown only to motivate the use of RK3 and is not used.
\subsection{Viscosity}\label{sec:visc_cfl}
The individual components of each tensor, and the scalars in the viscosity equations, are presented in Appendix~A. We take the perpendicular coefficient of $\boldsymbol{\Pi}^{(1)}$ to be $\nu_{\perp1}$ and that of $\boldsymbol{\Pi}^{(2)}$ to be $\nu_{\perp2}$; these are separate in the implementation, and equal in the strongly magnetised limit (Section~2). With only $x$-direction derivatives, and constant magnetic unit vector and density, the bulk, isotropic, parallel, and perpendicular parts of the viscosity reduce to
\begin{equation}
\partial_{t}\vec{v}=\boldsymbol{M}^{(\nu,\parallel,\perp)}\partial_{xx}\vec{v},
\end{equation}
with the real symmetric matrix $\boldsymbol{M}^{(\nu,\parallel,\perp)}$ given in Appendix~B, so that at the binding mode $k_{x}\Delta x=\pi$, $d\tilde{\vec{v}}/dt = -(4/\Delta x^{2})\boldsymbol{M}^{(\nu,\parallel,\perp)}\tilde{\vec{v}}$. Its entries at general $\phi$ are those at $\phi=0$ rotated by $\phi$ about $\hat{x}$, so every eigenvalue is independent of $\phi$ and depends on the orientation only through $u$. At $\phi=0$ every coupling to $v_{z}$ is proportional to $b_{z}=0$, so the matrix splits into an $xy$ block and $M_{zz}=\lambda_{\nu,\parallel,\perp,z}$, with
\begin{subequations}
 \begin{equation}
     \text{Tr}(\boldsymbol{M}^{(\nu,\parallel,\perp)}_{xy})= \frac{\zeta}{\rho} + \frac{7}{3}(1-x_{\text{ion}})\nu_{\text{neu}} + x_{\text{ion}}\left[ \nu_{\parallel}\left(u+{\frac{1}{3}}\right) + \nu_{\perp1}(1-u) + 4\nu_{\perp2} \right],
 \end{equation}
 \begin{align}
  \text{Det}(\boldsymbol{M}^{(\nu,\parallel,\perp)}_{xy}) =& \frac{\zeta}{\rho}(1-x_{\text{ion}})\nu_{\text{neu}} + \frac{4}{3}(1-x_{\text{ion}})^{2}\nu_{\text{neu}}^{2} \nonumber \\& + x_{\text{ion}}\Big[\frac{\zeta}{\rho}\Big[(3\nu_{\parallel}+\nu_{\perp1})u(1-u) + 4\nu_{\perp2}(2u-1)^{2}\Big] \nonumber \\& + (1-x_{\text{ion}})\nu_\text{neu}\Big[3\nu_{\parallel}(u-\frac{1}{3})^{2} + \nu_{\perp1}(1-u)^{2} + 16\nu_{\perp2}u(1-u)\Big] \\&  + \frac{4}{3}(1-x_{\text{ion}})\nu_{\text{neu}}\Big[(3\nu_{\parallel}+\nu_{\perp1})u(1-u) + 4\nu_{\perp2}(2u-1)^{2}\Big] \Big] \nonumber \\&  + \frac{4}{3}x_{\text{ion}}^{2}\Big[\nu_{\parallel}\nu_{\perp1}u(1-u) + \nu_{\parallel}\nu_{\perp2}(u+1)^{2} + 3\nu_{\perp1}\nu_{\perp2}(1-u)^{2}\Big] \nonumber,
 \end{align}
 \begin{align}
\lambda_{\nu,\parallel,\perp,z} = (1-x_{\text{ion}})\nu_{\text{neu}} + x_{\text{ion}} (\nu_{\perp1}(1-u) + 4\nu_{\perp2}u).
\end{align}
\end{subequations}
The $xy$-subsystem contains the largest eigenvalue of the entire system (Appendix~B). The eigenvalues of the $xy$-subsystem are
\begin{equation}
\lambda_{\nu \parallel,\perp,xy\pm}  = \frac{1}{2}\left[ \text{Tr}(\boldsymbol{M}^{(\nu,\parallel,\perp)}_{xy}) \pm \sqrt{\text{Tr}(\boldsymbol{M}^{(\nu,\parallel,\perp)}_{xy})^{2} - 4\text{Det}(\boldsymbol{M}^{(\nu,\parallel,\perp)}_{xy})} \right].
\end{equation}
Since $\boldsymbol{M}^{(\nu,\parallel,\perp)}$ is real and symmetric, its eigenvalues are real. For the system to be well-posed (dissipative) we require them to be positive, which for the $2\times2$ block holds if and only if $\text{Tr}(\boldsymbol{M}^{(\nu,\parallel,\perp)}_{xy})\geq0$ and $\text{Det}(\boldsymbol{M}^{(\nu,\parallel,\perp)}_{xy})\geq0$. For positive coefficients and $x_{\text{ion}},u\in[0,1]$, every term in both expressions is positive, so both conditions hold. We take the positive branch as the largest eigenvalue.

We now maximise $\lambda_{\nu,\parallel,\perp}$ to find the worst-case field orientation. Setting $d\lambda_{\nu,xy+}/du=0$ and squaring gives a polynomial condition which, as $\text{Tr}$ is linear and $\text{Det}$ quadratic in $u$, is a quadratic in $u$ with roots
\begin{subequations}
\begin{equation}
u^{*}_{1} = \frac{\nu_{\parallel} + \nu_{\perp1} - 8\nu_{\perp2}}{3\nu_{\parallel} + \nu_{\perp1} - 16\nu_{\perp2}},
\end{equation}
\begin{equation}
u^{*}_{2} = \frac{2(1-x_{\text{ion}})\nu_{\text{neu}} + x_{\text{ion}}(3\nu_{\perp1}-\nu_{\parallel}) + 6\zeta/\rho}{4(1-x_{\text{ion}})\nu_{\text{neu}} + x_{\text{ion}}(\nu_{\parallel}+3\nu_{\perp1}) + 12\zeta/\rho}.
\end{equation}
\end{subequations}
Squaring admits critical points of $\lambda_{\nu,xy-}$ as well as $\lambda_{\nu,xy+}$, but evaluating $\lambda_{\nu,xy+}$ at such a point cannot exceed its maximum, so including both roots is safe. $\lambda_{\nu,xy+}$ may have corners where the two eigenvalues cross, but as the larger of two analytic branches these cannot be local maxima. The worst-case orientation is therefore
\begin{equation}
\lambda_{\nu,\max} = \max\left(\lambda_{\nu,xy+}(0),~\lambda_{\nu,xy+}(1),~ \boldsymbol{1}_{[0,1]}(u^{*}_{1})\lambda_{\nu,xy+}(u^{*}_{1}),~ \boldsymbol{1}_{[0,1]}(u^{*}_{2})\lambda_{\nu,xy+}(u^{*}_{2})\right).
\end{equation}
Solving for the RK1 stability polynomial gives the timestep
\begin{equation}
    \Delta t_{\zeta,\nu,\parallel,\perp} = \text{CFL}\frac{\min(\Delta x)^{2}}{2N_{\text{Dim}}\lambda_{\nu,\max}}.
\end{equation}
\subsection{Gyroviscosity (Viscous Hall Effect)}\label{sec:gyro_cfl}
We now consider the sum of two wedge terms of viscosity. Taking a constant magnetic field unit vector and only the $x$-direction derivatives, the sum of $\boldsymbol{\Pi}^{(3)}$ and $\boldsymbol{\Pi}^{(4)}$ reduces to
\begin{equation}
\partial_{t}\begin{pmatrix}
    v_{x} \\ v_{y} \\ v_{z}
\end{pmatrix} =
\begin{pmatrix}
0 & M_{xy}^{(\nu,\wedge)} & M_{xz}^{(\nu,\wedge)} \\
-M_{xy}^{(\nu,\wedge)} & 0 & M_{yz}^{(\nu,\wedge)} \\
-M_{xz}^{(\nu,\wedge)} & -M_{yz}^{(\nu,\wedge)} & 0
\end{pmatrix}\cdot \partial_{xx}\begin{pmatrix} v_{x} \\ v_{y} \\ v_{z} \end{pmatrix},
\label{eq:gyrovisc-collected-general}
\end{equation}
with
\begin{subequations}
\begin{align}
M_{xy}^{(\nu,\wedge)} &= -M_{yx}^{(\nu,\wedge)} = x_{\text{ion}}b_{z}\left[\nu_{\wedge1}(1-u)+4\nu_{\wedge2}u\right], \\
M_{xz}^{(\nu,\wedge)} &= -M_{zx}^{(\nu,\wedge)} =  -x_{\text{ion}}b_{y}\left[\nu_{\wedge1}(1-u)+4\nu_{\wedge2}u\right], \\
M_{yz}^{(\nu,\wedge)} &= -M_{zy}^{(\nu,\wedge)} = x_{\text{ion}}b_{x}\left[\nu_{\wedge1}(1-u)+2\nu_{\wedge2}(2u-1)\right].
\end{align}
\end{subequations}
The matrix $\boldsymbol{M}^{(\nu,\wedge)}$ is skew-symmetric and, therefore, its eigenvalues are purely imaginary. Eliminating the $\phi$ dependence in the same manner as viscosity, the eigenvalues of the system at the binding mode are
\begin{equation}
    \lambda_{\nu,\wedge,z} = 0, \qquad
 \lambda_{\nu,\wedge,xy,\pm} = \pm x_{\text{ion}}\sqrt{-1}\sqrt{(\nu_{\wedge1}(1-u)+2\nu_{\wedge2}u)^{2} + 4\nu_{\wedge2}^{2}u(1-u)}.
\end{equation}
The magnitude of the positive branch has a single critical point
\begin{equation}
u^{*} = \frac{\nu_{\wedge1}^{2}-2\nu_{\wedge1}\nu_{\wedge2}-2\nu_{\wedge2}^{2}}
{\nu_{\wedge1}\left(\nu_{\wedge1}-4\nu_{\wedge2}\right)}, \qquad
    |\lambda_{\nu,\wedge+}(u^{*})| = \left|2x_{\text{ion}}\sqrt{\frac{\nu_{\wedge2}^{3}\left(2\nu_{\wedge1}+\nu_{\wedge2}\right)}{\nu_{\wedge1}\left(4\nu_{\wedge2}-\nu_{\wedge1}\right)}}\right|.
\end{equation}
As $u~\in~[0,1]$, if $u^{*}$ is outside this range, we must also consider the end points of the interval, $|\lambda_{\nu,\wedge+}(0)| = x_{\text{ion}}\nu_{\wedge1}$ and $|\lambda_{\nu,\wedge+}(1)| = 2x_{\text{ion}}\nu_{\wedge2}$.
Therefore we take our maximum eigenvalue to be
\begin{equation}
|\lambda_{\nu,\wedge+,~\max}|
= \max\left(
 x_{\text{ion}}\nu_{\wedge1},~
2x_{\text{ion}}\nu_{\wedge2},~
\boldsymbol{1}_{[0,1]}(u^{*}) \left|2x_{\text{ion}}\sqrt{\frac{\nu_{\wedge2}^{3}\left(2\nu_{\wedge1}+\nu_{\wedge2}\right)}{\nu_{\wedge1}\left(4\nu_{\wedge2}-\nu_{\wedge1}\right)}}\right|
\right).
\end{equation}
The stability parameter, $Z_{\nu,\wedge} = \pm\sqrt{-1}4\Delta t|\lambda_{\nu,\wedge}|/\Delta x^{2}$, is purely imaginary, and solving the RK3 stability polynomial gives
\begin{equation}
    \Delta t_{\nu,\wedge} = \text{CFL} \frac{\min(\Delta x)^{2}}{\frac{4}{\sqrt{3}}N_{\text{Dim}}|\lambda_{\nu,\wedge+,~\max}|}.
\end{equation}
If STS is disabled (unsplit viscosity), the operator is the sum of the dissipative and non-dissipative viscosity matrices, $\boldsymbol{M}^{(\nu,\parallel,\perp)}+\boldsymbol{M}^{(\nu,\wedge)}$. We find that this matrix can only be solved analytically using the full cubic formula. Therefore, with STS disabled, we conservatively bound the timestep by the inverse sum of the dissipative and non-dissipative timesteps
\begin{equation}
    \Delta t_{\nu} = \left( \frac{1}{\Delta t_{\zeta,\nu,\parallel,\perp}} + \frac{1}{\Delta t_{\nu,\wedge}}\right)^{-1}.
\end{equation}
In Appendix~E we compare this to numerically calculated timesteps with physical diffusion coefficients at RK3 time integration.
\subsection{Isolated Thermoelectric Effects}\label{sec:te_iso}
Thermoelectric effects are given by $\vec{q}_{J}$ and $\vec{F}_{T}$, and must both be active to form an Onsager--Casimir pair. Converting the total energy equation to internal energy ($\partial_{t}\varepsilon_{\text{int}} = \partial_{t}\varepsilon - \vec{B}\cdot\partial_{t}\vec{B}$) gives $\partial_{t}\varepsilon = \vec{\nabla}\cdot\vec{q}_{j} + \vec{J}\cdot\vec{F}_{T}$, where $\vec{J}\cdot\vec{F}_{T}$ is the second order Joule heating term from the Seebeck and Nernst effects. These describe the local conversion of magnetic energy to internal energy, rather than the flux of energy.

By linearising, only $\vec{q}_{J}$ remains in the energy equation. Taking a frozen magnetic unit vector, and the $\kappa_{J,*}/(P/\rho)$ terms as a constant (so that the induction equation only contains $\vec{F}_{T}$), the linearised system in Fourier space at the binding mode $k_{x}\Delta x = \pi$ is \begin{equation}
 \frac{d}{dt}(\tilde{\varepsilon}_{\text{int}},\tilde{B}_{y},\tilde{B}_{z})^{T} = -(4/\Delta x^{2})\boldsymbol{M}^{(J)}(\tilde{\varepsilon}_{\text{int}},\tilde{B}_{y},\tilde{B}_{z})^{T},
\end{equation} with
\begin{equation}
    \boldsymbol{M}^{(J)} = \begin{pmatrix}
        0 & \kappa_{J,w}b_{y}-\Delta\kappa_{J}b_{x}b_{z} & \kappa_{J,w}b_{z}+\Delta\kappa_{J}b_{x}b_{y} \\
        \dfrac{\kappa_{J,w}b_{y}+\Delta\kappa_{J}b_{x}b_{z}}{\varepsilon_{\text{int}}} & 0 & 0 \\
        \dfrac{\kappa_{J,w}b_{z}-\Delta\kappa_{J}b_{x}b_{y}}{\varepsilon_{\text{int}}} & 0 & 0
    \end{pmatrix}.
\end{equation}
Where $\Delta \kappa_{J} = x_{\text{ion}}(\kappa_{J,\perp} - \kappa_{J,\parallel})$, and $\kappa_{J,w} = x_{\text{ion}}\kappa_{J,\wedge}$. Each eigenvector of $\boldsymbol{M}^{(J)}$, with eigenvalue $\lambda_{\text{TE}}$, has the stability parameter $Z_{\text{TE}} = -4\Delta t\lambda_{\text{TE}}/\Delta x^{2}$, and the characteristic polynomial is \begin{equation}
-\lambda_{\text{TE}}(\lambda_{\text{TE}}^{2} - M^{(J)}_{\varepsilon y}M^{(J)}_{y\varepsilon} - M^{(J)}_{\varepsilon z}M^{(J)}_{z\varepsilon}) = 0,
\end{equation} where both products are differences of squares, e.g.\ $M^{(J)}_{\varepsilon y}M^{(J)}_{y\varepsilon}=(\kappa_{J,w}^{2}b_{y}^{2}-\Delta\kappa_{J}^{2}b_{x}^{2}b_{z}^{2})/\varepsilon_{\text{int}}$. Their sum depends on $b_{y}$ and $b_{z}$ only through $b_{y}^{2}+b_{z}^{2}=1-u$, the $\phi$ dependence cancels, and the eigenvalues are
\begin{equation}
    \lambda_{\text{TE},0}=0, \qquad
\lambda_{\text{TE},\pm} = \pm\frac{1}{\sqrt{\varepsilon_{\text{int}}}}\sqrt{\left(\kappa_{J,w}^{2} - \Delta\kappa_J^{2}u\right)(1-u)}.
\end{equation}
The radicand is quadratic in $u$ with its stationary point at
\begin{equation}
 u^{*}_{1} = \frac{\Delta\kappa_{J}^{2} + \kappa_{J,w}^{2}}{2 \Delta\kappa_{J}^{2}},
\end{equation}
and it vanishes at $u = 1$ and $u = \kappa_{J,w}^{2}/\Delta\kappa_{J}^{2}$, where the derivative of $\lambda_{\text{TE},\pm}$ is singular. These are branch points at which the eigenvalues change from real to imaginary. Evaluating $\lambda_{\text{TE},+}$ at the stationary point, the endpoints, and the branch points gives $\lambda_{\text{TE},+}(0) = \kappa_{J,w}/\sqrt{\varepsilon_{\text{int}}}$, $\lambda_{\text{TE},+}(u^{*}_{1}) =  \sqrt{-1}|\Delta\kappa_{J}^{2} - \kappa_{J,w}^{2}|/(2|\Delta\kappa_{J}|\sqrt{\varepsilon_{\text{int}}})$, and $\lambda_{\text{TE},+}(1) = \lambda_{\text{TE},+}(\kappa_{J,w}^{2}/\Delta\kappa_{J}^{2}) = 0$. We take our maximum eigenvalue to be
\begin{equation}
|\lambda_{\text{TE},\max}| = \frac{1}{\sqrt{\varepsilon_{\text{int}}}}\max\left(\kappa_{J,w},~\boldsymbol{1}_{[0,1]}(u^{*}_{1})\frac{\Delta\kappa_{J}^{2}-\kappa_{J,w}^{2}}{2|\Delta\kappa_{J}|}\right),
\end{equation}
with $\Delta\kappa_{J}$ and $\kappa_{J,w}$ as defined above.
If $\kappa_{J,\wedge}=0$ then the eigenvalue is always purely imaginary and conditionally stable to integrators with $r_{I}>0$. However, with $\kappa_{J,\wedge}>0$ the real negative branch of the eigenvalue forms a backwards heat equation and is ill-posed.

In a scenario where thermoelectric effects are modelled, we assume that a user likely has anisotropic thermal conduction and magnetic diffusion present. These terms will allow the combined system to be well-posed (which we will show in the next subsection). We take the isolated thermoelectric limit to be
\begin{equation}
    \Delta t_{\text{TE}} = \text{CFL}\frac{\min(\Delta x)^{2}}{\frac{4}{\sqrt{3}}N_{\text{Dim}}|\lambda_{\text{TE},\max}|},
\end{equation}
which is a valid stability limit only when $\kappa_{J,\wedge}=0$; with the Nernst and Ettingshausen effects enabled, the isolated system is unstable for any timestep.
\subsection{Well-Posed Criterion for Thermoelectric Effects} \label{sec:wellposed}
We now combine the full expressions for thermal conduction, magnetic diffusion, and thermoelectric effects into a single system. We will refer to this as the thermo-magnetic system for brevity. Its linearised operator $\boldsymbol{M}^{(\kappa,\eta,J)}$, acting on $(\tilde{\varepsilon}_{\text{int}},\tilde{B}_{y},\tilde{B}_{z})$ with $\psi=k_{x}\Delta x/2$, combines the conduction, magnetic-diffusion and thermoelectric operators of the preceding subsections, and is given in full in Appendix~B.
Deriving the CFL condition directly from this system, by solving the cubic and maximising over the three angles, is infeasible even with computer algebra. We instead find a sufficient criterion for the system to be well-posed. If every eigenvalue of $\boldsymbol{M}^{(\kappa,\eta,J)}$ has a negative real part, the system is well-posed; if any real part is positive, its magnitude scales as $\Delta t/\Delta x^{2}$, so the growth rate is unbounded as $\Delta x\to0$, the system is ill-posed, and no $\Delta t$ is stable for any integrator.

We make use of the eigenvalues of two similar matrices being invariant. For square matrices $\boldsymbol{A}$ and $\boldsymbol{B}$, and invertible $\boldsymbol{P}$ such that $\boldsymbol{B} = \boldsymbol{P}^{-1}\cdot\boldsymbol{A}\cdot\boldsymbol{P}$, $\boldsymbol{A}$ and $\boldsymbol{B}$ have the same eigenvalues \cite[chapter~6, theorem~12]{lax_2013}. By choosing $\boldsymbol{P} = \text{diag}(1,\varepsilon_{\text{int}}^{-1/2},\varepsilon_{\text{int}}^{-1/2})$, the $\kappa_{J,\wedge}$ entries of $\tilde{\boldsymbol{M}}^{(\kappa,\eta,J)} = \boldsymbol{P}^{-1}\cdot\boldsymbol{M}^{(\kappa,\eta,J)}\cdot\boldsymbol{P}$ become symmetric, and the $\Delta \kappa_{J}$ entries anti-symmetric, though analytically solving for the eigenvalue is still infeasible.

We now bound the real parts of the eigenvalues of $\tilde{\boldsymbol{M}}^{(\kappa,\eta,J)}$ by those of its Hermitian part \cite{Bendixson_1902}, 
\begin{equation}
  \boldsymbol{H}_{\tilde{M}} = \frac{1}{2}\left( \tilde{\boldsymbol{M}}^{(\kappa,\eta,J)} + \tilde{\boldsymbol{M}}^{(\kappa,\eta,J)\dagger} \right).  
\end{equation}
For a unit eigenvector $\vec{x}$ with eigenvalue $\lambda$, $\lambda=\vec{x}^{\dagger}\tilde{\boldsymbol{M}}^{(\kappa,\eta,J)}\vec{x}$, so $\text{Re}(\lambda)=\vec{x}^{\dagger}\boldsymbol{H}_{\tilde{M}}\vec{x}\in[\lambda_{\min}(\boldsymbol{H}_{\tilde{M}}),\lambda_{\max}(\boldsymbol{H}_{\tilde{M}})]$ for every eigenvalue. The same argument applies to every vector, so when $\boldsymbol{H}_{\tilde{M}}$ is negative semi-definite the quantity $|\tilde{\varepsilon}_{\text{int}}|^{2}+\varepsilon_{\text{int}}(|\tilde{B}_{y}|^{2}+|\tilde{B}_{z}|^{2})$ cannot grow, even though $\tilde{\boldsymbol{M}}^{(\kappa,\eta,J)}$ is not normal.

The Hermitian part of $\tilde{\boldsymbol{M}}^{(\kappa,\eta,J)}$ (with $M^{(\kappa,\eta,J)}_{T,\wedge}$ cancelling and maximising the bounding mode as $\sin^{2}(\psi)=1$) is given in Appendix~B. To eliminate the $\phi$ dependence we again make use of invariant eigenvalues, with the rotation matrix $\boldsymbol{Q}$, taking
\begin{equation}
 \boldsymbol{H}_{\tilde{F},\text{rot}} = \boldsymbol{Q}^{-1}\cdot \boldsymbol{H}_{\tilde{M}} \cdot \boldsymbol{Q} =  \frac{-4\Delta t}{\Delta x^{2}}\begin{pmatrix}(\gamma-1)(\Delta\kappa \cos^{2}{\left(\theta \right)} + \kappa_{0}) & \frac{\kappa_{J,w} \sin{\left(\theta \right)}}{\sqrt{\varepsilon_{int}}} & 0 \\
 \frac{\kappa_{J,w} \sin{\left(\theta \right)}}{\sqrt{\varepsilon_{int}}} & \eta + \eta_{AD} & 0\\0 & 0 & \eta + \eta_{AD} \cos^{2}{\left(\theta \right)}\end{pmatrix},
\end{equation}
we retrieve our criterion by solving $\lambda_{H_{\tilde{M}},\text{rot},\max} < 0$. The eigenvalues of this system now simplify greatly
\begin{subequations}
    \begin{equation}
    \lambda_{H_{\tilde{M}}1} = - 4\frac{ \Delta t}{\Delta x^{2}} \left(\eta + \eta_{AD} \cos^{2}{\left(\theta \right)}\right), 
    \end{equation}
    \begin{align}
       \lambda_{H_{\tilde{M}}2,3} &= - 2\frac{ \Delta t }{\Delta x^{2}}\left((\gamma-1)\Delta\kappa  \cos^{2}{\left(\theta \right)} +  \eta +  \eta_{AD} +  (\gamma-1)\kappa_{0} \right.\nonumber \\
       &\pm\left.\sqrt{\left((\gamma-1)\kappa_{0} + (\gamma-1)\Delta\kappa\cos^{2}(\theta) -\eta -\eta_{AD}\right)^{2} + 4 \frac{\kappa_{J,w}^{2}}{\varepsilon_{\text{int}}} \sin^{2}{\left(\theta \right)}}\right),
    \end{align}
\end{subequations}
with the negative sign for $\lambda_{H_{\tilde{M}}2}$. From these eigenvalues only $\lambda_{H_{\tilde{M}}3}$ can become positive.

We now construct an inequality between the negative and positive components of $\lambda_{H_{\tilde{M}}3}$
\begin{equation}
  (\gamma-1)\left(\kappa_{0}+ \Delta\kappa\cos^{2}{\left(\theta \right)}\right) +  \eta +  \eta_{AD}  > \sqrt{\left((\gamma-1)\kappa_{0} + (\gamma-1)\Delta\kappa\cos^{2}(\theta) -\eta -\eta_{AD}\right)^{2} + 4 \frac{\kappa_{J,w}^{2}}{\varepsilon_{\text{int}}} \sin^{2}{\left(\theta \right)}}.
\end{equation}
This is linear in $u$, and at $u=1$ it reduces to $(\gamma-1)(\kappa_{0}+\Delta\kappa)(\eta+\eta_{AD})\geq0$, which always holds, so the worst case is $u=0$:
\begin{equation}
    (\gamma-1)\kappa_{0}(\eta + \eta_{AD}) > \frac{\kappa_{J,w}^{2}}{\varepsilon_{\text{int}}}.
\end{equation}
Expanding to original coefficients gives
\begin{equation}
     (\gamma-1)((1-x_{\text{ion}})\kappa_{n} + x_{\text{ion}}\kappa_{\perp})(\eta + \eta_{AD}) > \frac{x_{\text{ion}}^{2}\kappa_{J\wedge}^{2}}{\varepsilon_{\text{int}}}.   
\end{equation}
This is a sufficient condition for well-posedness. When the Righi--Leduc, Hall, and Peltier/Seebeck terms are absent, $\tilde{\boldsymbol{M}}^{(\kappa,\eta,J)}$ is real and symmetric, $\boldsymbol{H}_{\tilde{M}}=\tilde{\boldsymbol{M}}^{(\kappa,\eta,J)}$, and the condition is also necessary. Physically, the product of the perpendicular thermal conduction and the magnetic diffusion must exceed the square of the Nernst/Ettingshausen effective diffusivity. In Appendix~E we test this criterion against physically determined coefficients and find that it always holds.

We take a conservative worst-case bound for the timestep using the inverse sum of each subsystem's timestep
\begin{equation}
    \Delta t_{\kappa,\eta,J} = \left( \frac{1}{\Delta t_{\kappa,\parallel\perp}} + \frac{1}{\Delta t_{\kappa,\wedge}}+\frac{1}{\Delta t_{{\eta},RK3}}  + \frac{1}{\Delta t_{\text{TE}}} \right)^{-1}.
\end{equation}
In Appendix~E we test this bound with RK3 time integration against physical diffusion coefficients.
\subsection{Dissipative Thermo-Magnetic System}
The Hermitian part of the scaled thermo-magnetic system keeps thermal conduction, magnetic diffusion and the Nernst/Ettingshausen coupling. After the scaling $\boldsymbol{P}$ it is symmetric, and it is exactly the Hermitian part $\boldsymbol{H}_{\tilde{M}}$ of Section~\ref{sec:wellposed}, so its stability parameters are the eigenvalues $\lambda_{H_{\tilde{M}}1,2,3}$ found there. These are real, so this part can be super-time-stepped (assuming well-posedness holds). The most negative is $\lambda_{H_{\tilde{M}}2}=-4\frac{\Delta t}{\Delta x^{2}}\lambda_{\text{TM,R},+}$, with
\begin{equation}
\lambda_{\text{TM,R},+} = \frac{1}{2}\left[(\gamma-1)(\kappa_{0}+\Delta\kappa u)+\eta+\eta_{AD}\right] + \sqrt{\frac{1}{4}\left[(\gamma-1)(\kappa_{0}+\Delta\kappa u)-\eta-\eta_{AD}\right]^{2} + \frac{\kappa_{J,w}^{2}}{\varepsilon_{\text{int}}}(1-u)}.
\end{equation}
Setting $d\lambda_{\text{TM,R},+}/du=0$ and squaring leaves a condition independent of $u$, so $\lambda_{\text{TM,R},+}$ has no interior critical point and is monotonic in $u$. Its maximum therefore lies at an endpoint, $\lambda_{\text{TM,R},+}(1) = \max\left((\gamma-1)(\kappa_{0}+\Delta\kappa),~\eta+\eta_{AD}\right)$ or
\begin{equation}
\lambda_{\text{TM,R},+}(0) = \frac{1}{2}\left[(\gamma-1)\kappa_{0}+\eta+\eta_{AD}\right] + \sqrt{\frac{1}{4}\left[(\gamma-1)\kappa_{0}-\eta-\eta_{AD}\right]^{2} + \frac{\kappa_{J,w}^{2}}{\varepsilon_{\text{int}}}}.
\end{equation}
The dissipative thermo-magnetic timestep is
\begin{equation}
\Delta t_{\text{TM,R}} = \text{CFL}\frac{\min(\Delta x)^{2}}{2N_{\text{Dim}}\max(\lambda_{\text{TM,R},+}(0),\lambda_{\text{TM,R},+}(1))}.
\end{equation}
\subsection{Non-Dissipative Thermo-Magnetic System}\label{sec:TM_I}
The skew-Hermitian part of the scaled thermo-magnetic system contains the Righi--Leduc, Hall, and Peltier/Seebeck terms. We first treat the Hall and Peltier/Seebeck terms, and combine them with the Righi--Leduc effect, which has a different Fourier factor, at the end. At the binding mode, scaling by $\boldsymbol{P}$ as in Section~\ref{sec:wellposed}, the Hall and Peltier/Seebeck part of the system is
\begin{equation}
\tilde{\boldsymbol{M}}^{(\text{HJ})} = -4\frac{\Delta t}{\Delta x^{2}}\cos\theta\begin{pmatrix}
0 & -\Delta\kappa_{J}\sin\theta\sin\phi/\sqrt{\varepsilon_{\text{int}}} & \Delta\kappa_{J}\sin\theta\cos\phi/\sqrt{\varepsilon_{\text{int}}} \\
\Delta\kappa_{J}\sin\theta\sin\phi/\sqrt{\varepsilon_{\text{int}}} & 0 & \eta_{H} \\
-\Delta\kappa_{J}\sin\theta\cos\phi/\sqrt{\varepsilon_{\text{int}}} & -\eta_{H} & 0
\end{pmatrix}.
\end{equation}
This is real and antisymmetric, so its eigenvalues are $0$ and $\pm\sqrt{-1}$ times the root of the sum of squares of its upper off-diagonal entries. Since $\sin\phi$ and $\cos\phi$ enter only as $\sin^{2}\phi+\cos^{2}\phi$, $\phi$ drops out, and, omitting the prefactor $-4\Delta t/\Delta x^{2}$,
\begin{equation}
\lambda_{\text{HJ},0} = 0, \qquad
\lambda_{\text{HJ},\pm} = \pm\sqrt{-1}\sqrt{u}\sqrt{\eta_{H}^{2} + (1-u)\frac{\Delta\kappa_{J}^{2}}{\varepsilon_{\text{int}}}}.
\end{equation}
The stability parameter $Z_{\text{HJ}}=-4\Delta t\lambda_{\text{HJ},\pm}/\Delta x^{2}$ is purely imaginary, so these terms require an integrator with $r_{\text{I}}>0$. The magnitude of the non-trivial eigenvalues is $|\lambda_{\text{HJ}}(u)| = \sqrt{u\eta_{H}^{2} + u(1-u)\Delta\kappa_{J}^{2}/\varepsilon_{\text{int}}}$, whose square is a concave quadratic in $u$ with derivative $\eta_{H}^{2} + (1-2u)\Delta\kappa_{J}^{2}/\varepsilon_{\text{int}}$. This vanishes at $u^{*} = 1/2 + \eta_{H}^{2}\varepsilon_{\text{int}}/(2\Delta\kappa_{J}^{2})$. At the endpoints $|\lambda_{\text{HJ}}(0)| = 0$ and $|\lambda_{\text{HJ}}(1)| = \eta_{H}$, so our final eigenvalue for the Hall Peltier/Seebeck regime is
\begin{equation}
    |\lambda_{\eta,J,H}| = \max\left(\eta_{H},\frac{1}{2}\boldsymbol{1}_{[0,1]}(u^{*})\left(\frac{\eta_{H}^{2}\sqrt{\varepsilon_{\text{int}}}}{|\Delta\kappa_{J}|} + \frac{|\Delta\kappa_{J}|}{\sqrt{\varepsilon_{\text{int}}}}\right)\right),
\end{equation}
where the second argument is $|\lambda_{\text{HJ}}(u^{*})|$.
Without the Hall term this reduces to $|\Delta\kappa_{J}|/(2\sqrt{\varepsilon_{\text{int}}})$, the isolated Peltier/Seebeck result, and without Peltier/Seebeck to $\eta_{H}$, the Hall result. The RK3 timestep for the non-dissipative thermo-magnetic system is
\begin{equation}
 \Delta t_{\text{HJ}} = \text{CFL}\frac{\min(\Delta x)^{2}}{\frac{4}{\sqrt{3}}N_{\text{Dim}}|\lambda_{\text{HJ},\max}|}, \qquad
    \Delta t_{\text{TM,I}} = \left(\frac{1}{\Delta t_{\kappa,\wedge}} + \frac{1}{\Delta t_{\text{HJ}}}\right)^{-1}.
\end{equation}
\subsection{Relationship with Ideal Timestep}
Linearising the ideal induction equation with the Hall term gives the whistler dispersion relation, whose phase speed at the grid scale can exceed the fast-magnetosonic speed $c_{f}$; the fastest signal is then the larger of the two (we will re-derive this in the next section). Since $c_{w}(k)=\frac{1}{2}\eta_{H}k+\sqrt{\frac{1}{4}\eta_{H}^{2}k^{2}+c_{A}^{2}}\leq c_{f}+\eta_{H}k$, the combined speed is bounded by the sum of the ideal and non-dissipative speeds, and the same holds for the other non-dissipative terms. We therefore combine the ideal limit $\Delta t_{\text{ideal}}$ with the non-ideal limits through inverse sums of their rates, which is conservative. This modifies the current Athena++ implementation which takes $\Delta t = \min(\Delta t_{\text{ideal}}, \Delta t_{\text{diff}})$.

After linearisation the viscous operator acts only on $\vec{v}$, and the thermo-magnetic operator only on $(\varepsilon_{\text{int}},\vec{B})$, since viscous and Joule heating are quadratic in the perturbation. The combined non-ideal operator is therefore block-diagonal, its eigenvalues are those of the two blocks, and the binding rate is the larger of the two blocks' rates.

When STS is enabled, the dissipative parts are super-time-stepped with the parabolic limit $\Delta t_{\text{par}}=\min(\Delta t_{\nu,\parallel\perp},\Delta t_{\text{TM,R}})$, and the hyperbolic update is limited by
\begin{equation}
\Delta t_{\text{hyp}} = \left(1/\Delta t_{\text{ideal}} + 1/\Delta t_{\kappa,\wedge} + \max(1/\Delta t_{\nu,\wedge} , 1/\Delta t_{\text{HJ}})  \right)^{-1}.
\end{equation}
When STS is disabled we take in the inverse sum of the ideal timestep, and the most restrictive unsplit timestep
\begin{equation}
    \Delta t_{\text{unsplit}} = \left(1/\Delta t_{\text{ideal}} + \max(1/\Delta t_{\zeta,\nu,\parallel,\perp} + 1/\Delta t _{\nu_{\wedge}} ,~1/\Delta t_{\kappa,\parallel,\perp} + 1/\Delta t_{\kappa,\wedge} + 1/\Delta t_{\eta} + 1/\Delta t_{TE}) \right)^{-1}.
\end{equation}
The ideal timestep is given by the standard result
\begin{equation}
    \Delta t_{\text{ideal}} = \text{CFL}\min_{i\in x,y,z}\frac{\Delta i}{N_{\text{dim}}(v_{i}+c_{f})}.
\end{equation}

\section{HLLE-Mod Riemann Solver}\label{sec:hlle_mod}
The Hall effect suffers from non-linear grid-scale instabilities even at RK3 time integration \cite{iwasaki_2025}. One solution to this is Hall-HLLE-Mod, whereby the fastest whistler wave (which scales $\propto 1/\Delta x$) replaces the fast-magnetosonic speed as the magnetic field phase speed if it exceeds it \cite{marchand_2019}. We therefore opt to implement the same logic for every other non-dissipative term and its appropriate variable in the HLLE Riemann solver.

Throughout we write $u = \cos^{2}\theta$ for the wave-normal parameter (as before), $E = \eta_{H}^{2}k_{x}^{2}$, $D = \frac{(\Delta\kappa_{J}k_{x})^{2}}{\varepsilon_{\text{int}}} = \frac{(\gamma-1)(\Delta\kappa_{J}k_{x})^{2}}{\rho_{0}c_{\text{iso}}^{2}}$, $c_{\text{iso}} = c_{s}/\sqrt{\gamma}$ for the isothermal sound speed, and $c_{f}^{2} = c_{a}^{2}+c_{s}^{2}$ for the fast magnetosonic speed at perpendicular propagation, which bounds it at all orientations. In each case, the dispersion relation is obtained by substituting $-\lambda^{2}/k_{x}^{2} = \Omega$, so that $\sqrt{\Omega}$ is a phase speed, and the characteristic polynomial is monic and cubic in $\Omega$. We choose $k_{x}=1/\Delta x$ following \cite{Lesur_2014}, which \cite{iwasaki_2025} found sufficient to suppress the Hall instability.
For each derivation we convert to the primitive variables $\vec{W} = (\rho,v_{x},v_{y},v_{z},\varepsilon_{\text{int}},B_{x},B_{y},B_{z})$, We linearise about a uniform state at rest, $(\rho_{0},\vec{0},\varepsilon_{0},\vec{B})$, and assume the perturbation of $\vec{W}$ has the form $\delta\vec{W} = \delta\tilde{\vec{W}}\exp(\sqrt{-1}(k_{x}x - \omega t))$.
Each cubic can be solved by Cardano's formula, but maximising its largest root over $u$ cannot, in general, be done in closed form. We instead bound the largest root using the following test: let $F(\Omega)=\Omega^{3}+A_{2}\Omega^{2}+A_{1}\Omega+A_{0}$ have real roots $0\leq r_{1}\leq r_{2}\leq r_{3}$, and let $\Lambda$ satisfy $F(\Lambda)\geq0$ and $2\Lambda\geq r_{1}+r_{2}+r_{3}=-A_{2}$. Then $r_{3}\leq\Lambda$, otherwise $F(\Lambda)\geq0$ with $\Lambda<r_{3}$ forces $r_{1}\leq\Lambda\leq r_{2}$, so $r_{2}+r_{3}>2\Lambda\geq r_{1}+r_{2}+r_{3}\geq r_{2}+r_{3}$, a contradiction. The roots are real and positive for every system considered here.
\subsection{Hall Effect (Whistler Waves)} \label{sec:whist}
First we derive the fastest whistler wave phase speed. We consider Eq.~(\ref{non-ideal-mhd-eq}) with only the ideal and Hall terms 
   \begin{equation}
   \partial_{t}\delta\vec{W} = \boldsymbol{M}^{(\text{I},\eta_{H})}\delta\partial_{xx}\vec{W}.
   \end{equation}
The full matrix is given in Appendix~C (the $B_{x}$ equation, which is always zero, is dropped).
The characteristic polynomial for this equation is degree 7 and of form $\lambda_{\text{I},\eta_{H}}$ times a sextic. Dropping the entropy ($\lambda_{\text{I},\eta_{H}}$=0) mode the polynomial contains only even powers of $\lambda_{\text{I},\eta_{H}}$. In terms of $\Omega$
\begin{equation}
\label{eq:F-Hall}
F(\Omega, u) =  - c_{a}^{4} c_{s}^{2} u^{2} + u \left(-\Omega^{2}(E+c_{a}^{2}) + \Omega(Ec_{s}^{2}  + c_{a}^{4} + 2 c_{a}^{2} c_{s}^{2})\right) + \Omega^{2}(\Omega - c_{a}^{2} - c_{s}^{2}).
\end{equation}
Here, the dependence $\phi$ has dropped out as it only enters as $b_{y}^{2} + b_{z}^{2} = 1-u$.

Since $\partial^{2}F/\partial u^{2} = -2c_{s}^{2}c_{a}^{4} < 0$, $F(\Omega,u)$ is always concave in $u\in[0,1]$, independent of $\Omega$, and as it is quadratic in $u$
\begin{equation}
\label{eq:chord}
F(\Omega,u) = (1-u)F(\Omega,0) + uF(\Omega,1)  + c_{a}^{4}c_{s}^{2}u(1-u).
\end{equation}
Let $\Omega_{3}(u)$ denote the largest real root of $F(\Omega,u)$ and set
$\Omega_{3,\max} = \max\left(\Omega_{3}(0),\Omega_{3}(1)\right)$. As $F$ is
monic in $\Omega$ it is positive above its largest real root, so for any
$\Omega\geq\Omega_{3,\max}$ both $F(\Omega,0)\geq0$ and $F(\Omega,1)\geq0$.
Eq (\ref{eq:chord}) then gives $F(\Omega,u)>0$ throughout $u\in(0,1)$. Hence
$F(\Omega,u)$ has no root in $[\Omega_{3,\max},\infty)$ for interior $u$, and $\max_{u\in[0,1]}\Omega_{3}(u) = \max\left(\Omega_{3}(0),\Omega_{3}(1)\right)$ is attained only at an endpoint.

The square roots of the roots of $F$ give the phase speeds of the system. At $u=0$ they are $c_{\eta_{H},0} = 0$ and $c_{\eta_{H},1} = \sqrt{c_{s}^{2} + c_{a}^{2}} = c_{f}$: the Alfv\'en and slow speeds (a double root), which vanish for $\hat{b}\perp\hat{x}$, and the fast-magnetosonic speed. At $u=1$ they are, after some radical denesting,
\begin{equation}
\label{eq:Hall-bound}
 c_{\eta_{H},2} = c_{s}, \qquad
 c_{\eta_{H},3,4} =  \mp\frac{|\eta_{H}|}{2\Delta x} + \frac{\sqrt{\eta_{H}^{2}/\Delta x^{2} + 4 c_{a}^{2}}}{2}.
\end{equation}
These are the adiabatic sound speed, ion-cyclotron speed, and whistler phase speed, with $c_{\eta_{H},3}c_{\eta_{H},4} = c_{a}^{2}$. Writing $c_{w}\equiv c_{\eta_{H},4}$, and since $c_{f}\geq c_{s}$, $\max_{u}\Omega_{3}(u)=\max(c_{f}^{2},c_{w}^{2})$. The signal speed of the magnetic field with the Hall effect is therefore $\max(c_{f},c_{w})$.
\subsection{Peltier/Seebeck (Thermoelectric Wave)}\label{sec:hlle_ps}
We now consider the Peltier/Seebeck effects with ideal terms to retrieve the thermoelectric phase speed. We take Eq.~(\ref{non-ideal-mhd-eq}) with only the ideal and Peltier/Seebeck terms ($\vec{q}_{J,\parallel}$, $\vec{q}_{J,\perp}$, $\vec{F}_{T,\parallel}$, and $\vec{F}_{T,\perp}$).
   \begin{equation}
   \partial_{t}\delta\vec{W} = \boldsymbol{M}^{(\text{I},J)}\delta\partial_{xx}\vec{W}.
   \end{equation}
The entries for matrix $\boldsymbol{M}^{(\text{I},J)}$ given in Appendix~C.

The characteristic polynomial is
\begin{equation}\label{eq:F-te}
F(\Omega,u) = \left(\Omega - c_{a}^{2}u\right)
 \left[\Omega^{2} - c_{f}^{2}\Omega + c_{a}^{2}c_{s}^{2}u\right]
 - Du(1-u)\left[\Omega^{2}
   - \left(c_{a}^{2}+c_{\text{iso}}^{2}\right)\Omega
   + c_{a}^{2}c_{\text{iso}}^{2}u\right].
\end{equation}
The first term describes ideal MHD and is adiabatic; substituting $D=0$ and/or $u=0$, $u=1$ retrieves the fast-magnetosonic, Alfv\'en and sound speeds. The second term describes the Peltier/Seebeck effects and is isothermal.

This cubic is solvable only through the general cubic formula, and maximising over $u$ then yields a quintic with no closed form. We instead construct an upper bound, which becomes appreciably larger than $c_{f}$ only in strong Peltier/Seebeck regimes.

The roots of $F$ are real. Defining the weight $\boldsymbol{S}$ by
\begin{equation}\label{eq:S}
\delta\vec{W}^{\dagger}\boldsymbol{S}\delta\vec{W} = \rho_{0}|\delta\vec{v}|^{2} + |\delta\vec{B}|^{2} + \frac{\gamma-1}{\gamma}\frac{|\delta\varepsilon_{\text{int}}|^{2}}{\varepsilon_{0}} + \frac{\varepsilon_{0}}{\gamma}\left|\frac{\delta\varepsilon_{\text{int}}}{\varepsilon_{0}} - \gamma\frac{\delta\rho}{\rho_{0}}\right|^{2},
\end{equation}
the linearised wave energy (kinetic, magnetic, compressional and entropy parts) is positive definite for $\gamma>1$. The ideal and Peltier/Seebeck parts of $\boldsymbol{M}^{(\text{I},J)}$ each satisfy $\boldsymbol{S}\boldsymbol{M}+\boldsymbol{M}^{\dagger}\boldsymbol{S}=0$. For the Peltier/Seebeck part this reads $S_{\varepsilon\varepsilon}M^{(J)}_{\varepsilon B_{y}} = -M^{(J)}_{B_{y}\varepsilon}$ and $S_{\rho\varepsilon}M^{(J)}_{\varepsilon B_{y}} = -M^{(J)}_{B_{y}\rho}$, and likewise for $B_{z}$. Hence $\boldsymbol{S}^{1/2}\boldsymbol{M}\boldsymbol{S}^{-1/2}$ is skew-Hermitian, so every $\omega$ is real and every root of $F$ is real and positive.

Before bounding the full cubic we isolate the thermoelectric contribution alone. Setting $c_{a}=c_{s}=0$ removes the ideal term, leaving 
\begin{equation}
F(\Omega,u)|_{c_{a}=c_{s}=0} = \Omega^{2}(\Omega - Du(1-u)).    
\end{equation}
The Peltier/Seebeck thermoelectric phase speed is 
\begin{equation}
 c_{\kappa J} = \sqrt{Du(1-u)} \leq \sqrt{D}/2.   
\end{equation}
This is maximised at $u=1/2$, $u(1-u) = 1/4$, where $c_{\kappa J,\max} = |\Delta \kappa_{J}| k_{x}/(2\sqrt{\varepsilon_{\text{int}}})$.
The ideal maximum is attained at $u=0$ giving $c_{f}$, where $u(1-u)$ vanishes; the two maxima are therefore attained at different wave-normal angles, and we take the root of the sum
of their squares,
\begin{equation}\label{eq:Lambda}
\Lambda(u) = c_{f}^{2} + Du(1-u),
\qquad c_{\text{TE}}(u) = \sqrt{\Lambda(u)},
\end{equation}
as the candidate bound. We show that no root of $F$ exceeds $\Lambda(u)$.

The Peltier/Seebeck cubic has $-A_{2}=c_{f}^{2}+c_{a}^{2}u+Du(1-u)\leq2\Lambda(u)$, since $c_{a}^{2}u\leq c_{f}^{2}$. By the test of Section~\ref{sec:hlle_mod} it therefore suffices that
\begin{equation}
    F(\Lambda) =  Du(1-u)\Lambda\left[c_{a}^{2}(1-u)+c_{\text{iso}}^{2}\right]
    + c_{a}^{2}c_{s}^{2}u\left[c_{f}^{2} - c_{a}^{2}u + Du(1-u)\frac{\gamma-1}{\gamma}\right] \geq 0,
\end{equation}
which holds term by term, so no root of $F$ exceeds $\Lambda(u)$ for any $u\in[0,1]$.

Maximising $u(1-u)=1/4$, the upper bound for the thermoelectric wave is (with $k_{x} = 1/\Delta x$)
\begin{equation}
 c_{\text{TE}} = \sqrt{c_{f}^{2} + \frac{\Delta \kappa_{J}^{2}}{4\varepsilon_{\text{int}}\Delta x^{2}}}.
\end{equation}
Since $c_{\text{TE}}\geq c_{f}$, with equality when $\Delta\kappa_{J}=0$, the signal speed $\max(c_{f},c_{\text{TE}})=c_{\text{TE}}$. Following \cite{marchand_2019}, we apply it to the internal energy and magnetic field, the variables on which the Peltier/Seebeck terms act directly.
\subsection{Combined Hall and Peltier/Seebeck Wave}
Taking Eq.~(\ref{non-ideal-mhd-eq}) now with the ideal, Hall, and Peltier/Seebeck terms, with a 1D gradient, converting to primitives and linearising, $\boldsymbol{M}^{(\text{I},\eta_{H},J)} = \boldsymbol{M}^{(\text{I})} + \boldsymbol{M}^{(\eta_{H})} + \boldsymbol{M}^{(J)}$ and the characteristic polynomial is
$F(\Omega,u) = \Omega^{3} + A_{2}\Omega^{2} + A_{1}\Omega + A_{0}$ with
\begin{subequations}\label{eq:Acoeffs}
\begin{equation}
A_{2} = -\left[c_{f}^{2} + u\left(c_{a}^{2}+E\right) + Du(1-u)\right]
 \leq 0,
\end{equation}
\begin{equation}
A_{1} = c_{a}^{2}u\left(c_{a}^{2}+2c_{s}^{2}\right) + Ec_{s}^{2}u
 + u(1-u)\left[D\left(c_{a}^{2}+c_{\text{iso}}^{2}\right)
   + 2sc_{a}c_{\text{iso}}\sqrt{(\gamma-1)DE}\right]
 \geq 0,
\end{equation}
\begin{equation}
A_{0} = -c_{a}^{2}c_{\text{iso}}^{2}u^{2}
 \left[\gamma c_{a}^{2} + D(1-u)\right] \leq 0 ,
\end{equation}
\end{subequations}
where $s = -\operatorname{sgn}(\eta_{H}\Delta\kappa_{J}) = \pm1$; the cross term is odd in both $\eta_{H}$ and $\Delta\kappa_{J}$, and every step below holds for either sign. The signs of the coefficients follow from Vieta's formulas, as the roots are real and positive by Eq.~(\ref{eq:S}). Collecting terms,
\begin{align}\label{eq:F-full}
F(\Omega,u) = &\left(\Omega - c_{a}^{2}u\right)
   \left[\Omega^{2} - c_{f}^{2}\Omega + c_{a}^{2}c_{s}^{2}u\right]
   \nonumber \\
&- Eu\Omega\left[\Omega - c_{s}^{2}\right] \nonumber \\
&- Du(1-u)\left[\Omega^{2}
   - \left(c_{a}^{2}+c_{\text{iso}}^{2}\right)\Omega
   + c_{a}^{2}c_{\text{iso}}^{2}u\right] \nonumber \\
&+ 2sc_{a}c_{\text{iso}}\sqrt{(\gamma-1)DE}u(1-u)\Omega .
\end{align}
The term carrying neither $D$ nor $E$ describes ideal adiabatic MHD, the $D$ term the isothermal Peltier/\allowbreak Seebeck effects, the $E$ term the Hall effect, and the $\sqrt{DE}$ term the first-order interaction between them.

Equivalently, the polynomial decomposes as
\begin{equation}\label{eq:decomp}
F(\Omega,u) = G(\Omega,u) - u(1-u)\mathcal{T}(\Omega,u),
\end{equation}
where $G$ is the ideal + Hall polynomial of Eq.~(\ref{eq:F-Hall}) and $\mathcal{T}$ collects the $D$-dependent terms (both are given in full in Appendix~C). Every $D$-dependent term therefore carries a factor $u(1-u)$, which vanishes at both $u=0$ and $u=1$ and peaks at $u=1/2$.

This polynomial is solvable only through the general cubic formula, and maximising its largest root over $u$ has no closed form. We therefore construct an upper bound. It must recover the individual phase speeds when the other terms vanish: $c_{f}$ for $D=E=0$, $\max(c_{f},c_{w})$ for $D=0$, and $c_{\text{TE}}$ for $E=0$. Writing $\Psi = \max(c_{f},c_{w})$, we take
\begin{equation}
\label{eq:cTEw}
c_{\text{TEw}}^{2} = \Psi^{2} + \frac{D}{4} + \frac{\sqrt{DE}}{2}.
\end{equation}
This reduces correctly in all three limits, since $\Psi = c_{f}$ when $E=0$ and the $D$-dependent terms vanish when $D=0$. As in Eq.~(\ref{eq:Lambda}), the $\Psi^{2}$ and $D/4$ contributions are attained at different wave-normal angles: the ideal + Hall maximum lies at $u=0$ or $u=1$ by Eq.~(\ref{eq:chord}), where the $u(1-u)$ prefactor of Eq.~(\ref{eq:decomp}) vanishes.

In Appendix~C we show that $-A_{2}\leq 2c_{\text{TEw}}^{2}$ and $F(c_{\text{TEw}}^{2},u)\geq0$ for all $u\in[0,1]$, either sign of $s$, and all $\gamma>1$. By our bounding test, no root of $F$ exceeds $c_{\text{TEw}}^{2}$ for any $D,E\geq0$. 

With $k_{x}=1/\Delta x$,
\begin{equation}
c_{\text{TEw}} = \sqrt{\max\left(c_{f},c_{w}\right)^{2}
 + \frac{\Delta\kappa_{J}^{2}}{4\varepsilon_{\text{int}}\Delta x^{2}}
 + \frac{|\eta_{H}\Delta\kappa_{J}|}
        {2\Delta x^{2}\sqrt{\varepsilon_{\text{int}}}}}.\footnote{This proof was found by Claude Opus 5.5 (Anthropic) and confirmed with SymPy.}
\end{equation}

\subsection{Gyro-Viscous Waves}\label{sec:hlle_gv}
Starting from the ideal MHD equations and adding the gyro-viscous tensors $\boldsymbol{\Pi}^{(3)}$ and $\boldsymbol{\Pi}^{(4)}$, the linearised operator $\boldsymbol{M}^{(\text{I},\nu_{\wedge})}$ (given in Appendix~C) contains the gyro-viscous coefficients
\begin{equation}
g_{1} = x_{\text{ion}}\left[\nu_{\wedge1}(1-u) + 4\nu_{\wedge2}u\right],
\qquad
g_{2} = g_{1} - 2x_{\text{ion}}\nu_{\wedge2}.
\end{equation}
The gyro-viscous entries act only on the velocity, $\partial_{t}\delta\vec{v} = k_{x}^{2}\vec{w}\times\delta\vec{v}$ with $\vec{w} = (g_{2}b_{x}, g_{1}b_{y}, g_{1}b_{z})$, and
\begin{equation}
\label{eq:Gnu}
G_{\nu}(u) \equiv |\vec{w}|^{2} = g_{1}^{2}(1-u) + g_{2}^{2}u
= x_{\text{ion}}^{2}\left[\nu_{\wedge1}^{2}(1-u)^{2} + 4\nu_{\wedge1}\nu_{\wedge2}u(1-u) + 4\nu_{\wedge2}^{2}u\right].
\end{equation}
As before, $\phi$ drops out, and the characteristic polynomial is $F(\Omega,u) = \Omega^{3} + A_{2}\Omega^{2} + A_{1}\Omega + A_{0}$ with
\begin{subequations}
    \begin{equation}
       A_{2} = -\left[c_{f}^{2} + c_{a}^{2}u + k_{x}^{2}G_{\nu}(u)\right],
    \end{equation}
    \begin{equation}
        A_{1} = c_{a}^{2}u\left(c_{a}^{2} + 2c_{s}^{2}\right) + x_{\text{ion}}^{2}k_{x}^{2}u\left[4\nu_{\wedge2}^{2}c_{a}^{2}(1-u) + c_{s}^{2}\left(\nu_{\wedge1}(1-u) + 2\nu_{\wedge2}(2u-1)\right)^{2}\right],
    \end{equation}
    \begin{equation}
        A_{0} = -c_{a}^{4}c_{s}^{2}u^{2}.
    \end{equation}
\end{subequations}
At the endpoints the characteristic polynomial factorises,
\begin{subequations}
    \begin{equation}
        F(\Omega,0) = \Omega^{2}\left(\Omega - c_{f}^{2} - x_{\text{ion}}^{2}\nu_{\wedge1}^{2}k_{x}^{2}\right),
    \end{equation}
    \begin{equation}
        F(\Omega,1) = \left(\Omega - c_{s}^{2}\right)\left[\Omega^{2} - 2\left(c_{a}^{2} + 2x_{\text{ion}}^{2}\nu_{\wedge2}^{2}k_{x}^{2}\right)\Omega + c_{a}^{4}\right],
    \end{equation}
\end{subequations}
giving the phase speeds 
\begin{subequations}
    \begin{equation}
    c_{gv,0}(0) = 0, \qquad  c_{gv,0}(1) = c_{s} 
    \end{equation}
    \begin{equation}
     c_{gv,1}(0) = \sqrt{c_{f}^{2} + x_{\text{ion}}^{2}\nu_{\wedge1}^{2}k_{x}^{2}},  , \qquad  c_{gv,1,2}(1) = \sqrt{x_{\text{ion}}^{2}\nu_{\wedge2}^{2}k_{x}^{2} + c_{a}^{2}} \mp x_{\text{ion}}|\nu_{\wedge2}|k_{x}.   
    \end{equation}
\end{subequations}
Apart from the Alfv\'en and slow speeds, which vanish at $u=0$, and the sound speed at $u=1$; the $u=0$ endpoint gives a fast-magnetosonic speed raised by gyroviscosity, of the same form as $c_{\text{TE}}$; while the $u=1$ endpoint gives a whistler-like pair whose product is $c_{a}^{2}$, Hall effect-like.

Unlike the thermoelectric case, a root-sum-of-squares bound fails here: at $u=1$ with $c_{s}=0$, $c_{\text{gv},2}^{2}(1)=c_{a}^{2}+2(x_{\text{ion}}\nu_{\wedge2}k_{x})^{2}+2x_{\text{ion}}\nu_{\wedge2}k_{x}\sqrt{(x_{\text{ion}}\nu_{\wedge2}k_{x})^{2}+c_{a}^{2}}$, which exceeds $c_{f}^{2}+k_{x}^{2}G_{\nu}(1)=c_{a}^{2}+4(x_{\text{ion}}\nu_{\wedge2}k_{x})^{2}$. 

We have therefore opted to use the Weyl bound. The ideal and gyroviscous parts of $\boldsymbol{M}^{(\text{I},\nu_{\wedge})}$ each satisfy $\boldsymbol{S}\boldsymbol{M}+\boldsymbol{M}^{\dagger}\boldsymbol{S}=0$ with $\boldsymbol{S}$ of Eq.~(\ref{eq:S}), so $\boldsymbol{S}^{1/2}\boldsymbol{M}^{(\text{I},\nu_{\wedge})}\boldsymbol{S}^{-1/2}$ is a sum of skew-Hermitian matrices, and its spectral radius is at most the sum of theirs. The ideal part has phase speed at most $c_{f}$, and gyroviscosity alone has phase speed $k_{x}|\vec{w}| = k_{x}\sqrt{G_{\nu}(u)}$, so every phase speed satisfies $c\leq c_{f}+k_{x}\sqrt{G_{\nu}(u)}$. Therefore with $k_{x}=1/\Delta x$, we take the gyroviscous phase speed upper bound as
\begin{equation}
    c_{\text{gv}} = c_{f} + \frac{|\lambda_{\nu,\wedge,\max}|}{\Delta x},
\end{equation}
where $\lambda_{\nu,\wedge,\max}$ is the same eigenvalue as the gyroviscous CFL condition of section \ref{sec:gyro_cfl}. Since $c_{\text{gv}}\geq c_{f}$, the signal speed is $\max(c_{f},c_{\text{gv}})=c_{\text{gv}}$, which, following \cite{marchand_2019}, we apply to the momentum, on which the gyroviscous terms act.
\subsection{Righi--Leduc Velocity}
The Righi--Leduc effect is a first-order advective operator acting on internal energy. We therefore simply add this to the phase speed for the internal energy, with the characteristic speed being
\begin{equation}
    c_{\kappa\wedge} = \frac{\gamma-1}{\rho}\left|\left[\vec{\nabla}\times\left(x_{\text{ion}}\rho\kappa_{\wedge}\hat{b}\right)\right]\cdot\hat{n}\right|.
\end{equation}
Instead of upper bounding the curl of the unit vector in the normal direction (as we do with the CFL condition), we evaluate it with centred differences at the cell centres on either side of the face, including the metric factors of curvilinear coordinates. Along an unresolved direction the field is uniform and only the derivative of the scale factor contributes, which retains, for example, the $\cot\theta b_{\phi}/r$ term of one-dimensional spherical problems.
\subsection{HLLE-Mod Signal Speed Summary}
Following \cite{marchand_2019}, each variable is assigned the fastest phase speed of the modes in which it participates, and the signal speeds are formed as $S_{L}=\min(\tilde{v}_{x,L}-\tilde{c}_{L,*},v_{x,L}-c_{L,*})$ and $S_{R}=\max(\tilde{v}_{x,R}+\tilde{c}_{R,*},v_{x,R}+c_{R,*})$ (where a tilde denotes a Roe averaged variable), with
\begin{equation}
        c_{\rho} = c_{f}, \qquad
        c_{v} = c_{\text{gv}}, \qquad
        c_{\varepsilon_{\text{int}}} = c_{\kappa\wedge} + c_{\text{TE}}, \qquad
        c_{B} = c_{\text{TEw}}.
\end{equation}
Each speed reduces to $c_{f}$ when the corresponding non-ideal effects are absent, so the scheme reverts to the standard HLLE solver (the whistler-modified speed enters the internal energy only through the thermoelectric coupling, so we assign $c_{\text{TE}}$ rather than $c_{\text{TEw}}$ to $\varepsilon_{\text{int}}$; likewise, gyroviscous waves are assigned to the momentum only).

$c_{f}$ in $c_{\rho}$, $c_{v}$, and $c_{\varepsilon_{\text{int}}}$ is the fast magnetosonic speed in the face-normal direction at the local field orientation, as in the standard HLLE solver. The gyroviscous bound holds at each orientation, since the Weyl argument applies for fixed $u$, and the thermoelectric bound evaluated with the local fast speed was verified numerically over $3\times10^{5}$ random parameter sets. The local evaluation of $c_{w}$ and $c_{\text{TEw}}$ are not a bound, undershooting the largest root by up to $21\%$ in the same test, so $c_{B}$ uses the orientation-maximised fast and Alfv\'en speeds, $c_{f}^{2}=(\gamma P+|\vec{B}|^{2})/\rho$ and $c_{a}=|\vec{B}|/\sqrt{\rho}$, as assumed in its derivation, whenever the Hall or Peltier/Seebeck terms are enabled.

The Riemann solver in Athena++ evolves the total energy rather than the internal energy, so we split the ideal energy flux into internal, kinetic, and magnetic parts,
\begin{equation}
\mathcal{F}_{\text{int}}=(\varepsilon_{\text{int}}+P)v_{x}, \qquad
\mathcal{F}_{\text{kin}}=\frac{1}{2}\rho|\vec{v}|^{2}v_{x}, \qquad
\mathcal{F}_{\text{mag}}=|\vec{B}|^{2}v_{x}-B_{x}(\vec{B}\cdot\vec{v}),
\end{equation}
which sum to the ideal energy flux, with densities $\varepsilon_{\text{int}}$, $\varepsilon_{\text{kin}}=\frac{1}{2}\rho|\vec{v}|^{2}$, and $\varepsilon_{\text{mag}}=\frac{1}{2}|\vec{B}|^{2}$. Each part is computed as an HLLE flux with the signal speeds of the variable that carries it,
\begin{equation}
\label{eq:energy-split}
\mathcal{F}^{\text{HLL}}(\varepsilon) = \mathcal{F}^{\text{HLL}}\left(\varepsilon_{\text{int}};S_{\varepsilon_{\text{int}}}\right) + \mathcal{F}^{\text{HLL}}\left(\varepsilon_{\text{kin}};S_{v}\right) + \mathcal{F}^{\text{HLL}}\left(\varepsilon_{\text{mag}};S_{B}\right),
\end{equation}
with
\begin{equation}
\mathcal{F}^{\text{HLL}}(q;S) = \frac{S_{R}\mathcal{F}_{q,L} - S_{L}\mathcal{F}_{q,R} + S_{L}S_{R}\left(q_{R}-q_{L}\right)}{S_{R}-S_{L}}.
\end{equation}
The non-ideal fluxes are computed separately and added at the faces. Each sub-flux reduces to its physical flux for equal left and right states, so the total is consistent and the total energy remains conserved. Since the HLL flux is linear in $(q,\mathcal{F}_{q})$ at fixed signal speeds, when no non-dissipative term is enabled the sub-fluxes sum to a single HLLE flux and the scheme reduces to the standard HLLE solver. The split gives each part of the energy flux the same numerical dissipation as the variable it derives from: it removes the mismatch between the dissipation of the magnetic field and of the magnetic energy that arises when only the field flux is modified, and extends the same treatment to the internal and kinetic energy, since the non-dissipative terms act on every variable except the density. The cost is that the energy flux is no longer a single HLLE flux for one pair of signal speeds, but rather the sum of three sub-fluxes.
\section{Implementation}
Following standard conventions, cell centred quantities carry integer indices, and a subscript $i-1/2$ denotes the $x$-face bounding cells $i-1$ and $i$. Two spacings are required: $\Delta x_{i}$ is the width of cell $i$, while $dx_{i}\equiv x_{i+1}-x_{i}$ is the centre-to-centre spacing, so that $dx_{i-1}$ separates centres $i-1$ and $i$. The two coincide only on a uniform Cartesian grid.
The density, ionisation mass fraction and diffusion coefficients are taken directly at the arithmetic mean across the cell face, e.g. $\rho_{i-1/2,j,k} = \frac{1}{2}(\rho_{i,j,k} + \rho_{i-1,j,k})$. 
Magnetic fields are stored at cell faces. The cell centred magnetic fields are a linear interpolation of the adjacent cell centred fields, $B_{x,i,j,k} = \frac{1}{2}(B_{x,i+1/2,j,k} + B_{x,i-1/2,j,k})$, and likewise for $B_{y}$ and $B_{z}$.
\subsection{Thermal Conduction}
The spatial derivatives of the temperature are taken as the central difference of adjacent cell centred values. For the $x$-face (normal) this is 
\begin{equation}
\partial_{x}T_{i-1/2,j,k} = (T_{i,j,k} - T_{i-1,j,k})/dx_{i-1}. 
\end{equation} 
For the $y$-face (transverse)
\begin{equation}
        \partial_{y}T_{i-1/2,j,k} = \frac{1}{2}((T_{i,j+1,k} + T_{i-1,j+1,k}) - (T_{i,j-1,k} + T_{i-1,j-1,k}))/(dy_{j-1}+dy_{j}),
\end{equation}
and likewise for $\partial_{z}T$. 

In non-Cartesian coordinates, the appropriate metric factors are present to allow for curvilinear coordinates. The fluxes are then calculated directly at $(i-1/2,j,k)$ using Eq.~(\ref{kappa-aniso}).
With STS enabled the wedge flux will be sent to the main hydro flux, while the neutral, parallel, and perpendicular fluxes will be sent to the STS solver. Without STS enabled, all fluxes are sent to the main hydro solver.

To prevent heat from flowing from cold cells into hot cells in highly anisotropic regimes (which would violate the second law of thermodynamics) we limit the transverse temperature gradients following the asymmetric scheme of \cite{Sharma_2007}. On the $x$-face $(i-1/2,j,k)$ the transverse gradients are
\begin{equation}
\label{eq:transverse_limited}
\partial_{y}T_{i-1/2,j,k} = \text{MC}\left[\text{MC}\left(\partial^{+}_{y}T_{i,j,k},\partial^{-}_{y}T_{i,j,k}\right),\ \text{MC}\left(\partial^{+}_{y}T_{i-1,j,k},\partial^{-}_{y}T_{i-1,j,k}\right)\right],
\end{equation}
and likewise for $\partial_{z}T_{i-1/2,j,k}$. These are
built from the one-sided differences in each of the two cells sharing the face,
\begin{equation}
\partial^{+}_{y}T_{i,j,k} = \frac{T_{i,j+1,k} - T_{i,j,k}}{dy_{j}},
\qquad
\partial^{-}_{y}T_{i,j,k} = \frac{T_{i,j,k} - T_{i,j-1,k}}{dy_{j-1}},
\end{equation}
and likewise for $z$ with $dz_{k}$ and $dz_{k-1}$. The divisors are the centre-to-centre spacings, multiplied by the metric factors in curvilinear coordinates. $\text{MC}$ is the monotonized-central limiter,
\begin{equation}\label{eq:MC}
\text{MC}(a,b) =
\begin{cases}
\operatorname{sign}(a)\min\left(2|a|,2|b|,\frac{1}{2}|a+b|\right) & ab>0\\
0 & ab\leq0
\end{cases}.
\end{equation}
The normal gradient, $\partial_{x}T_{i-1/2,j,k} = (T_{i,j,k}-T_{i-1,j,k})/dx_{i-1}$, is not limited, and the wedge flux keeps the centred transverse gradients.

By default (though the user can turn it off) we also apply the saturation flux limiter of \cite{Mignone_2012}. The saturated flux is
\begin{equation}
q_{\text{sat}} = 5\phi_{\text{sat}}\rho c_{\text{iso}}^{3}, \qquad \phi_{\text{sat}} = 0.3,
\end{equation}
evaluated in the cell on the upwind side of the face, i.e. the side the heat flows from. The parallel and perpendicular fluxes are each multiplied by their own factor, $\vec{q}_{\parallel}\rightarrow f_{\text{sat},\parallel}\vec{q}_{\parallel}$ and $\vec{q}_{\perp}\rightarrow f_{\text{sat},\perp}\vec{q}_{\perp}$, with
\begin{equation}
\label{eq:fsat}
f_{\text{sat},\parallel} = \frac{q_{\text{sat}}}{q_{\text{sat}} + \kappa_{\parallel}\rho\left|\hat{b}\cdot\nabla T\right|},
\qquad
f_{\text{sat},\perp} = \frac{q_{\text{sat}}}{q_{\text{sat}} + \kappa_{\perp}\rho\left|\nabla T - \hat{b}\left(\hat{b}\cdot\nabla T\right)\right|}.
\end{equation}
\subsection{Viscosity}
The primitive variables and viscosity coefficients at the interface are taken as the arithmetic mean of the adjacent cell centres, as is done with thermal conduction. The velocity derivatives at the $x$ face use the same stencils as the temperature derivatives, with $T$ replaced by each velocity component $v_{m}$. The appropriate metric factors are in place to allow this to function in curvilinear coordinates.

The fluxes at face $(i-1/2,j,k)$ are then directly calculated using the formulas provided in Appendix~A and Eq~(\ref{visc-aniso}).

We then limit the parallel viscosity by capping the pressure anisotropy at the firehose and mirror instability thresholds, following \cite{Kempski_2019}. The pressure anisotropy produced by the parallel viscosity is
\begin{equation}
\Delta P = P_{\perp} - P_{\parallel} = 3\rho x_{\text{ion}}\nu_{\parallel}\left(\hat{b}\otimes\hat{b} - \frac{1}{3}\boldsymbol{I}\right):\vec{\nabla}[\vec{v}],
\end{equation}
which we clip to the range in which neither instability is excited, $-B^{2} \leq \Delta P \leq B^{2}/2$. With the magnetic pressure $B^{2}/2$, the lower bound is the firehose instability threshold, $P_{\parallel}-P_{\perp} = B^{2}$, and the upper bound the mirror instability threshold, $P_{\perp}-P_{\parallel} = B^{2}/2$. The clipped $\Delta P$ is then used in the parallel viscous stress $\boldsymbol{\Pi}^{(0)}$. This can be disabled by the user, as with the conduction saturation limiter.

Anisotropic viscosity can also transport momentum in the wrong direction across steep velocity gradients, analogous to heat flowing from cold to hot cells in anisotropic thermal conduction. Following \cite{ZuHone_2015}, who limited the anisotropic viscous fluxes with the MC limiter, we limit the transverse velocity derivatives in the dissipative stresses $\boldsymbol{\Pi}^{(0)}$--$\boldsymbol{\Pi}^{(2)}$ in the same way as the transverse temperature gradients. On the $x$-face $(i-1/2,j,k)$, each transverse derivative $\partial_{y}v_{m}$ and $\partial_{z}v_{m}$ of the velocity components $m=x,y,z$ is given by Eq.~(\ref{eq:transverse_limited})--(\ref{eq:MC}) with $T$ replaced by $v_{m}$. The same as conduction, the normal derivatives $\partial_{x}v_{m}$ are not limited. The gyro-viscous stresses $\boldsymbol{\Pi}^{(3)}$ and $\boldsymbol{\Pi}^{(4)}$ keep the centred derivatives.

 $\boldsymbol{\Pi}^{(3)}$ and $\boldsymbol{\Pi}^{(4)}$ are always sent to the main hydro solver, while bulk, neutral, $\boldsymbol{\Pi}^{(0)}$, $\boldsymbol{\Pi}^{(1)}$, and $\boldsymbol{\Pi}^{(2)}$ are sent to the STS solver if it is enabled, otherwise they are sent to the main hydro solver.

\subsection{Hall Effect}
The edge-centred current density is calculated from the face centred magnetic field in the follow way
\begin{equation}\label{eq:edge-current}
J_{x,i,j-1/2,k-1/2} = \frac{B_{z,i,j,k}-B_{z,i,j-1,k}}{dy_{j-1}}
                     - \frac{B_{y,i,j,k}-B_{y,i,j,k-1}}{dz_{k-1}},
\end{equation}
and similarly for $J_{y}$ and $J_{z}$. These are written for Cartesian coordinates, and appropriate metric factors are included for curvilinear coordinates. The transverse current densities at the $x$ edge are taken as the arithmetic mean of the four surrounding transverse currents 
\begin{equation}
 J_{y,i,j-1/2,k-1/2} = \frac{1}{4}(J_{y,i-1/2,j,k-1/2}+J_{y,i+1/2,j,k-1/2} + J_{y,i-1/2,j-1,k-1/2}+J_{y,i+1/2,j-1,k-1/2}).
\end{equation}
The same logic is applied to calculate the edge magnetic fields and the Hall coefficient: the normal field $B_{x}$ and $\eta_{H}$ are averaged over the four cells sharing the edge,
\begin{equation}
\label{edge-field}
        B_{x,i,j-1/2,k-1/2} = \frac{1}{4}(B_{x,i,j,k}+B_{x,i,j-1,k} + B_{x,i,j,k-1}+B_{x,i,j-1,k-1}),
\end{equation}
and the transverse fields over the two faces that share it
\begin{equation}
 B_{y,i,j-1/2,k-1/2} = \frac{1}{2}(B_{y,i,j-1/2,k}+B_{y,i,j-1/2,k-1}). 
\end{equation}
The Hall EMF at $(i,j-1/2,k-1/2)$ is then calculated directly as
\begin{equation}
    \vec{\mathcal{E}} = -\eta_{H}\hat{b}\times\vec{J}.
\end{equation}
\subsection{Thermoelectric Effects}
For what we term thermoelectric conduction; the Peltier and  Ettingshausen effects ($\vec{q}_{J}$), the flux is determined at the cell face in a similar manner to thermal conduction. The normal magnetic field is already stored at the face, and the transverse field and coefficients are arithmetically averaged in the same manner as the previous subsections. The current density at the cell faces are retrieved from the current at the cell centre. The cell centred current is averaged from the four surrounding edges,
\begin{equation}
\label{eq:jcc}
 J_{x,i,j,k} = \frac{1}{4}(J_{x,i,j-1/2,k-1/2}+J_{x,i,j+1/2,k-1/2} + J_{x,i,j-1/2,k+1/2}+J_{x,i,j+1/2,k+1/2}),
\end{equation}
and likewise for $J_{y}$ and $J_{z}$. The face centred current densities are then retrieved from the two adjacent cells, $\vec{J}_{i-1/2,j,k} = \frac{1}{2}(\vec{J}_{i,j,k}+\vec{J}_{i-1,j,k})$, and likewise for the other faces.
The double averaging of Eq.~(\ref{eq:jcc}) applies to the transverse current densities an additional $\frac{1}{4}(1,2,1)$ filter in the face-normal direction. The normal component does not. Both remain second-order accurate and correctly centred. The common cell-centred intermediate ensures that all three components at a face derive from a single consistently sampled vector.

The thermoelectric Peltier and Ettingshausen fluxes at the $x$ face $(i-1/2,j,k)$ are then calculated from Eq~(\ref{therm-elec-cond}).

For the thermoelectric force - the Seebeck and Nernst effects, the edge centred fields require the edge-centred temperature, which are taken as the arithmetic mean of the four surrounding cell centres 
\begin{equation}
 T_{i,j-1/2,k-1/2} = \frac{1}{4}(T_{i,j,k} + T_{i,j,k-1} + T_{i,j-1,k} + T_{i,j-1,k-1}).
\end{equation} 
At the $x$-edge the normal temperature gradient is
\begin{equation}
\partial_{x}T_{i,j-1/2,k-1/2} = \frac{T_{i+1,j-1/2,k-1/2} - T_{i-1,j-1/2,k-1/2}}{dx_{i-1}+dx_{i}}.
\end{equation}
The transverse temperature gradient is
\begin{equation}
  \partial_{y}T_{i,j-1/2,k-1/2} = \frac{T_{i,j+1/2,k-1/2} - T_{i,j-3/2,k-1/2}}{\Delta y_{j-1}+\Delta y_{j}},  
\end{equation}
and likewise for $\partial_{z}T$.

The edge centred magnetic fields are then calculated in the same way as Eq.~(\ref{edge-field}), and the thermoelectric EMF is calculated using Eq.~(\ref{therm-elec-force}).

The perpendicular factor $f_{\text{sat}\perp}$ of Eq.~(\ref{eq:fsat}) is applied to the Nernst EMF $\vec{F}_{T,\wedge}$ and the Ettingshausen heat flux $\vec{q}_{J,\wedge}$. \cite{Brodrick_2018} recommend limiting Nernst advection by the same factor as the perpendicular heat flux. We apply the same factor to the Ettingshausen flux. The Ettingshausen flux through the $x$-face reuses the factor of the conduction flux on that face,
\begin{equation}
\label{eq:fperp_face}
f_{\text{sat}\perp,i-1/2,j,k} = \frac{q_{\text{sat}}}{q_{\text{sat}} + \kappa_{\perp,i-1/2,j,k}\rho_{i-1/2,j,k}\sqrt{\left|\nabla T\right|^{2}_{i-1/2,j,k} - \left(\hat{b}\cdot\nabla T\right)^{2}_{i-1/2,j,k}}},
\end{equation}
The $x$-component of the Nernst EMF lives on the $x$-edge $(i,j-1/2,k-1/2)$ for the constrained-transport update, where the factor is recomputed as
\begin{equation}
\label{eq:fperp_edge}
f_{\text{sat}\perp,i,j-1/2,k-1/2} = \frac{\bar{q}_{\text{sat}}}{\bar{q}_{\text{sat}} + \bar{\kappa}_{\perp}\bar{\rho}\sqrt{\left|\nabla T\right|^{2}_{i,j-1/2,k-1/2} - \left(\hat{b}\cdot\nabla T\right)^{2}_{i,j-1/2,k-1/2}}}.
\end{equation}
Limiting $\kappa_{\perp}$ and $\kappa_{J,\wedge}$ by a common factor $f\leq1$ preserves the well-posedness criterion of Section~\ref{sec:wellposed}, since its left-hand side decreases at most linearly in $f$ while its right-hand side decreases as $f^{2}$.
\newpage
\section{Tests\protect\footnote{The tests below were run, and the figures produced, by Bash and Python scripts respectively. These were written with the assistance of Claude Opus 5.5 (Anthropic). All figures generated from these scripts directly use the simulation output.}}
For all tests we use piecewise-linear reconstruction and the HLLE Riemann solver (HLLE-Mod where stated), with a CFL number of 0.9. The dissipative tests use RK1; the non-dissipative tests are run with RK1, RK2, and RK3. With STS enabled, the dissipative terms are integrated with RKL1. With STS the super-step is capped at 50 parabolic timesteps, otherwise STS runs finish in a single ideal hydro timestep. We have modified Athena++'s ideal timestep so that the number of dimensions is accounted for before the CFL number is applied for consistency with the non-ideal limits. Unless stated otherwise, the background state is $\rho_{0}=1$, $P_{0}=1$, $\vec{B}_{0}=(1,0,0)$, on uniform Cartesian meshes of $16^{3}$, $32^{3}$, and $64^{3}$ cells spanning $[-10,10]^{3}$ with periodic boundaries, decomposed into meshblocks of $8^{3}$, $16^{3}$, and $32^{3}$ cells respectively. All runs use 8 MPI ranks on an AMD Ryzen 7 9800X3D. Serial runs of the isotropic and neutral ($x_{\text{ion}}=0$) versions of both thermal conduction and viscosity were bitwise identical.

The analytical solutions are derived for frozen flow, retaining only the non-ideal terms under test. The simulations evolve the full MHD equations, so we choose coefficients for which the non-ideal timescales are much shorter than the sound and Alfv\'en crossing times, making the ideal dynamics negligible over each test. The saturation and pressure-anisotropy are disabled, while the slope limiters are enabled. We measure convergence with
\begin{equation}
  \|e_{N}(t)\|_{2} = \left(\frac{1}{V}\sum_{i} e_{i}^{2}(t)\Delta V_{i}\right)^{1/2}, \qquad
  E_{N} = \max_{t}\|e_{N}(t)\|_{2}, \qquad
  p = \log_{2}\frac{E_{N/2}}{E_{N}}.
\end{equation}
The convergence orders of all tests are collected in Tables~\ref{tab:conv-diss} and~\ref{tab:conv-nondiss}.
\subsection{Thermal Conduction}
To test the dissipative thermal conduction we initialise a temperature perturbation given by
\begin{equation}
    T(0,x,y,z) = T_{0}\left(1+A_{T}(0)\cos\left(k_{x}x+k_{y}y\right)\right),
\end{equation}
with $k_{x}= 2\pi N_{x}/L_{x}$, $k_{y}= 2\pi N_{y}/L_{y}$, and $k_{z}= 2\pi N_{z}/L_{z}$ - one wavelength of the box in $x$ and $y$.

For our anisotropic thermal conduction implementation, this has the analytical solution
\begin{subequations}
\begin{equation}
T(t,x,y,z) = T_{0}(A_{T}(0)\exp(-\Gamma_{\kappa}t)\cos(k_{x}x+k_{y}y+k_{z}z) + 1),
\end{equation}
\begin{equation}
    \Gamma_{\kappa} = (\gamma-1)(k_{\parallel}^{2}(x_{\text{ion}}\kappa_{\parallel}+(1-x_{\text{ion}})\kappa_{n}) + k_{\perp}^{2}(x_{\text{ion}}\kappa_{\perp}+(1-x_{\text{ion}})\kappa_{n})),
\end{equation}
\end{subequations}
with parallel and perpendicular wavenumbers $\vec{k}_{\parallel} = (\vec{k}\cdot\hat{b})\hat{b}$ and $\vec{k}_{\perp} =  \vec{k} - \vec{k}_{\parallel}$.
We test neutral conduction only, parallel only, perpendicular only, and a mixed system with neutral, parallel, and perpendicular conduction. We set $k_{x} = 2\pi/L_{x}$, $k_{y} = 2\pi/L_{y}$, $k_{z} = 0$, and $A_{T}(0)=10^{-6}$.

\begin{figure}[H]
\centering
    \includegraphics[scale=0.95]{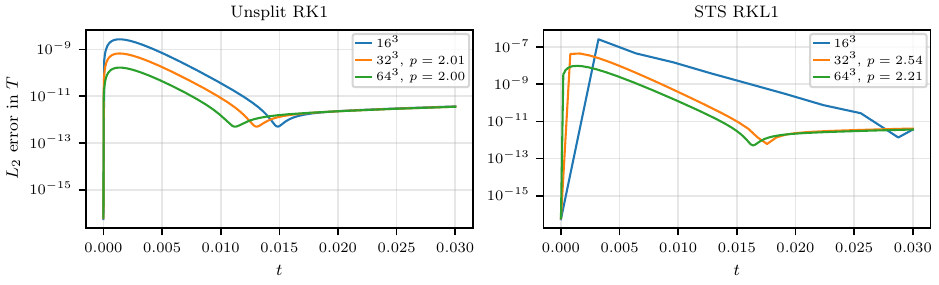}
    \caption{Thermal conduction test: combined neutral, parallel, and perpendicular. $\kappa_{n}=\kappa_{\parallel}=\kappa_{\perp}=5500$, $x_{\text{ion}}=0.5$.}
    \label{fig:cond-test}
\end{figure}
Figure \ref{fig:cond-test} shows the dissipative thermal conduction L2 error over time for RK1 and RKL1 time integration. The error peaks between $t\approx1.1\times10^{-3}$ and $t\approx1.5\times10^{-3}$, close to the decay time of the perturbation, $\Gamma_{\kappa}^{-1}\approx1.4\times10^{-3}$. RK1 falls by a factor of four with each doubling of resolution. The RKL1 errors are larger, by a factor of 60 at $64^{3}$, this is because RKL1 is first order in time and each super-step spans up to 50 parabolic timesteps. At late times all runs tend towards a single L2 error of order $10^{-12}$. This is the point at which the temperature gradient has diffused throughout the entire domain, leaving an essentially constant temperature throughout, so the discretisation error has decayed with the perturbation. We attribute the remaining error (which is independent of resolution and integrator and grows slowly with time) to the ideal MHD response to the initial pressure perturbation. This is omitted from the frozen-flow analytical solution.
\subsection{Viscosity}
We initialise our velocity perturbation with a wavevector $\vec{k}=(k_{x},0,0)$, and magnetic field unit vector $\hat{b} = (0,\cos(\theta),\sin(\theta))$ with $\theta = \pi/3$, and with the velocity field as
\begin{equation}
 \vec{v}(x,y,z,0)  =  \begin{pmatrix}v_{0} a_{1}{\left(0 \right)} \sin{\left(k_{x} x \right)}\\
 v_{0} \left(a_{2}{\left(0 \right)} \cos{\left(\theta \right)} - a_{3}{\left(0 \right)} \sin{\left(\theta \right)}\right) \sin{\left(k_{x} x \right)}\\
 v_{0} \left(a_{2}{\left(0 \right)} \sin{\left(\theta \right)} + a_{3}{\left(0 \right)} \cos{\left(\theta \right)}\right) \sin{\left(k_{x} x \right)}\end{pmatrix}.
\end{equation}
This has the analytical solution
\begin{equation}
\vec{v}(x,y,z,t) = v_{0} \sin{\left(k_{x} x \right)} \begin{pmatrix} a_{1}{\left(0 \right)} \exp\left(- \Gamma_{1} t\right)\\
a_{2}{\left(0 \right)} \exp\left(- \Gamma_{2} t\right) \cos{\left(\theta \right)} - a_{3}{\left(0 \right)} \exp\left(- \Gamma_{3} t\right) \sin{\left(\theta \right)}\\
a_{2}{\left(0 \right)} \exp\left(- \Gamma_{2} t\right) \sin{\left(\theta \right)} + a_{3}{\left(0 \right)} \exp\left(- \Gamma_{3} t\right) \cos{\left(\theta \right)}\end{pmatrix},
\end{equation}
where
\begin{subequations}
\begin{equation}
\Gamma_{1} = k_{x}^{2} \left(- \frac{4 \nu_{n} \left(x_{\text{ion}} - 1\right)}{3} + \frac{x_{\text{ion}} \left(\nu_{\parallel} + 3 \nu_{\perp1}\right)}{3} + \frac{\zeta}{\rho_{0}}\right),
\end{equation}
\begin{equation}
\Gamma_{2} = k_{x}^{2} \left(4 \nu_{\perp2} x_{\text{ion}} + \nu_{n} \left(1 - x_{\text{ion}}\right)\right),
\end{equation}
\begin{equation}
\Gamma_{3} = k_{x}^{2} \left(\nu_{\perp1} x_{\text{ion}} + \nu_{n} \left(1 - x_{\text{ion}}\right)\right).
\end{equation}
\end{subequations}
We set $|B_{0}|=1$, and $v_{0}=10^{-6}$. We only activate the modes that are diffused by the active viscosity components, otherwise we find our error dominated by undampened ideal effects. We set $a_{1}(0) = a_{2}(0) = a_{3}(0) = 1$ for the runs in which they are active, and zero otherwise. The active modes are: bulk $a_{1}$; isotropic, neutral, perpendicular 1+2, and mixed $a_{1}$, $a_{2}$, and $a_{3}$; parallel and perpendicular 1 $a_{1}$ and $a_{3}$; and perpendicular 2 $a_{2}$.
\begin{figure}[H]
\centering
    \includegraphics[scale=0.95]{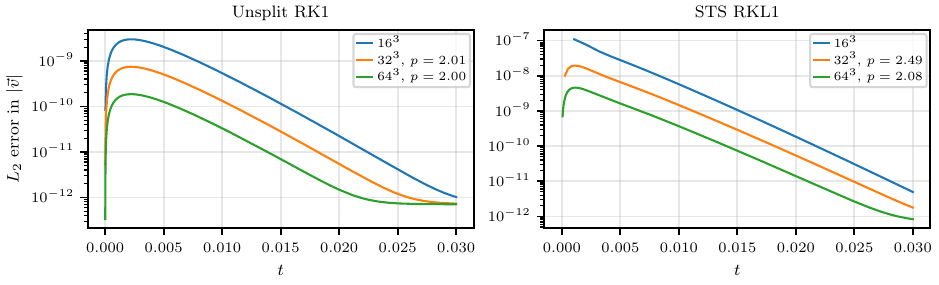}
    \caption{Viscosity test: combined bulk, neutral, parallel ($\boldsymbol{\Pi}^{(0)}$), and perpendicular ($\boldsymbol{\Pi}^{(1)} + \boldsymbol{\Pi}^{(2)}$). $\zeta=5200$, $\nu_{n}=\nu_{\parallel}=\nu_{\perp1}=\nu_{\perp2}=3900$, $x_{\text{ion}}=0.5$.}
    \label{fig:visc-test}
\end{figure}
Figure \ref{fig:visc-test} shows the L2 error in dissipative viscosity over time for RK1 and RKL1 time integration. The RK1 error peaks at $t\approx2\times10^{-3}$ and then decreases continuously, as the velocity perturbation decays towards $\vec{0}$ and the discretisation error decays with it. As for thermal conduction, the RK1 errors fall by a factor of four with each doubling of resolution, and the RKL1 errors are larger because of the first-order temporal error of the super-steps. By $t\approx0.025$ the $32^{3}$ and $64^{3}$ RK1 errors level off at $\approx7\times10^{-13}$, independent of resolution, which we attribute to the same ideal MHD response, here to the compressive $a_{1}$ mode.
\subsection{Hall Effect (+ Magnetic Diffusion)}
We consider the induction equation with frozen flow. Only Ohmic, ambipolar, and Hall terms are present. We take our field to be $|\vec{B}_{0}|=1$ and unit vector $\hat{b}=(\cos(\theta),\sin(\theta),0)$ with $\theta=\pi/6$. Our initial field is $\vec{B}(x,y,z,0) = (B_{0}\cos\theta, B_{0}\sin\theta, \delta B_{0}\cos(k_{x}x))$, with $\delta B_{0}=10^{-6}$. this has the analytical solution
\begin{equation}
    \vec{B}(x,y,z,t) = \begin{pmatrix}
        B_{0}\cos(\theta) \\
        B_{0}\sin(\theta) - \frac{\beta \delta B_{0}}{D}\exp(-\bar{a}k_{x}^{2}t)\sinh(Dk_{x}^{2}t)\cos(k_{x}x) \\
        \delta B_{0}\left(\cosh(Dk_{x}^{2}t) + \frac{\delta_{a}}{D}\sinh(Dk_{x}^{2} t)\right)\exp(-\bar{a}k_{x}^{2}t)\cos(k_{x}x)
    \end{pmatrix},
\end{equation}
with
\begin{equation}
        \bar{a} = \eta + \frac{1}{2}\eta_{AD}(1+\cos^{2}(\theta)), \qquad
        \delta_{a} = \frac{1}{2}\eta_{AD}\sin^{2}(\theta), \qquad
        \beta = \eta_{H}\cos(\theta), \qquad
        D^{2} = \delta_{a}^{2} - \beta^{2}.
\end{equation}
\begin{figure}[H]
\centering
    \includegraphics[scale=0.95]{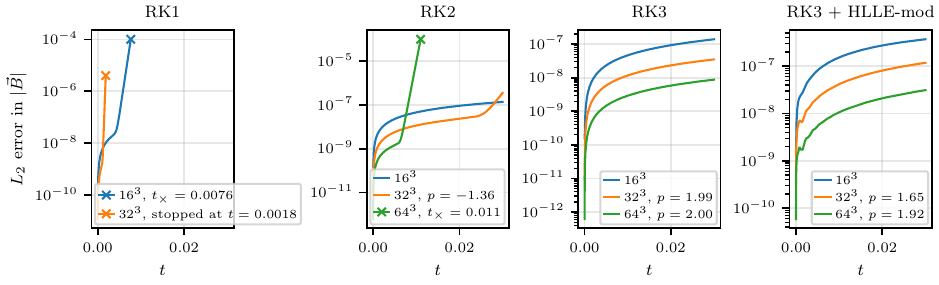}
    \caption{Hall effect test. $\eta_{H}=6000$.}
    \label{fig:hall-test}
\end{figure}

Figure \ref{fig:hall-test} shows the magnetic field test L2 errors with only the Hall effect active (non-dissipative), across RK1, RK2, and RK3 (with the standard and modified HLLE Riemann solver) time integration. The Hall effect fails completely at RK1. Mass, momentum, and energy conservation fail, the timestep becomes zero, and infinities and NaNs propagate throughout the grid. The blow-up time decreases with resolution, scaling as $O(\Delta x^{2})$. RK2 suffers from the same failure, only surviving longer. This can be seen in the L2 error sharply rising towards the end of the simulation at $32^{3}$. With RK3 the simulation becomes stable, converging at the expected order of two. As the simulation only contains smooth fields, we do not expect grid scale instabilities to cause it to fail. Using the modified Riemann solver hurts convergence, particularly at lower resolution. This is due to the increase in numerical diffusion. It also raises the L2 errors by a factor of about three.
\subsection{Nernst/Ettingshausen Effects}
We consider the ideal MHD equations with only neutral and perpendicular thermal conduction; Ohmic and ambipolar magnetic diffusion; and the Nernst/Ettingshausen effects. We take $B_{0}=1$, $\rho_{0}=1$, and unit vector $\hat{b}=(0,1,0)$. We initialise our temperature and field as $T(x,y,z,0) = T_{0} + \delta T_{0} \cos(k_{x}x)$ and $\vec{B}(x,y,z,0) = (0, B_{0} + \delta B_{0} \cos(k_{x}x), 0)$, with $\delta T_{0} = \delta B_{0} = 10^{-6}$.

this has the analytical solution
\begin{subequations}
    \begin{equation}
        T(x,y,z,t) = T_{0} + \left( \delta T_{0} \cosh(Dk_{x}^{2}t) - \frac{\delta B_{0}D_{TB} + d \delta T_{0}}{D}\sinh(Dk_{x}^{2}t) \right)\exp(-\bar{a}k_{x}^{2}t)\cos(k_{x}x),
    \end{equation}
    \begin{equation}
        \vec{B}(x,y,z,t) = \begin{pmatrix}
            0 \\
        B_{0} + \left( \delta B_{0} \cosh(Dk_{x}^{2}t) + \frac{d\delta B_{0} - \delta T_{0} D_{BT}}{D}\sinh(Dk_{x}^{2}t) \right)\exp(-\bar{a}k_{x}^{2}t)\cos(k_{x}x) \\
        0
        \end{pmatrix}.
    \end{equation}
\end{subequations}
With
\begin{align}
        &D_{T} = (\gamma -1)((1-x_{\text{ion}})\kappa_{n} + x_{\text{ion}}\kappa_{\perp}), \qquad
        &D_{B} = \eta + \eta_{AD}, \qquad
        &D_{TB} = \frac{(\gamma-1)x_{\text{ion}}\kappa_{J\wedge}}{\rho_{0}}, \\
        &D_{BT} = \frac{x_{\text{ion}}\kappa_{J\wedge}}{T_{0}}, \qquad
        &\bar{a} = \frac{1}{2}(D_{T}+D_{B}), \qquad
        &d = \frac{1}{2}(D_{T}-D_{B}), \\
        &D^{2} = D_{TB}D_{BT} + d^{2}.
\end{align}
This also demonstrates the well-posedness criterion in a model: when $D>\bar{a}$ the functions \\
$\cosh(Dk_{x}^{2}t)\exp(-\bar{a}k_{x}^{2}t)$ and $\sinh(Dk_{x}^{2}t)\exp(-\bar{a}k_{x}^{2}t)$ grow at a rate $\exp((D-\bar{a})k_{x}^{2}t)$. Since $\bar{a}^{2}-d^{2}=D_{T}D_{B}$, this happens when $D_{TB}D_{BT}>D_{T}D_{B}$. The rate of growth scales $k_{x}^{2}$, unbounded by the wavenumber. The system forms a backward diffusion equation. Refinement of the mesh shortens rather than extends the interval over which the solution is meaningful, and the initial-value problem is ill-posed.
\begin{figure}[H]
\centering
    \begin{subfigure}{0.95\linewidth}
    \centering
    \includegraphics[scale=0.8]{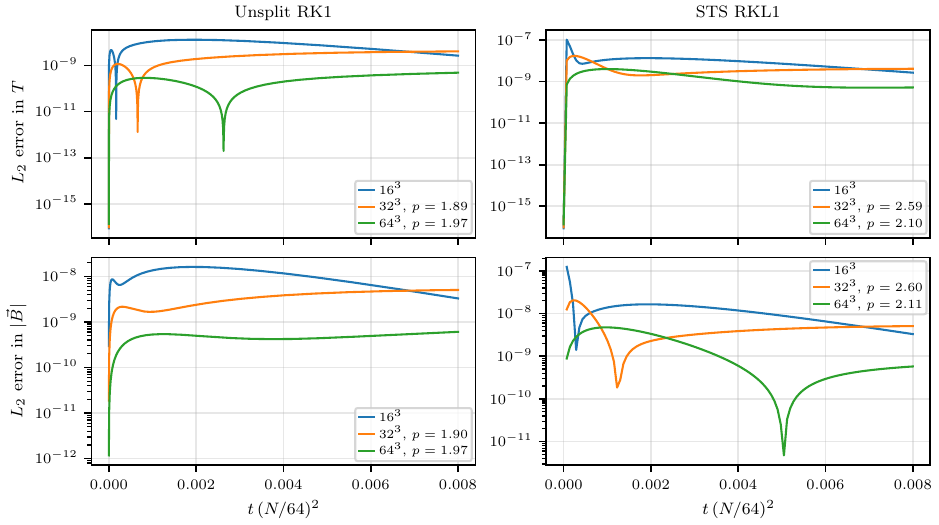}
          \caption{Well-posed thermo-magnetic system. $\kappa_{\perp}=7800$, $\eta_{AD}=5200$, $\kappa_{J\wedge}=6070$, $x_{\text{ion}}=1$.}
          \label{fig:NE-tests-wp}
        \end{subfigure}
    \begin{subfigure}{\linewidth}
    \centering
    \includegraphics[scale=0.8]{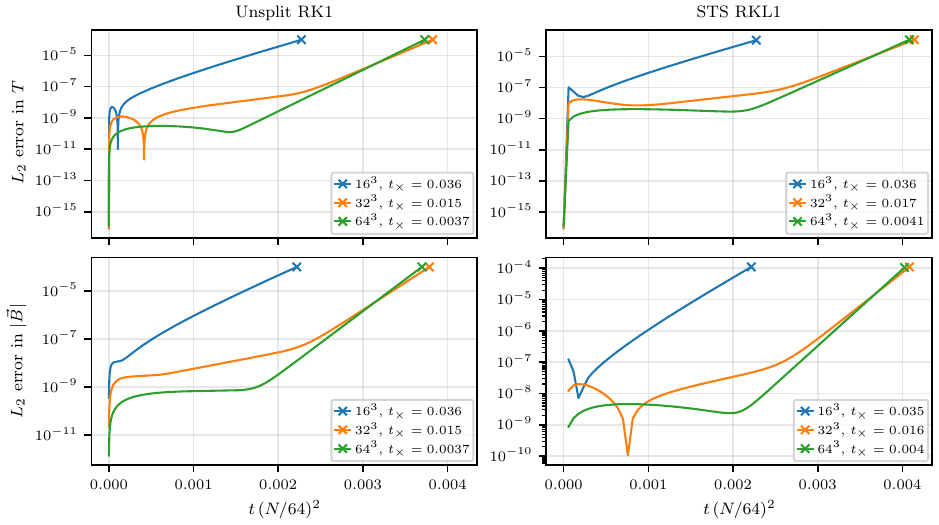}
          \caption{Ill-posed thermo-magnetic system. $\kappa_{\perp}=7800$, $\eta_{AD}=5200$, $\kappa_{J\wedge}=9010$, $x_{\text{ion}}=1$.}
          \label{fig:NE-tests-ip}
        \end{subfigure}
    \caption{Nernst/Ettingshausen tests.}
    \label{fig:NE-tests}
\end{figure}
Figure \ref{fig:NE-tests} shows the dissipative thermo-magnetic system (perpendicular thermal conduction, ambipolar diffusion, and Nernst/Ettingshausen effects) L2 errors at RK1 and RKL1 time integration. Subfigures \ref{fig:NE-tests-wp} and \ref{fig:NE-tests-ip} are simulations where the well-posedness condition of Section~\ref{sec:wellposed} holds and fails, respectively. The time axis is scaled by $(N/64)^{2}$, so that each resolution covers the same number of steps. As expected the ill-posed runs fail in the same manner as the Hall effect test does, with the error growing exponentially until blow-up. However, unlike the Hall effect, no integrator can stabilise them. Ill-posed runs fail earlier at higher resolution, with blow-up times scaling approximately as $O(\Delta x^{2})$. The well-posed runs converge at second order, displaying the same convergence behaviour of the other dissipative systems.
\subsection{Righi--Leduc Effect}
We set up a Beltrami magnetic field with $|B_{0}|=10^{-8}$, and unit vector $\hat{b}=(\cos(k_{B}z),\sin(k_{B}z),0)$. We initialise a transverse temperature gradient 
as 
\begin{equation}
T(x,y,z,0) = T_{0} + \delta T_{0}\cos(k_{T}x).    
\end{equation}
This has the analytical solution
\begin{equation}
    T(x,y,z,t) = T_{0} + \delta T \cos(D_{\wedge}k_{B}k_{T}\cos(k_{B}z)t - k_{T}x),
\end{equation}
with $D_{\wedge}=(\gamma-1)x_{\text{ion}}\kappa_{\wedge}$.
We take $T_{0}=1$ and $\delta T = 10^{-6}$. Our wavenumbers are $k_{B} = 2\pi/L_{z}$ and $k_{T}=2\pi/L_{x}$.
\begin{figure}[H]
    \centering
    \includegraphics[width=0.95\linewidth]{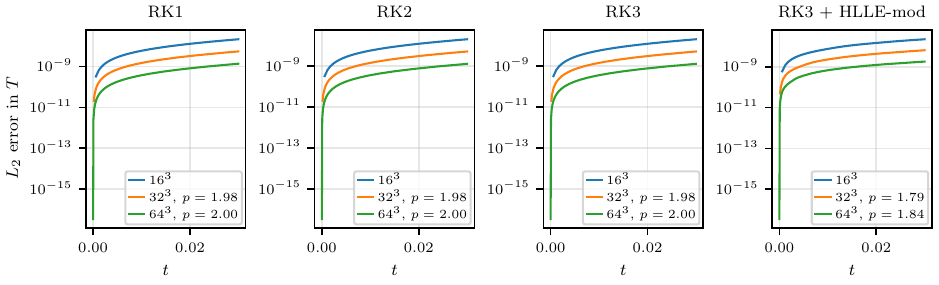}
    \caption{Righi--Leduc effect.}
    \label{fig:RL-test}
\end{figure}
Figure \ref{fig:RL-test} shows the Righi--Leduc L2 errors across RK1, RK2, and RK3 (with the standard and modified HLLE Riemann solver) time integration. The Righi--Leduc effect remains stable across all integrators and converges at second order (with some hindrance from HLLE-Mod). The errors are nearly identical for all three integrators (those at RK1 are about 10 per cent larger), so they are dominated by the spatial discretisation. Despite this, we do not expect this term to be stable at RK1 or RK2 in general. The instability is present in the RK1 and RK2 test, however, the growth rate of errors is limited by $|\vec{\nabla}\times\hat{b}|$, with error growth rates simply being too small to capture in this test. The temperature smoothly advects across the Beltrami field. In the HLLE-Mod white noise tests we show that the Righi--Leduc effect can fail even at RK3.
\newpage
\subsection{Gyroviscosity}
We take ideal MHD with only the $\boldsymbol{\Pi}^{(3)}$ and $\boldsymbol{\Pi}^{(4)}$ viscosity tensors, covering the gyro-viscous effects. The magnetic field is set along the $z$-axis ($\hat{b}=(0,0,1)$) and $|B_{0}|=1$. We start with zero background velocity and initialise the velocity perturbation as $\vec{v}(x,y,z,0) = v_{0}(\sin(k_{z}z), \sin(k_{x}x), 0)$, with $v_{0} = 10^{-6}$. this has the analytical solution
\begin{equation}
\vec{v}(x,y,z,t) = 
    v_{0}\begin{pmatrix}- \sin{\left(x_{\text{ion}}\nu_{\wedge 1}k_{x}^{2} t \right)} \sin{\left(k_{x} x \right)} + \sin{\left(k_{z} z \right)} \cos{\left(2x_{\text{ion}}\nu_{\wedge 2}k_{z}^{2} t \right)}\\
    \sin{\left(2x_{\text{ion}}\nu_{\wedge 2}k_{z}^{2} t \right)} \sin{\left(k_{z} z \right)} + \sin{\left(k_{x} x \right)} \cos{\left(x_{\text{ion}}\nu_{\wedge 1}k_{x}^{2} t \right)}\\
    0\end{pmatrix}.
\end{equation}
\begin{figure}[H]
    \centering
    \includegraphics[scale=0.95]{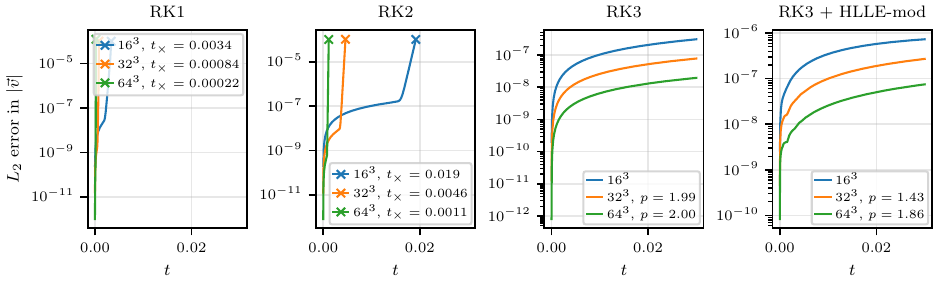}
          \caption{Gyro-viscosity test ($\boldsymbol{\Pi}^{(3)} + \boldsymbol{\Pi}^{(4)}$). $\nu_{\wedge1}=\nu_{\wedge2}=5200$, $x_{\text{ion}}=1$.}
    \label{fig:GV-test}
\end{figure}
Figure \ref{fig:GV-test} shows the gyroviscosity ($\boldsymbol{\Pi}^{(3)} + \boldsymbol{\Pi}^{(4)}$) L2 errors across RK1, RK2, and RK3 (with the standard and modified HLLE Riemann solver) time integration. Gyroviscosity behaves similarly to the Hall effect. It fails at both RK1 and RK2, with failure times scaling $O(\Delta x^{2})$. In a simulation with smooth fields, RK3 alone is enough to stabilise it, as no grid-scale instabilities form. HLLE-Mod once again hurts convergence rates at lower resolution due to the increased presence of numerical diffusion, and raises the errors by a factor of 2--4. 
\subsection{Peltier/Seebeck Effects (with Hall Effect)}
We initialise the background temperature and magnetic field as $T(x,y,z,0) = T_{0} + \delta T(0)\cos(k_{x}x)$ and $\vec{B}(x,y,z,0) = |B_{0}|(\cos(\theta),\sin(\theta),0)$. We take $T_{0}=1$, $|B_{0}| = 1$, $\delta T(0) = 10^{-6}$, and $\theta = \pi/4$ (maximum of Peltier/Seebeck effect).

With only the Hall effect and the parallel and perpendicular Peltier/Seebeck effects, this has the analytical solution
\small
\begin{subequations}
    \begin{equation}
       T(x,y,z,t) = T_{0} + \left(\delta{T}_{0} + \frac{1}{T_{0} \Omega^{2} \rho_{0}}\left(\delta{T}_{0} x_{\text{ion}}^{2} \left(\gamma - 1\right) \left(\kappa_{J\parallel} - \kappa_{J\perp}\right)^{2} \left(\cos{\left(\Omega k_{x}^{2} t \right)} - 1\right) \sin^{2}{\left(\theta \right)} \cos^{2}{\left(\theta \right)}\right)\right) \cos{\left(k_{x} x \right)},
    \end{equation}
    \begin{equation}
        \vec{B}(x,y,z,t) = \begin{pmatrix}B_{0} \cos{\left(\theta \right)}\\
        
        B_{0} \sin{\left(\theta \right)} - \frac{1}{T_{0} \Omega^{2}}\left(\delta{T}_{0} \eta_{H} x_{\text{ion}} \left(\kappa_{J\parallel} - \kappa_{J\perp}\right) \left(\cos{\left(\Omega k_{x}^{2} t \right)} - 1\right) \sin{\left(\theta \right)} \cos^{2}{\left(\theta \right)} \cos{\left(k_{x} x \right)}\right)\\
        
        - \frac{1}{T_{0} \Omega}\left(\delta{T}_{0} x_{\text{ion}} \left(\kappa_{J\parallel} - \kappa_{J\perp}\right) \sin{\left(\theta \right)} \sin{\left(\Omega k_{x}^{2} t \right)} \cos{\left(\theta \right)} \cos{\left(k_{x} x \right)} \right)\end{pmatrix},
    \end{equation}
\end{subequations}
\normalsize
with
\begin{align}
    \Omega^{2} = \frac{\cos^{2}{\left(\theta \right)}}{T_{0} \rho_{0}}
   & \left(T_{0} \eta_{H}^{2} \rho_{0} + \gamma \kappa_{J\parallel}^{2} x_{\text{ion}}^{2} \sin^{2}{\left(\theta \right)} - 2 \gamma \kappa_{J\parallel} \kappa_{J\perp} x_{\text{ion}}^{2}\sin^{2}{\left(\theta \right)} + \gamma \kappa_{J\perp}^{2} x_{\text{ion}}^{2} \sin^{2}{\left(\theta \right)} \right. \nonumber \\
  & \left.   - \kappa_{J\parallel}^{2} x_{\text{ion}}^{2} \sin^{2}{\left(\theta \right)} + 2 \kappa_{J\parallel} \kappa_{J\perp} x_{\text{ion}}^{2} \sin^{2}{\left(\theta \right)} - \kappa_{J\perp}^{2} x_{\text{ion}}^{2} \sin^{2}{\left(\theta \right)}\right).
\end{align}
We run with and without the Hall effect enabled to verify the CFL condition holds at RK3, and to confirm the Peltier/Seebeck physics alone is correctly implemented.
\begin{figure}[H]
\centering
        \begin{subfigure}{\linewidth}
    \centering
    \includegraphics[scale=0.95]{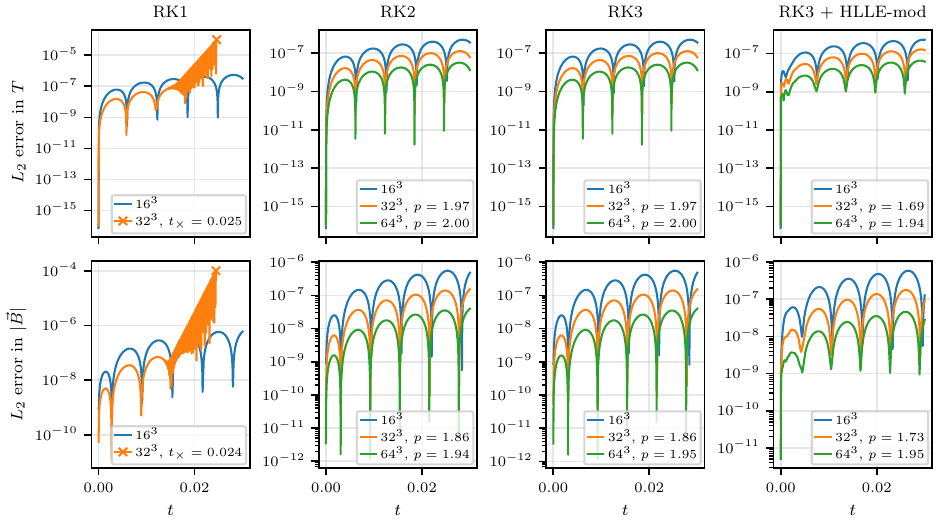}
          \caption{Peltier/Seebeck ($\vec{q}_{J,\parallel,\perp} + \vec{F}_{T,\parallel,\perp}$). $\kappa_{J\parallel}=19050$, $\kappa_{J\perp}=6350$, $x_{\text{ion}}=1$.}
          \label{fig:sub-ps-test}
        \end{subfigure}
    \begin{subfigure}{\linewidth}
    \centering
    \includegraphics[scale=0.95]{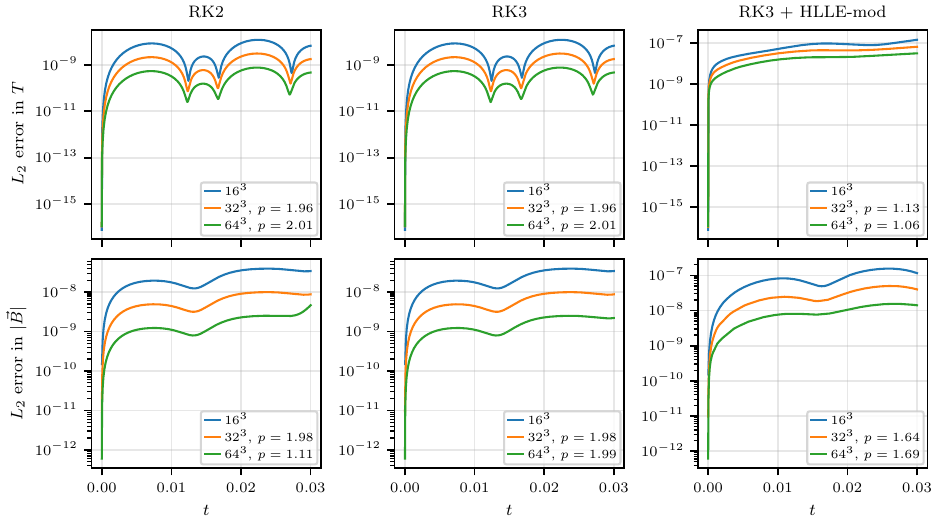}
          \caption{Combined Peltier/Seebeck ($\vec{q}_{J,\parallel,\perp} + \vec{F}_{T,\parallel,\perp}$) with Hall effect. $\eta_{H}=5200$, $\kappa_{J\parallel}=9000$, $\kappa_{J\perp}=6350$, $x_{\text{ion}}=1$.}
          \label{fig:hps-test}
        \end{subfigure}
    \caption{Peltier/Seebeck (+Hall) tests.}
    \label{fig:ps-test}
\end{figure}
Figure \ref{fig:ps-test} shows the Peltier/Seebeck (+Hall) L2 errors across RK1 (without the Hall effect only), RK2, and RK3 (with the standard and modified HLLE Riemann solver) time integration. Subfigures \ref{fig:sub-ps-test} and \ref{fig:hps-test} are without and with the Hall effect, respectively. The Peltier/Seebeck tests fail at RK1 as expected, blow-up times scaling with $O(\Delta x^{2})$. The RK2 runs converge, despite the stability analysis predicting failure. This is specific to the test case itself, and does not hold generally. With $\eta_{H}=0$ the analytical solution reduces to $T = T_{0} + \delta T_{0}\cos(\Omega k_{x}^{2}t)\cos(k_{x}x)$, with $\Omega^{2}=(\gamma-1)x_{\text{ion}}^{2}\Delta\kappa_{J}^{2}\sin^{2}\theta\cos^{2}\theta/(\rho_{0}T_{0})$, so without the Hall effect the temperature simply oscillates and the growth per step under RK2 is too small to appear within the run. 

With the Hall effect the growth does appear by the end of the $64^{3}$ run, where the order of the RK2 error in $|\vec{B}|$ falls to 1.11. At RK3 the systems stabilise and converge at second-order. The increased numerical diffusion from HLLE-Mod once again lowers convergence rates, most strongly for the temperature with the Hall effect, where the order falls to $\approx1$ and the error is 12--41 times that of the standard solver.

The RK1 one runs are omitted from Peltier/Seebeck + Hall as the timestep collapses after the first few steps, causing the simulation to effectively stall.
\begin{table}[H]
\centering
\small
\caption{Convergence order of the dissipative tests (and the Nernst/Ettingshausen tests) for unsplit RK1 and STS RKL1: $p_{32}$ and $p_{64}$ compare the $32^{3}$ and $64^{3}$ runs with the next coarser resolution, or give $t_{\times}$ where the finer run blew up at time $t$. For the Nernst/Ettingshausen runs every resolution runs for the same number of steps, to $t=0.128$, $0.032$ and $0.008$ at $16^{3}$, $32^{3}$ and $64^{3}$, and $p$ compares the largest errors up to $t=0.008$.}
\label{tab:conv-diss}
\begin{tabular}{lp{4.6cm}cccc}
\hline
 & & \multicolumn{2}{c}{Unsplit RK1} & \multicolumn{2}{c}{STS RKL1} \\
Test & Parameters & $p_{32}$ & $p_{64}$ & $p_{32}$ & $p_{64}$ \\
\hline
\multicolumn{6}{l}{\textit{Thermal conduction}} \\
Isotropic (stock Athena++) & $\kappa_{\text{iso}}=5500$ & 2.01 & 2.00 & 2.54 & 2.21 \\
Neutral & $\kappa_{n}=5500$, $x_{\text{ion}}=0$ & 2.01 & 2.00 & 2.54 & 2.21 \\
Parallel & $\kappa_{\parallel}=11000$, $x_{\text{ion}}=1$ & 2.04 & 2.01 & 2.60 & 2.10 \\
Perpendicular & $\kappa_{\perp}=11000$, $x_{\text{ion}}=1$ & 2.04 & 2.01 & 2.60 & 2.10 \\
Combined & $\kappa_{n}=\kappa_{\parallel}=\kappa_{\perp}=5500$, $x_{\text{ion}}=0.5$ & 2.01 & 2.00 & 2.54 & 2.21 \\
\multicolumn{6}{l}{\textit{Viscosity}} \\
Isotropic (stock Athena++) & $\nu_{\text{iso}}=3900$ & 1.99 & 2.00 & 2.54 & 2.09 \\
Neutral & $\nu_{n}=3900$, $x_{\text{ion}}=0$ & 2.01 & 2.00 & 2.53 & 2.09 \\
Bulk & $\zeta=5200$ & 2.02 & 2.01 & 2.60 & 2.11 \\
Parallel & $\nu_{\parallel}=15600$, $x_{\text{ion}}=1$ & 2.00 & 2.00 & 2.11 & 2.03 \\
Perpendicular 1 & $\nu_{\perp1}=5200$, $x_{\text{ion}}=1$ & 2.04 & 2.01 & 2.60 & 2.10 \\
Perpendicular 2 & $\nu_{\perp2}=1300$, $x_{\text{ion}}=1$ & 2.04 & 2.01 & 2.60 & 2.10 \\
Perpendicular 1+2 & $\nu_{\perp1}=5200$, $\nu_{\perp2}=1300$, $x_{\text{ion}}=1$ & 2.01 & 2.00 & 2.45 & 2.07 \\
Combined & $\zeta=5200$, $\nu_{n}=\nu_{\parallel}=\nu_{\perp1}=\nu_{\perp2}=3900$, $x_{\text{ion}}=0.5$ & 2.01 & 2.00 & 2.29 & 2.06 \\
\multicolumn{6}{l}{\textit{Magnetic diffusion}} \\
Ohmic (stock Athena++) & $\eta=5200$ & 2.04 & 2.01 & 2.60 & 2.10 \\
Ambipolar (stock Athena++) & $\eta_{AD}=5900$ & 2.01 & 2.00 & 2.45 & 2.07 \\
Combined (stock Athena++) & $\eta=2600$, $\eta_{AD}=3000$ & 2.02 & 2.01 & 2.53 & 2.09 \\
\multicolumn{6}{l}{\textit{Nernst/Ettingshausen}} \\
Well-posed, $T$ & $\kappa_{\perp}=7800$, $\eta_{AD}=5200$, $\kappa_{J\wedge}=6070$, $x_{\text{ion}}=1$ & 1.97 & 2.10 & 2.59 & 2.10 \\
Well-posed, $|\vec{B}|$ &  & 1.97 & 2.11 & 2.60 & 2.11 \\
Ill-posed, $T$ & $\kappa_{\perp}=7800$, $\eta_{AD}=5200$, $\kappa_{J\wedge}=9010$, $x_{\text{ion}}=1$ & $\times 0.015$ & $\times 0.0037$ & $\times 0.017$ & $\times 0.0041$ \\
Ill-posed, $|\vec{B}|$ &  & $\times 0.015$ & $\times 0.0037$ & $\times 0.016$ & $\times 0.0040$ \\
\hline
\end{tabular}
\end{table}
Table \ref{tab:conv-diss} summarises the convergence rates across all dissipative physics tests. For well-posed tests, RK1 runs converge at second order for all configurations. Using RKL1 the convergence rate is initially greater than two on coarser runs, but approaches two as resolution increases. The super-step is capped at 50 parabolic timesteps, it therefore scales as $\Delta x^{2}$ and the first-order temporal error of RKL1 also converges at second order in $\Delta x$. We expect on longer runs, for the convergence rates to eventually decrease and approach order 1 once the spatial discretisation error is subdominant.
\begin{table}[H]
\centering
\small
\setlength{\tabcolsep}{5pt}
\caption{Convergence order of the non-dissipative tests for RK1, RK2, RK3, and RK3 with HLLE-Mod: $p_{32}$ and $p_{64}$ compare the $32^{3}$ and $64^{3}$ runs with the next coarser resolution, or give $t_{\times}$ where the finer run blew up at time $t$.}
\label{tab:conv-nondiss}
\begin{tabular}{l>{\raggedright\arraybackslash}p{3.4cm}cccccccc}
\hline
 & & \multicolumn{2}{c}{RK1} & \multicolumn{2}{c}{RK2} & \multicolumn{2}{c}{RK3} & \multicolumn{2}{c}{\begin{tabular}{@{}c@{}}RK3 +\\HLLE-mod\end{tabular}} \\
Test & Parameters & $p_{32}$ & $p_{64}$ & $p_{32}$ & $p_{64}$ & $p_{32}$ & $p_{64}$ & $p_{32}$ & $p_{64}$ \\
\hline
\multicolumn{10}{l}{\textit{Hall effect}} \\
Hall & $\eta_{H}=6000$ & $\times 0.0021$ & $\times 0.0006$ & $-1.36$ & $\times 0.011$ & 1.99 & 2.00 & 1.65 & 1.92 \\
\multicolumn{10}{l}{\textit{Gyroviscosity}} \\
Wedge 1 & $\nu_{\wedge1}=5200$, $x_{\text{ion}}=1$ & $\times 0.0019$ & $\times 0.00053$ & $\times 0.026$ & $\times 0.0072$ & 1.99 & 2.00 & 1.45 & 1.28 \\
Wedge 2 & $\nu_{\wedge2}=5200$, $x_{\text{ion}}=1$ & $\times 0.00099$ & $\times 0.00027$ & $\times 0.013$ & $\times 0.0037$ & 1.99 & 2.00 & 1.20 & 1.51 \\
Combined & $\nu_{\wedge1}=\nu_{\wedge2}=5200$, $x_{\text{ion}}=1$ & $\times 0.00084$ & $\times 0.00022$ & $\times 0.0046$ & $\times 0.0011$ & 1.99 & 2.00 & 1.43 & 1.86 \\
\multicolumn{10}{l}{\textit{Peltier/Seebeck}} \\
Combined, $T$ & $\kappa_{J\parallel}=19050$, $\kappa_{J\perp}=6350$, $x_{\text{ion}}=1$ & $\times 0.025$ & $\times 0.0066$ & 1.97 & 2.00 & 1.97 & 2.00 & 1.69 & 1.94 \\
Combined, $|\vec{B}|$ &  & $\times 0.024$ & $\times 0.0066$ & 1.86 & 1.94 & 1.86 & 1.95 & 1.73 & 1.95 \\
With Hall, $T$ & $\eta_{H}=5200$, $\kappa_{J\parallel}=9000$, $\kappa_{J\perp}=6350$, $x_{\text{ion}}=1$ & $\times 0.006$ & $\times 0.0016$ & 1.96 & 2.01 & 1.96 & 2.01 & 1.13 & 1.06 \\
With Hall, $|\vec{B}|$ &  & $\times 0.0051$ & $\times 0.0014$ & 1.98 & 1.11 & 1.98 & 1.99 & 1.64 & 1.69 \\
\hline
\end{tabular}
\end{table}
Table \ref{tab:conv-nondiss} shows the convergence and blow-up times for the non-dissipative tests. As discussed previously, several tests appear to show RK2 convergence, but this is an artefact of the test problem and is not general. The growing oscillations are far smaller at RK2 than at RK1, allowing the simulations to appear stable over the duration of the test. We are also able to capture the failure of RK2 for the Hall effect via its negative convergence rate. HLLE-Mod can lower the order of convergence down to one due to the increased numerical dissipation: the modified signal speeds scale as $1/\Delta x$, so the dissipation they add decreases one order more slowly with resolution than that of the standard solver \cite{Toth_2008}. We expect on longer runs, for the convergence rates to eventually increase and approach order 3 once the spatial discretisation error and numerical diffusion are subdominant.

Our convergence orders match those reported for similar tests against analytical solutions in other codes. \cite{Vaidya_2017} tested anisotropic (parallel) thermal conduction with a diffusing Gaussian in PLUTO, finding second-order convergence of the L1 error with RKL2, but first-order convergence with RKL1. This is because the super-step was scaled with $\Delta x$. Here it is capped at 50 parabolic timesteps and so scales with $\Delta x^{2}$.  \cite{Berlok_2020} found second-order convergence of the root-mean-square difference between the simulated and analytical internal energy for Braginskii (parallel) viscosity in AREPO with RKL2.

Tests with discontinuous initial conditions converge more slowly. \cite{Sharma_2007} tested anisotropic thermal conduction (parallel and perpendicular) with a hot patch diffusing along circular field lines. They report L2 errors that fall only from 0.025--0.063 at $50^{2}$ to 0.010--0.054 at $400^{2}$, depending on the scheme. \cite{Kannan_2016} find L1 convergence orders of 0.41--0.49 for a simillar test with parallel thermal conduction. 

For the Hall effect, \cite{Toth_2008} found relative L1 errors of 0.42, 0.085, 0.019, and 0.0052 after one whistler wave period with 16, 32, 64, and 128 cells per wavelength respectively. They used a second-order Runge--Kutta scheme with the MC limiter. Our RK3 Hall test converges at a similar rate, with an error of about $10^{-2}$ of the perturbation amplitude at $64^{3}$ at the end of the run. 

In Athena++ with RK3, \cite{iwasaki_2025} found second-order convergence of the L2 error of linear waves of relative amplitude $10^{-6}$ with HLLD. HLLE-Mod, which assigns a separate signal speed to each variable, gives 1.65 and 1.92 for the Hall test, approaching second order similarly to their Hall-HLLmod. 

We are not aware of comparable published tests for the thermoelectric, Righi--Leduc, or gyroviscous terms.
\newpage
\subsection{HLLE-Mod Tests}
To test HLLE-mod we initialise a white noise background for the relevant variables: temperature for Righi--Leduc, field for Hall, field and temperature for Peltier/Seebeck (+ Hall), velocity for gyroviscosity, and all variables for a test with all effects active.

We perform a similar white noise test to \cite{iwasaki_2025}. We seed Gaussian white noise in the temperature, velocity, and magnetic field, and compare RK3 with the standard HLLE solver against RK3 with HLLE-mod.

Each noise field is generated from independent Gaussian deviates in every cell, filtered in Fourier space, and scaled to an RMS amplitude; the magnetic noise is the discrete curl of a filtered vector potential. The procedure is given in full in Appendix~D.

We use a periodic box of side $L=4$ with $N=32^{3}$ ($\Delta x=1/8$), filled with gas at rest with $\rho_{0}=1$ and $P_{0}=1/\gamma$ (so $c_{s}=1$), and no background field, $\vec{B}_{0}=\vec{0}$, so the field is the noise alone and has zero mean.

We set the effective dispersivity of every term to $\alpha_{\text{eff}}=10^{4}$ (with $x_{\text{ion}}=1$): $\eta_{H}=10^{4}$; $(\gamma-1)\kappa_{\wedge}=10^{4}$; $|\kappa_{J\parallel}-\kappa_{J\perp}|/(2\sqrt{\varepsilon_{\text{int}}})=10^{4}$; and; with the Braginskii closure (strongly magnetised limit) $\nu_{\wedge1}=\nu_{\wedge2}$, $2\nu_{\wedge2}=10^{4}$, the rate of the faster gyro-viscous mode. The combined Hall and Peltier/Seebeck test splits $\alpha_{\text{eff}}$ equally between the two terms.

We run to $20t_{w}$, where $t_{w} = 2\Delta x^{2}/(\pi \alpha_{\text{eff}}) \approx 9.95\times10^{-7}$ is the period of a wave with $\omega=\alpha_{\text{eff}}k^{2}$ at the Nyquist wavenumber. For every test except Hall alone we set the RMS of the field noise to $\sigma_{B}=10^{-6}$, otherwise the Lorentz forces accelerate the gas, with the HLLE numerical diffusion converting some of the magnetic energy into heat. This completely dominates the non-ideal effects. The Hall test keeps $\sigma_{B}=1$, as in \cite{iwasaki_2025}, since only the field is seeded and measured there. We set $\sigma_{T}=10^{-6}$ for the thermal conduction and thermoelectric tests, and $\sigma_{v}=10^{-6}$ for the gyroviscosity test. 
\begin{figure}[H]
    \centering
\begin{subfigure}{0.49\linewidth}
    \centering
        \includegraphics[scale=0.95]{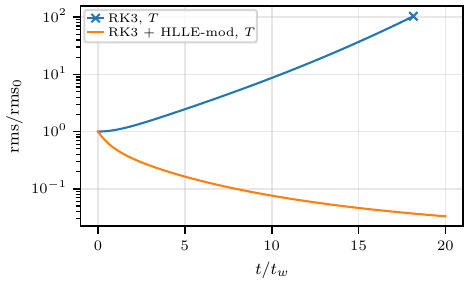}
        \caption{Righi--Leduc.}
    \label{fig:RL-Noise-RMS}
\end{subfigure}
\begin{subfigure}{0.49\linewidth}
    \centering
        \includegraphics[scale=0.95]{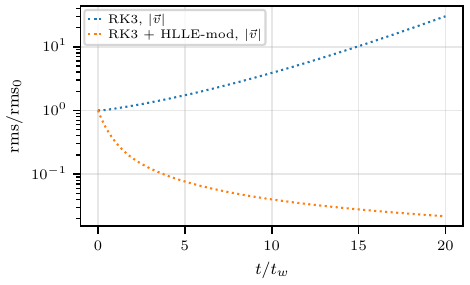}
        \caption{Gyroviscosity.}
    \label{fig:GV-Noise-RMS}
\end{subfigure}
\begin{subfigure}{0.49\linewidth}
    \centering
        \includegraphics[scale=0.95]{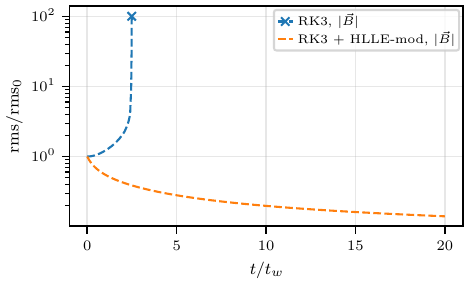}
        \caption{Hall Effect.}
    \label{fig:Hall-Noise-RMS}
\end{subfigure}
\begin{subfigure}{0.49\linewidth}
    \centering
        \includegraphics[scale=0.95]{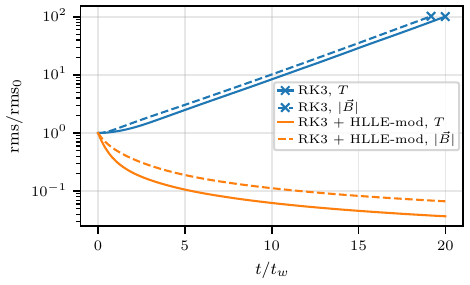}
        \caption{Peltier/Seebeck.}
    \label{fig:PS-Noise-RMS}
\end{subfigure}
    \caption{RMS of variables for each non-dissipative effect.}
    \label{fig:Noise-RMS}
\end{figure}
 
\begin{figure}[H]
    \ContinuedFloat
    \centering
\begin{subfigure}{0.49\linewidth}
    \centering
        \includegraphics[scale=0.95]{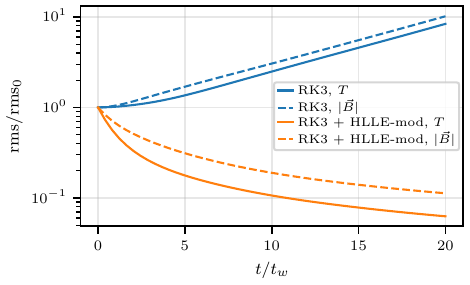}
        \caption{Peltier/Seebeck + Hall.}
    \label{fig:HPS-Noise-RMS}
\end{subfigure}
\begin{subfigure}{0.49\linewidth}
    \centering
        \includegraphics[scale=0.95]{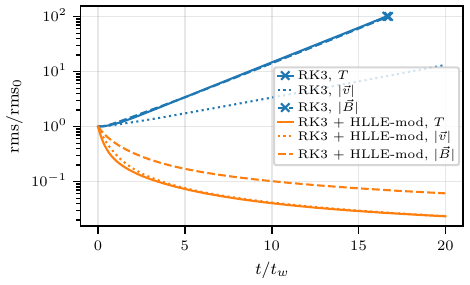}
        \caption{All effects.}
    \label{fig:All-Noise-RMS}
\end{subfigure}
    \caption[]{RMS of variables for each non-dissipative effect (continued).}
\end{figure}
Figure \ref{fig:Noise-RMS} shows the RMS of each seeded variable divided by its initial value. Using the standard HLLE Riemann solver the noise grows in every run under RK3. The Righi--Leduc, Peltier/Seebeck, and all-effects runs blow up between $17t_{w}$ and $19t_{w}$. The Hall run fails at $2.5t_{w}$, though strong ideal effects most likely speed up this failure time. The gyroviscous and combined Peltier/Seebeck and Hall runs survive to $20t_{w}$, but with the noise amplified by factors of about 30 and 10, respectively. This happens although the timesteps satisfy the limits of Section~3, which treat $\hat{b}$ as locally uniform. White noise in the field makes $\hat{b}$ vary from cell to cell, introducing non-linear instabilities. Under HLLE-Mod every seeded variable decays, to between 2 and 14 per cent of its initial RMS by $20t_{w}$.

\begin{figure}[H]
    \centering
        \includegraphics[scale=0.95]{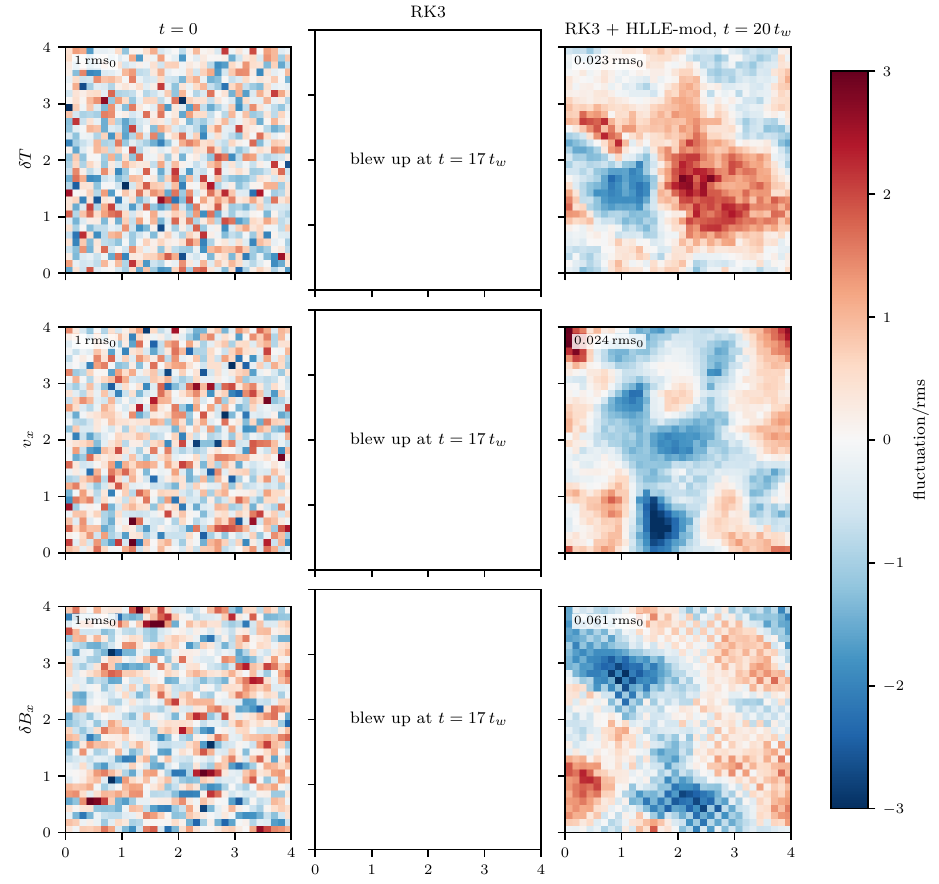}
    \caption{Slices through the noise fields at $t=0$ and $t=20t_{w}$ (or the time of blow-up) for the test with all effects active. Each field is shown normalised by its own RMS. Slices for the individual effects are shown in Appendix~D.}
    \label{fig:All-Noise-slice}
\end{figure}
Figure \ref{fig:All-Noise-slice} shows the $z=0$, $xy$-plane slice of the white noise test in which velocity, temperature, and field are all seeded with white noise, with all non-dissipative effects active. Going left to right the panels show $t=0$, RK3 with the standard solver (empty, as this run blew up at $17t_{w}$), and RK3 with HLLE-Mod at $t=20t_{w}$. Going top to bottom the panels show $\delta T$, $\delta v_{x}$, and $\delta B_{x}$. With HLLE-Mod the grid-scale noise is removed within $20t_{w}$ while larger-scale structure survives. Since white noise carries most of its power near the grid scale, the RMS falls to between 2 and 6 per cent of its initial value.
\newpage
\section{Conclusions}
We have implemented anisotropic thermal conduction, bulk and anisotropic viscosity, the Hall effect, and the thermoelectric effects of the Braginskii closure in Athena++. The thermal conduction comprises an isotropic neutral component and the parallel, perpendicular, and Righi--Leduc components of the ionised gas. The viscosity comprises the molecular bulk viscosity, an isotropic neutral component, and all five Braginskii ion viscosity tensors. In the neutral and ionised parts are weighted by the ionisation fraction. The thermoelectric terms are the Peltier, Seebeck, Nernst, and Ettingshausen effects. All terms enter through the face fluxes and the edge-centred electric fields of Athena++, and the dissipative parts can be integrated with super-time-stepping.

We have derived timestep limits for every term, valid for any Runge--Kutta integrator through the extent of its stability region along the real and imaginary axes. The dissipative terms (thermal conduction, viscosity, and magnetic diffusion) need only the negative real axis, so forward Euler suffices. The non-dissipative terms (the Hall, Righi--Leduc, gyroviscous, and Peltier/Seebeck terms) require a stability region that includes part of the imaginary axis, such as that of RK3. The Nernst/Ettingshausen terms are ill-posed on their own. We have derived the criterion for well-posedness: $(\gamma-1)\left((1-x_{\text{ion}})\kappa_{n}+x_{\text{ion}}\kappa_{\perp}\right)(\eta+\eta_{AD})\geq x_{\text{ion}}^{2}\kappa_{J\wedge}^{2}/\varepsilon_{\text{int}}$, which requires sufficient perpendicular thermal conduction and magnetic diffusion. The tests confirm second-order convergence for every term and the predicted well-posedness threshold. RK1 and RK2 are unstable for the non-dissipative terms, although with smooth fields the growth can be too slow to appear within a test.

The new signal speeds in HLLE-Mod keep runs with grid-scale noise bounded where RK3 with the standard solver fails. The cost is excess numerical dissipation when the non-dissipative terms are strong, which lowers the convergence order to between 1.1 and 1.9. To reduce this without resorting to hyper-resistivity, the approach could be extended to lower-dissipation HLL-type solvers, such as HLLD or LDHLLD \cite{MINOSHIMA_2021}. This would require a derivation for the intermediate states of a Riemann fan containing the new non-dissipative waves.

The Biermann battery, the one Ohm's-law term of the single-fluid Braginskii closure we have not implemented, is harder to treat than the others. First, a direct discretisation generates spurious field at shocks that grows with resolution, the `Biermann catastrophe' \cite{Graziani_2015}. Graziani et al. obtained convergence by writing the term in terms of the electron temperature, which is continuous through a shock only when it is evolved separately from the ion temperature. Second, the battery term linearised on its own is nilpotent, with only zero eigenvalues, so its stability limit comes from its second-order coupling to the internal-energy equation. Once coupled, the term supports thermal--magnetic waves whose speed depends on the background gradients, requiring new signal speeds and a further stability analysis. A single-fluid, single-temperature implementation exists \cite{Attia_2021}; adapting it to Athena++ is left to future work.
\section*{Code availability}
Athena-Aniso is available at \url{https://github.com/AlexandraAlbery/Athena-Aniso} DOI: 10.5281/zenodo.23887593, and the repository documentation describes the input parameters and the interface for supplying spatially varying coefficients.

The scripts used to verify the analytical and numerical results in this paper are available at \url{https://github.com/AlexandraAlbery/athena-aniso-python-scripts}
\section*{Data availability}
Data files for the tests of Section~6 are available upon a reasonable request to the corresponding author.
\section*{CRediT Authorship Contribution Statement}
\textbf{Alexandra F. Albery:} Conceptualisation, Methodology, Software, Validation, Formal analysis, Investigation, Writing - Original Draft, Visualisation. \\
\textbf{Julian M. Pittard:} Writing - Review \& Editing, Supervision, Project administration, Funding acquisition.
 
\section*{Declaration of Competing Interest}
The authors declare that they have no known competing financial interests or
personal relationships that could have appeared to influence the work reported
in this paper.
 
\section*{Acknowledgements}
This work was supported by the United Kingdom Research and Innovation (UKRI) Science and Technology Facilities Council (STFC) grant
ST/Y509711/1.

\section*{Declaration of Generative AI and AI-assisted Technologies in the Manuscript Preparation Process}
 During the preparation of this work, the authors used Claude Opus 5.5 (Anthropic) for proof reading and formatting of \LaTeX. After using this service, the authors reviewed and edited the content as needed and take full responsibility for the content of the published article.
 \newpage
\section*{Appendix~A - 1D Viscosity Terms}
    \begin{equation}
        \boldsymbol{W} = \begin{pmatrix}
            \frac{4}{3}\partial_{x}v_{x} & \partial_{x}v_{y} & \partial_{x}v_{z} \\
            \partial_{x}v_{y} & -\frac{2}{3}\partial_{x}v_{x} & 0 \\
            \partial_{x}v_{z} & 0 & -\frac{2}{3}\partial_{x}v_{x}
        \end{pmatrix},
    \end{equation}
    \begin{equation}
        \hat{b}\cdot\boldsymbol{W}\cdot\hat{b} = b_{x}\left[\frac{4}{3}b_{x}\partial_{x}v_{x} + b_{y}\partial_{x}v_{y} + b_{z}\partial_{x}v_{z} \right] 
        +b_{y}\left[ b_{x}\partial_{x}v_{y} - \frac{2}{3}b_{y}\partial_{x}v_{x} \right] + b_{z}\left[ b_{x}\partial_{x}v_{z} - \frac{2}{3}b_{z}\partial_{x}v_{x} \right],
    \end{equation}
    \begin{equation}
        \hat{b}\otimes\hat{b}-\frac{1}{3}\boldsymbol{I} = \begin{pmatrix}
            b_{x}^{2}-\frac{1}{3} & b_{x}b_{y} & b_{x}b_{z} \\
            b_{y}b_{x} & b_{y}^{2}-\frac{1}{3} & b_{y}b_{z} \\
            b_{z}b_{x} & b_{z}b_{y} & b_{z}^{2} - \frac{1}{3}
        \end{pmatrix},
    \end{equation}
    \begin{equation}
        \boldsymbol{I}_{\perp} = \begin{pmatrix}
            1 - b_{x}^{2} & -b_{x}b_{y} & -b_{x}b_{z} \\
            -b_{y}b_{x} & 1 - b_{y}^{2} & -b_{y}b_{z} \\ 
            -b_{z}b_{x} & -b_{z}b_{y} & 1-b_{z}^{2}
        \end{pmatrix},
    \end{equation}
    \begin{subequations}
    \begin{align}
       (\boldsymbol{W}_{\perp})_{xx} &= \left(1-b_{x}^{2}\right) \left(\frac{2}{3}(2-3b_{x}^{2})\partial_{x} v_{x} - 2b_{x} b_{y}\partial_x v_{y} - 2b_{x} b_{z}\partial_{x} v_{z}\right),\\
        (\boldsymbol{W}_{\perp})_{yy} &= \left[2b_{x}^{2} b_{y}^{2} - \frac{2}{3}(1-b_{y}^{2})\right]\partial_{x} v_{x} - 2b_{x} b_{y}(1-b_y^2)\partial_x v_y + 2b_{x} b_{y}^{2} b_{z}\partial_{x} v_{z},\\
    (\boldsymbol{W}_{\perp})_{zz} &= \left[2b_{x}^2 b_{z}^2 - \frac{2}{3}(1-b_{z}^2)\right]\partial_{x} v_{x} + 2b_{x} b_{y} b_{z}^{2}\partial_{x} v_{y} - 2b_{x} b_{z}(1-b_{z}^{2})\partial_{x} v_{z},\\
        (\boldsymbol{W}_{\perp})_{xy} = (\boldsymbol{W}_{\perp})_{yx} &= 2b_{x} b_{y}\left(b_{x}^{2}-\frac{2}{3}\right)\partial_{x} v_{x} + \left(b_{z}^{2} + 2b_{x}^{2} b_{y}^{2}\right)\partial_{x} v_{y} + b_{y} b_{z}\left(2b_{x}^{2}-1\right)\partial_{x} v_{z},\\
        (\boldsymbol{W}_{\perp})_{xz} = (\boldsymbol{W}_{\perp})_{zx} &= 2b_{x} b_{z}\left(b_{x}^{2}-\frac{2}{3}\right)\partial_{x} v_{x} + b_{y} b_{z}\left(2b_{x}^{2}-1\right)\partial_{x} v_{y} + \left(b_{y}^{2} + 2b_{x}^{2} b_{z}^{2}\right)\partial_{x} v_{z},\\
        (\boldsymbol{W}_{\perp})_{yz} = (\boldsymbol{W}_\perp)_{zy} &= 2b_{y} b_{z}\left(b_{x}^{2}+\frac{1}{3}\right)\partial_{x} v_{x} + b_{x} b_{z}\left(2b_{y}^{2}-1\right)\partial_{x} v_{y} + b_{x} b_{y}\left(2b_{z}^{2}-1\right)\partial_{x} v_{z}.
    \end{align}
    \end{subequations}
    
    \begin{subequations}
    \begin{align}
    (\boldsymbol{M}_{1})_{xx} &= 4b_{x}^{2}(1-b_{x}^{2})\partial_{x} v_{x} + 2b_{x} b_{y}(1-2b_{x}^{2})\partial_{x} v_{y} + 2b_{x} b_{z}(1-2b_{x}^{2})\partial_{x} v_{z},\\
        (\boldsymbol{M}_{1})_{yy} &= -4b_{x}^{2} b_{y}^{2}\partial_{x} v_{x} + 2b_{x} b_{y}(1-2b_{y}^{2})\partial_{x} v_{y} - 4b_{x} b_{y}^{2} b_{z}\partial_{x} v_{z},\\
        (\boldsymbol{M}_{1})_{zz} &= -4b_{x}^{2} b_{z}^{2}\partial_{x} v_{x} - 4b_{x} b_{y} b_{z}^{2}\partial_{x} v_{y} + 2b_{x} b_{z}(1-2b_{z}^{2})\partial_{x} v_{z},\\
        (\boldsymbol{M}_{1})_{xy} = (\boldsymbol{M}_{1})_{yx} &= 2b_{x} b_{y}(1-2b_{x}^2)\partial_{x} v_{x} + (b_{x}^2 + b_{y}^2 - 4b_{x}^2 b_{y}^2)\partial_{x} v_{y} + b_{y} b_{z}(1-4b_{x}^2)\partial_{x} v_{z},\\
        (\boldsymbol{M}_{1})_{xz} = (\boldsymbol{M}_{1})_{zx} &= 2b_{x} b_{z}(1-2b_{x}^{2})\partial_{x} v_{x} + b_{y} b_{z}(1-4b_{x}^2)\partial_{x} v_{y} + (b_{x}^{2} + b_{z}^{2} - 4b_{x}^{2} b_{z}^{2})\partial_{x} v_{z}, \\
        (\boldsymbol{M}_{1})_{yz} = (\boldsymbol{M}_{1})_{zy} &= -4b_{x}^{2} b_{y} b_{z}\partial_{x} v_{x} + b_{x} b_{z}(1-4b_{y}^{2})\partial_{x} v_{y} + b_{x} b_{y}(1-4b_{z}^{2})\partial_{x} v_{z},
    \end{align}
        \end{subequations}
    where: $\boldsymbol{I}_{\perp}\cdot\boldsymbol{W}\cdot(\hat{b}\otimes\hat{b}) + (\hat{b}\otimes\hat{b})\cdot\boldsymbol{W} \cdot \boldsymbol{I}_{\perp}  = \boldsymbol{M}$.

\begin{subequations}
\begin{align}
(\boldsymbol{W}\cdot\boldsymbol{I}_{\perp})_{xx} &= \frac{4}{3}\partial_{x}v_{x}(1-b_{x}^{2}) - b_{x}b_{y}\partial_{x}v_{y} - b_{x}b_{z}\partial_{x}v_{z}, \\
(\boldsymbol{W}\cdot\boldsymbol{I}_{\perp})_{yy} &= -\frac{2}{3}(1-b_{y}^{2})\partial_{x}v_{x} - b_{x}b_{y}\partial_{x}v_{y}, \\
(\boldsymbol{W}\cdot\boldsymbol{I}_{\perp})_{zz} &= -\frac{2}{3}(1-b_{z}^{2})\partial_{x}v_{x} - b_{x}b_{z}\partial_{x}v_{z}, \\
(\boldsymbol{W}\cdot\boldsymbol{I}_{\perp})_{xy} &= - \frac{4}{3}b_{x}b_{y}\partial_{x}v_{x} + (1-b_{y}^{2})\partial_{x}v_{y} - b_{y}b_{z}\partial_{x}v_{z}, \\
(\boldsymbol{W}\cdot\boldsymbol{I}_{\perp})_{xz} &= - \frac{4}{3}\partial_{x}v_{x}b_{x}b_{z} - b_{y}b_{z}\partial_{x}v_{y} + (1-b_{z}^{2})\partial_{x}v_{z},  \\
(\boldsymbol{W}\cdot\boldsymbol{I}_{\perp})_{yx} &=  \frac{2}{3}b_{x}b_{y}\partial_{x}v_{x} + (1-b_{x}^{2})\partial_{x}v_{y}, \\
(\boldsymbol{W}\cdot\boldsymbol{I}_{\perp})_{yz} &= \frac{2}{3}b_{y}b_{z}\partial_{x}v_{x} -b_{x}b_{z}\partial_{x}v_{y}, \\
(\boldsymbol{W}\cdot\boldsymbol{I}_{\perp})_{zx} &= \frac{2}{3}b_{x}b_{z}\partial_{x}v_{x} + (1-b_{x}^{2})\partial_{x}v_{z},  \\
(\boldsymbol{W}\cdot\boldsymbol{I}_{\perp})_{zy} &=  \frac{2}{3}b_{y}b_{z}\partial_{x}v_{x} -b_{x}b_{y}\partial_{x}v_{z} .
\end{align}
\end{subequations}

\begin{subequations}
\begin{align}
(\boldsymbol{I}_{\perp}\cdot\boldsymbol{W})_{xx} &= \frac{4}{3}\partial_{x}v_{x}(1-b_{x}^{2}) - b_{x}b_{y}\partial_{x}v_{y} - b_{x}b_{z}\partial_{x}v_{z}, \\
(\boldsymbol{I}_{\perp}\cdot\boldsymbol{W})_{yy} &= -\frac{2}{3}(1-b_{y}^{2})\partial_{x}v_{x} - b_{x}b_{y}\partial_{x}v_{y}, \\
(\boldsymbol{I}_{\perp}\cdot\boldsymbol{W})_{zz} &= -\frac{2}{3}(1-b_{z}^{2})\partial_{x}v_{x} - b_{x}b_{z}\partial_{x}v_{z}, \\
(\boldsymbol{I}_{\perp}\cdot\boldsymbol{W})_{xy} &=  \frac{2}{3}b_{x}b_{y}\partial_{x}v_{x} + (1-b_{x}^{2})\partial_{x}v_{y}, \\
(\boldsymbol{I}_{\perp}\cdot\boldsymbol{W})_{xz} &= \frac{2}{3}b_{x}b_{z}\partial_{x}v_{x} + (1-b_{x}^{2})\partial_{x}v_{z},  \\
(\boldsymbol{I}_{\perp}\cdot\boldsymbol{W})_{yx} &= - \frac{4}{3}b_{x}b_{y}\partial_{x}v_{x} + (1-b_{y}^{2})\partial_{x}v_{y} - b_{y}b_{z}\partial_{x}v_{z}, \\
(\boldsymbol{I}_{\perp}\cdot\boldsymbol{W})_{yz} &=  \frac{2}{3}b_{y}b_{z}\partial_{x}v_{x} -b_{x}b_{y}\partial_{x}v_{z}, \\
(\boldsymbol{I}_{\perp}\cdot\boldsymbol{W})_{zx} &= - \frac{4}{3}\partial_{x}v_{x}b_{x}b_{z} - b_{y}b_{z}\partial_{x}v_{y} + (1-b_{z}^{2})\partial_{x}v_{z},  \\
(\boldsymbol{I}_{\perp}\cdot\boldsymbol{W})_{zy} &= \frac{2}{3}b_{y}b_{z}\partial_{x}v_{x} -b_{x}b_{z}\partial_{x}v_{y}.
\end{align}
\end{subequations}

\begin{subequations}
    \begin{align}
(\hat{b}\times\boldsymbol{W}\cdot\boldsymbol{I}_{\perp})_{xx} &= (1-b_{x}^{2})\bigl(b_{y}\partial_{x}v_{z} - b_{z}\partial_{x}v_{y}\bigr), \\
(\hat{b}\times\boldsymbol{W}\cdot\boldsymbol{I}_{\perp})_{xy} &= \frac{2}{3}b_{z}\partial_{x}v_{x} + b_{x}b_{y}b_{z}\partial_{x}v_{y} - b_{x}b_{y}^{2}\partial_{x}v_{z}, \\
(\hat{b}\times\boldsymbol{W}\cdot\boldsymbol{I}_{\perp})_{xz} &= -\frac{2}{3}b_{y}\partial_{x}v_{x} + b_{x}b_{z}^{2}\partial_{x}v_{y} - b_{x}b_{y}b_{z}\partial_{x}v_{z}, \\
(\hat{b}\times\boldsymbol{W}\cdot\boldsymbol{I}_{\perp})_{yx} &=\left(\frac{4}{3}-2b_{x}^{2}\right)b_{z}\partial_{x}v_{x} - b_{x}b_{y}b_{z}\partial_{x}v_{y} - b_{x}(b_{y}^{2}+2b_{z}^{2})\partial_{x}v_{z}, \\
(\hat{b}\times\boldsymbol{W}\cdot\boldsymbol{I}_{\perp})_{yy} &= -2b_{x}b_{y}b_{z}\partial_{x}v_{x} + (1-b_{y}^{2})b_{z}\partial_{x}v_{y} + (b_{x}^{2}-b_{z}^{2})b_{y}\partial_{x}v_{z}, \\
(\hat{b}\times\boldsymbol{W}\cdot\boldsymbol{I}_{\perp})_{yz} &= \left(\frac{2}{3}-2b_{z}^{2}\right)b_{x}\partial_{x}v_{x} - b_{y}b_{z}^{2}\partial_{x}v_{y} + (2b_{x}^{2}+b_{y}^{2})b_{z}\partial_{x}v_{z}, \\
(\hat{b}\times\boldsymbol{W}\cdot\boldsymbol{I}_{\perp})_{zx} &= -\left(\frac{4}{3}-2b_{x}^{2}\right)b_{y}\partial_{x}v_{x} + b_{x}(2b_{y}^{2}+b_{z}^{2})\partial_{x}v_{y} + b_{x}b_{y}b_{z}\partial_{x}v_{z}, \\
(\hat{b}\times\boldsymbol{W}\cdot\boldsymbol{I}_{\perp})_{zy} &=\left(2b_{y}^{2}-\frac{2}{3}\right)b_{x}\partial_{x}v_{x} - b_{y}(2b_{x}^{2}+b_{z}^{2})\partial_{x}v_{y} + b_{y}^{2}b_{z}\partial_{x}v_{z}, \\
(\hat{b}\times\boldsymbol{W}\cdot\boldsymbol{I}_{\perp})_{zz} &= 2b_{x}b_{y}b_{z}\partial_{x}v_{x} + (b_{y}^{2}-b_{x}^{2})b_{z}\partial_{x}v_{y} - (1-b_{z}^{2})b_{y}\partial_{x}v_{z}.
    \end{align}
\end{subequations}

\begin{subequations}
    \begin{align}
        (\boldsymbol{I}_{\perp}\cdot\boldsymbol{W}\times\hat{b})_{xx} &= b_{x}^{2} b_{y} \partial_{x}v_{z} - b_{x}^{2} b_{z} \partial_{x}v_{y} - b_{y} \partial_{x}v_{z} + b_{z} \partial_{x}v_{y}, \\
        (\boldsymbol{I}_{\perp}\cdot\boldsymbol{W}\times\hat{b})_{xy} &= - b_{x}^{3} \partial_{x}v_{z} + 2 b_{x}^{2} b_{z} \partial_{x}v_{x} + b_{x} b_{y} b_{z} \partial_{x}v_{y} + b_{x} b_{z}^{2} \partial_{x}v_{z} + b_{x} \partial_{x}v_{z} - \frac{4 b_{z} \partial_{x}v_{x}}{3}, \\
        (\boldsymbol{I}_{\perp}\cdot\boldsymbol{W}\times\hat{b})_{xz} &=b_{x}^{3} \partial_{x}v_{y} - 2 b_{x}^{2} b_{y} \partial_{x}v_{x} - b_{x} b_{y}^{2} \partial_{x}v_{y} - b_{x} b_{y} b_{z} \partial_{x}v_{z} - b_{x} \partial_{x}v_{y} + \frac{4 b_{y} \partial_{x}v_{x}}{3}, \\
        (\boldsymbol{I}_{\perp}\cdot\boldsymbol{W}\times\hat{b})_{yx} &= b_{x} b_{y}^{2} \partial_{x}v_{z} - b_{x} b_{y} b_{z} \partial_{x}v_{y} - \frac{2 b_{z} \partial_{x}v_{x}}{3}, \\
        (\boldsymbol{I}_{\perp}\cdot\boldsymbol{W}\times\hat{b})_{yy} &= - b_{x}^{2} b_{y} \partial_{x}v_{z} + 2 b_{x} b_{y} b_{z} \partial_{x}v_{x} + b_{y}^{2} b_{z} \partial_{x}v_{y} + b_{y} b_{z}^{2} \partial_{x}v_{z} - b_{z} \partial_{x}v_{y}, \\
        (\boldsymbol{I}_{\perp}\cdot\boldsymbol{W}\times\hat{b})_{yz} &= b_{x}^{2} b_{y} \partial_{x}v_{y} - 2 b_{x} b_{y}^{2} \partial_{x}v_{x} + \frac{2 b_{x} \partial_{x}v_{x}}{3} - b_{y}^{3} \partial_{x}v_{y} - b_{y}^{2} b_{z} \partial_{x}v_{z} + b_{y} \partial_{x}v_{y}, \\
        (\boldsymbol{I}_{\perp}\cdot\boldsymbol{W}\times\hat{b})_{zx} &= b_{x} b_{y} b_{z} \partial_{x}v_{z} - b_{x} b_{z}^{2} \partial_{x}v_{y} + \frac{2 b_{y} \partial_{x}v_{x}}{3}, \\
        (\boldsymbol{I}_{\perp}\cdot\boldsymbol{W}\times\hat{b})_{zy} &= - b_{x}^{2} b_{z} \partial_{x}v_{z} + 2 b_{x} b_{z}^{2} \partial_{x}v_{x} - \frac{2 b_{x} \partial_{x}v_{x}}{3} + b_{y} b_{z}^{2} \partial_{x}v_{y} + b_{z}^{3} \partial_{x}v_{z} - b_{z} \partial_{x}v_{z}, \\
        (\boldsymbol{I}_{\perp}\cdot\boldsymbol{W}\times\hat{b})_{zz} &=b_{x}^{2} b_{z} \partial_{x}v_{y} - 2 b_{x} b_{y} b_{z} \partial_{x}v_{x} - b_{y}^{2} b_{z} \partial_{x}v_{y} - b_{y} b_{z}^{2} \partial_{x}v_{z} + b_{y} \partial_{x}v_{z}.
    \end{align}
\end{subequations}
\newpage
\section*{Appendix~B - Details of the Eigenvalue Derivations}
\subsection*{B.1 - Viscosity}
With only $x$-direction derivatives, and constant magnetic unit vector and density, the bulk, isotropic, parallel, and perpendicular parts of the viscosity of Section~\ref{sec:visc_cfl} reduce to
\begin{equation}
\partial_{t}\vec{v} = \begin{pmatrix}
M_{xx}^{(\nu,\parallel,\perp)} & M_{xy}^{(\nu,\parallel,\perp)} & M_{xz}^{(\nu,\parallel,\perp)} \\
M_{xy}^{(\nu,\parallel,\perp)} & M_{yy}^{(\nu,\parallel,\perp)} & M_{yz}^{(\nu,\parallel,\perp)} \\
M_{xz}^{(\nu,\parallel,\perp)} & M_{yz}^{(\nu,\parallel,\perp)} & M_{zz}^{(\nu,\parallel,\perp)}
\end{pmatrix}  \partial_{xx}\vec{v},
\end{equation}
with
\begin{subequations}
\begin{align}
M_{xx}^{(\nu,\parallel,\perp)} &= \frac{\zeta}{\rho} + \frac{4}{3}(1-x_{\text{ion}})\nu_{\text{neu}} + x_{\text{ion}}\left[3\nu_{\parallel}\left(u-\frac{1}{3}\right)^{2} + \nu_{\perp1}(1-u)^{2} + 16\nu_{\perp2}u(1-u)\right], \\
M_{yy}^{(\nu,\parallel,\perp)} &= (1-x_{\text{ion}})\nu_{\text{neu}} + x_{\text{ion}}\left[\nu_{\perp1}(1-u) + 4\nu_{\perp2}u\right] + x_{\text{ion}}b_{y}^{2}\left[3\nu_{\parallel}u - \nu_{\perp1}(1-u) + 4\nu_{\perp2}(1-4u)\right], \\
M_{zz}^{(\nu,\parallel,\perp)} &= (1-x_{\text{ion}})\nu_{\text{neu}} + x_{\text{ion}}\left[\nu_{\perp1}(1-u) + 4\nu_{\perp2}u\right] + x_{\text{ion}}b_{z}^{2}\left[3\nu_{\parallel}u - \nu_{\perp1}(1-u) + 4\nu_{\perp2}(1-4u)\right], \\
M_{xy}^{(\nu,\parallel,\perp)} &=  M_{yx}^{(\nu,\parallel,\perp)} = x_{\text{ion}}b_{x}b_{y}\left[3\nu_{\parallel}\left(u-\frac{1}{3}\right) - \nu_{\perp1}(1-u) + 8\nu_{\perp2}(1-2u)\right], \\
M_{xz}^{(\nu,\parallel,\perp)} &= M_{zx}^{(\nu,\parallel,\perp)} = x_{\text{ion}}b_{x}b_{z}\left[3\nu_{\parallel}\left(u-\frac{1}{3}\right) - \nu_{\perp1}(1-u) + 8\nu_{\perp2}(1-2u)\right], \\
M_{yz}^{(\nu,\parallel,\perp)} &= M_{zy}^{(\nu,\parallel,\perp)} = x_{\text{ion}}b_{y}b_{z}\left[3\nu_{\parallel}u - \nu_{\perp1}(1-u) + 4\nu_{\perp2}(1-4u)\right].
\end{align}
\end{subequations}
The eigenvalues of this system form the characteristic polynomial
\begin{equation}
    -\lambda_{\nu \parallel,\perp}^{3} + \text{Tr}(\boldsymbol{M}^{(\nu,\parallel,\perp)})\lambda_{\nu \parallel,\perp}^{2} - \frac{1}{2}(\text{Tr}\left(\boldsymbol{M}^{(\nu,\parallel,\perp)}\right)^{2} - \text{Tr}(\boldsymbol{M}^{(\nu)^{2}}))\lambda_{\nu \parallel,\perp} + \text{det}(\boldsymbol{M}^{(\nu,\parallel,\perp)}) =0.
\end{equation}
Substituting the trigonometric form of the magnetic unit vector, the coefficients of this polynomial are lengthy, but the $\phi$-dependence can be removed exactly. The entries above are those at $\phi=0$ rotated by $\phi$ about $\hat{x}$,
\begin{equation}
    \boldsymbol{M}^{(\nu,\parallel,\perp)}(\theta,\phi)=\boldsymbol{Q}(\phi)\boldsymbol{M}^{(\nu,\parallel,\perp)}(\theta,0)\boldsymbol{Q}(\phi)^{T}, \qquad \boldsymbol{Q}(\phi)=\begin{pmatrix}1&0&0\\0&\cos\phi&-\sin\phi\\0&\sin\phi&\cos\phi\end{pmatrix}.
\end{equation}
The characteristic polynomial, and hence every eigenvalue, is independent of $\phi$ and depends on the orientation only through $u$. At $\phi=0$ every coupling to $v_{z}$ is proportional to $b_{z}=0$, so the matrix splits into an $xy$ block and $M_{zz}=\lambda_{\nu,\parallel,\perp,z}$, and the characteristic polynomial factorises as
\begin{equation}
    (\lambda_{\nu \parallel,\perp}-\lambda_{\nu \parallel,\perp,z})(-\lambda_{\nu \parallel,\perp}^{2}+Tr(\boldsymbol{M}^{(\nu,\parallel,\perp)}_{xy})\lambda_{\nu \parallel,\perp}-\text{Det}(\boldsymbol{M}^{(\nu,\parallel,\perp)}_{xy})) = 0.
\end{equation}
The coefficients are given in Section~\ref{sec:visc_cfl}. The $xy$-subsystem contains the largest eigenvalue of the entire system. Taking $\hat{w}=(\sin\theta,-\cos\theta)$, the in-plane direction perpendicular to $\hat{b}$ is
\begin{equation}
    \lambda_{\nu \parallel,\perp,xy+}\geq\hat{w}\cdot\boldsymbol{M}^{(\nu,\parallel,\perp)}_{xy}\cdot\hat{w}=\lambda_{\nu,\parallel,\perp,z}+(1-u)\left[\frac{\zeta}{\rho}+\frac{(1-x_{\text{ion}})\nu_{\text{neu}}+x_{\text{ion}}\nu_{\parallel}}{3}\right]\geq\lambda_{\nu,\parallel,\perp,z}.
\end{equation}
Finding the critical point of the positive branch, we first obtain
\begin{equation}
    \frac{d}{du}\left[\lambda_{\nu \parallel,\perp,xy+}(u)\right] =  \frac{\frac{1}{2}\text{Tr}(\boldsymbol{M}^{(\nu,\parallel,\perp)}_{xy}) \frac{d}{du}\left[\text{Tr}(\boldsymbol{M}^{(\nu,\parallel,\perp)}_{xy})\right] - \frac{d}{du}\left[\text{Det}(\boldsymbol{M}^{(\nu,\parallel,\perp)}_{xy})\right] }{\sqrt{\text{Tr}(\boldsymbol{M}^{(\nu,\parallel,\perp)}_{xy})^{2}-4\text{Det}(\boldsymbol{M}^{(\nu,\parallel,\perp)}_{xy})}}+\frac{1}{2}\frac{d}{du}\left[\text{Tr}(\boldsymbol{M}^{(\nu,\parallel,\perp)}_{xy})\right].
\end{equation}
The derivatives of the two components are
\begin{subequations}
    \begin{equation}
        \frac{d}{du}\left[\text{Tr}(\boldsymbol{M}^{(\nu,\parallel,\perp)}_{xy})\right] = x_{\text{ion}}(\nu_{\parallel} - \nu_{\perp1}),
    \end{equation}
    \begin{align}
\frac{d}{du}\left[\text{Det}(\boldsymbol{M}^{(\nu,\parallel,\perp)}_{xy})\right] =\;& x_{\text{ion}}(1-2u)\left(3\nu_{\parallel}+\nu_{\perp1}-16\nu_{\perp2}\right)\left[\frac{\zeta}{\rho}+\frac{4}{3}(1-x_{\text{ion}})\nu_{\text{neu}}\right] \nonumber\\
&+2x_{\text{ion}}(1-x_{\text{ion}})\nu_{\text{neu}}\left[\nu_{\parallel}(3u-1)-\nu_{\perp1}(1-u)+8\nu_{\perp2}(1-2u)\right] \nonumber\\
&+\frac{4}{3}x_{\text{ion}}^{2}\left[\nu_{\parallel}\nu_{\perp1}(1-2u)+2\nu_{\parallel}\nu_{\perp2}(u+1)-6\nu_{\perp1}\nu_{\perp2}(1-u)\right].
\end{align}
\end{subequations}
Setting $d\lambda_{\nu,xy+}/du=0$ and squaring gives the polynomial condition
\begin{equation}
\frac{d}{du}\left[\text{Det}^{2}\right]-\text{Tr}\frac{d}{du}\left[\text{Tr}\right]\frac{d}{du}\left[\text{Det}\right]+\frac{d}{du}\left[\text{Tr}\right]^{2}\text{Det}=0.
\end{equation}
As $\text{Tr}$ is linear and $\text{Det}$ quadratic in $u$, this is a quadratic in $u$, whose roots are the $u^{*}_{1}$ and $u^{*}_{2}$ of Section~\ref{sec:visc_cfl}.

\subsection*{B.2 - Gyroviscosity}
In Fourier space, maximising over the binding mode, Eq.~(\ref{eq:gyrovisc-collected-general}) becomes
\begin{equation}
    \frac{d}{dt}\begin{pmatrix}
        \tilde{v}_{x} \\
        \tilde{v}_{y} \\
        \tilde{v}_{z}
    \end{pmatrix} = 
    \begin{pmatrix}
    0 & M^{(\nu,\wedge)}_{xy} & M^{(\nu,\wedge)}_{xz} \\ 
   - M^{(\nu,\wedge)}_{xy} & 0 & M^{(\nu,\wedge)}_{yz} \\ 
    -M^{(\nu,\wedge)}_{xz} & -M^{(\nu,\wedge)}_{yz} & 0 \\ 
    \end{pmatrix} \cdot \frac{-4}{\Delta x^{2}} \begin{pmatrix}
        \tilde{v}_{x} \\
        \tilde{v}_{y} \\
        \tilde{v}_{z}
    \end{pmatrix},
\end{equation}
The matrix $\boldsymbol{M}^{(\nu,\wedge)}$ is skew-symmetric and, therefore, its eigenvalues are purely imaginary.

Eliminating the $\phi$ dependence in the same manner as viscosity, the characteristic polynomial becomes
\begin{equation}
 \lambda_{\nu,\wedge}(-\lambda_{\nu,\wedge}^{2}+\Delta(\boldsymbol{M}^{(\nu,\wedge)}_{xy})) = 0,  
\end{equation}
where
\begin{equation}
 \Delta(\boldsymbol{M}^{(\nu,\wedge)}_{xy}) =  x_{\text{ion}}^{2}\left(- \nu_{\wedge1}^{2}u^{2} + 2\nu_{\wedge1}^{2}u - \nu_{\wedge1}^{2} + 4\nu_{\wedge1}\nu_{\wedge2}u^{2} - 4\nu_{\wedge1}\nu_{\wedge2}u - 4\nu_{\wedge2}^{2}u\right).
\end{equation}
Considering the magnitude of the positive branch eigenvalue, the $u$ derivative of its magnitude is
\begin{equation}
    \frac{d}{du}\left[ |\lambda_{\nu,\wedge+}(u)| \right] = \frac{x_{\text{ion}}\left(\nu_{\wedge1}^{2} u - \nu_{\wedge1}^{2} - 4 \nu_{\wedge1} \nu_{\wedge2} u + 2 \nu_{\wedge1}\nu_{\wedge2} + 2 \nu_{\wedge2}^{2}\right)}
{\sqrt{(\nu_{\wedge1}(1-u)+2\nu_{\wedge2}u)^{2} + 4\nu_{\wedge2}^{2}u(1-u)}}.
\end{equation}
This vanishes at the critical point $u^{*}$ given in Section~\ref{sec:gyro_cfl}.

\subsection*{B.3 - Isolated thermoelectric effects}
For the eigenvalues $\lambda_{\text{TE},\pm}$ of Section~\ref{sec:te_iso}, where $\left(\kappa_{J,w}^{2} - \Delta\kappa_J^{2}u\right)(1-u)>0$, the derivative of the positive branch is
\begin{equation}
\frac{d}{du}\left[\lambda_{\text{TE},+}\right] = \frac{2 \Delta\kappa_{J}^{2} u - \Delta\kappa_{J}^{2} - \kappa_{J,w}^{2}}{2 \sqrt{\varepsilon_{\text{int}}} \sqrt{\left(\kappa_{J,w}^{2} - \Delta\kappa_J^{2}u\right)(1-u)}},
\end{equation}
and where $\left(\kappa_{J,w}^{2} - \Delta\kappa_J^{2}u\right)(1-u)<0$,
\begin{equation}
\frac{d}{du}\left|\lambda_{\text{TE},\pm}\right| = -\frac{2 \Delta\kappa_{J}^{2} u - \Delta\kappa_{J}^{2} - \kappa_{J,w}^{2}}{2 \sqrt{\varepsilon_{\text{int}}} \sqrt{-\left(\kappa_{J,w}^{2} - \Delta\kappa_J^{2}u\right)(1-u)}}.
\end{equation}
Both vanish at the stationary point $u^{*}_{1}$ and are singular where the radicand vanishes.

\subsection*{B.4 - Thermo-magnetic system}
For the thermo-magnetic system of Section~\ref{sec:wellposed} we use $k_{x} \Delta x/2 = \psi$ and define the following matrix entries
\begin{subequations}
\begin{equation}
    M^{(\kappa,\eta,J)}_{T} = -4(\gamma-1)\frac{\Delta t}{\Delta x^{2}}(\kappa_{0} + \Delta\kappa\cos^{2}(\theta))\sin^{2}(\psi),
\end{equation}
\begin{equation}
    M^{(\kappa,\eta,J)}_{T,\wedge} = 4(\gamma-1) \sqrt{-1}\frac{ \Delta t}{\Delta_{\perp} \Delta x} \kappa_{w} \sin(\psi)\cos(\psi),
\end{equation}
\begin{equation}
    M^{(\kappa,\eta,J)}_{M,y} =  -4\frac{\Delta t}{\Delta x^{2}}\left(\eta + \eta_{AD}(1- \sin^{2}(\theta)\sin^{2}(\phi))\right)\sin^{2}(\psi),
\end{equation}
\begin{equation}
M^{(\kappa,\eta,J)}_{M,z} = -4\frac{\Delta t}{\Delta x^{2}}\left(\eta + \eta_{AD}(1- \sin^{2}(\theta)\cos^{2}(\phi))\right)\sin^{2}(\psi),
\end{equation}
\begin{equation}
M^{(\kappa,\eta,J)}_{M,AD} = -4\frac{\Delta t}{\Delta x^{2}}\eta_{AD}\sin^{2}(\theta)\cos(\phi)\sin(\phi)\sin^{2}(\psi),
\end{equation}
\begin{equation}
    M^{(\kappa,\eta,J)}_{H} = 4\frac{ \Delta t}{\Delta x^{2}} \eta_{H} \cos(\theta) \sin^{2}(\psi),
\end{equation}
\begin{equation}
   M^{(\kappa,\eta,J)}_{TE,\varepsilon y} = - 4\frac{ \Delta t}{\Delta x^{2}} \left(- \Delta\kappa_{J} \sin(\phi) \cos(\theta) \sin(\theta) + \kappa_{J,w}\cos(\phi) \sin(\theta)\right)\sin^{2}(\psi),
\end{equation}
\begin{equation}
  M^{(\kappa,\eta,J)}_{TE,\varepsilon z} = - 4\frac{ \Delta t}{\Delta x^{2}}\left(\Delta\kappa_{J} \cos(\phi) \cos(\theta) \sin(\theta) + \kappa_{J,w} \sin(\phi) \sin(\theta)\right)\sin^{2}(\psi),
\end{equation}
\begin{equation}
M^{(\kappa,\eta,J)}_{TE,By} = - 4\frac{ \Delta t}{\Delta x^{2}} \frac{1}{\varepsilon_{\text{int}}}\left(\Delta\kappa_{J} \cos(\theta) \sin(\phi) \sin(\theta) + \kappa_{J,w} \cos(\phi) \sin(\theta)\right)\sin^{2}(\psi),
\end{equation}
\begin{equation}
M^{(\kappa,\eta,J)}_{TE,Bz} = - 4\frac{\Delta t  }{\Delta x^{2}}\frac{1}{\varepsilon_{\text{int}}}\left(- \Delta\kappa_{J} \cos(\phi) \cos(\theta) \sin(\theta) + \kappa_{J,w} \sin(\phi) \sin(\theta)\right)\sin^{2}(\psi).
\end{equation}
\end{subequations}
The matrix equation then becomes
\begin{equation}
        \frac{d}{dt}\begin{pmatrix}
            \tilde{\varepsilon}_{\text{int}} \\
            \tilde{B}_{y} \\
            \tilde{B}_{z}
        \end{pmatrix} = \begin{pmatrix}
            M^{(\kappa,\eta,J)}_{T}+M^{(\kappa,\eta,J)}_{T,\wedge} & M^{(\kappa,\eta,J)}_{TE,\varepsilon y} & M^{(\kappa,\eta,J)}_{TE,\varepsilon z}\\
            M^{(\kappa,\eta,J)}_{TE,By} & M^{(\kappa,\eta,J)}_{My} & M^{(\kappa,\eta,J)}_{M,AD} - M^{(\kappa,\eta,J)}_{H} \\
            M^{(\kappa,\eta,J)}_{TE,Bz} & M^{(\kappa,\eta,J)}_{M,AD} + M^{(\kappa,\eta,J)}_{H} & M^{(\kappa,\eta,J)}_{Mz} 
        \end{pmatrix} \cdot\begin{pmatrix}
            \tilde{\varepsilon}_{\text{int}} \\
            \tilde{B}_{y} \\
            \tilde{B}_{z}
        \end{pmatrix}.
\end{equation}
With $\boldsymbol{P}$ chosen as
\begin{equation}
    \boldsymbol{P} = \begin{pmatrix}
        1 & 0 & 0 \\
        0& 1/\sqrt{\varepsilon_{\text{int}}} & 0 \\
        0 & 0 &   1/\sqrt{\varepsilon_{\text{int}}}
    \end{pmatrix},
\end{equation}
we obtain the scaled matrix
\begin{align}
    \tilde{\boldsymbol{M}}^{(\kappa,\eta,J)} &= \boldsymbol{P}^{-1}\cdot\boldsymbol{M}^{(\kappa,\eta,J)}\cdot\boldsymbol{P} \\ &=
\begin{pmatrix}
            M^{(\kappa,\eta,J)}_{T}+M^{(\kappa,\eta,J)}_{T,\wedge} & M^{(\kappa,\eta,J)}_{TE,\varepsilon y}/\sqrt{\varepsilon_{\text{int}}} & M^{(\kappa,\eta,J)}_{TE,\varepsilon z}/\sqrt{\varepsilon_{\text{int}}}\\
            M^{(\kappa,\eta,J)}_{TE,By}\sqrt{\varepsilon_{\text{int}}} & M^{(\kappa,\eta,J)}_{My} & M^{(\kappa,\eta,J)}_{M,AD} - M^{(\kappa,\eta,J)}_{H} \\
            M^{(\kappa,\eta,J)}_{TE,Bz}\sqrt{\varepsilon_{\text{int}}} & M^{(\kappa,\eta,J)}_{M,AD} + M^{(\kappa,\eta,J)}_{H} & M^{(\kappa,\eta,J)}_{Mz} 
        \end{pmatrix}.
\end{align}
This allows for the $\kappa_{J,\wedge}$ entries to be symmetric, and the $\Delta \kappa_{J}$ entries to be anti-symmetric.
The Hermitian part of $\tilde{\boldsymbol{M}}^{(\kappa,\eta,J)}$ is
(with $M^{(\kappa,\eta,J)}_{T,\wedge}$ cancelling and maximising the bounding mode as $\sin^{2}(\psi)=1$)
\begin{equation}
\boldsymbol{H}_{\tilde{M}} = \frac{-4\Delta t}{\Delta x^{2}}\begin{pmatrix}
H_{\varepsilon\varepsilon} & H_{\varepsilon y} & H_{\varepsilon z} \\
H_{\varepsilon y} & H_{yy} & H_{yz} \\
H_{\varepsilon z} & H_{yz} & H_{zz}
\end{pmatrix},
\end{equation}
with
\begin{subequations}
\begin{align}
H_{\varepsilon\varepsilon} &= (\gamma-1)(\kappa_{0}+\Delta\kappa\cos^{2}\theta), \\
H_{\varepsilon y} &= \kappa_{J,w}\sin\theta\cos\phi/\sqrt{\varepsilon_{\text{int}}}, \\
H_{\varepsilon z} &= \kappa_{J,w}\sin\theta\sin\phi/\sqrt{\varepsilon_{\text{int}}}, \\
H_{yy} &= \eta+\eta_{AD}(1-\sin^{2}\theta\sin^{2}\phi), \\
H_{zz} &= \eta+\eta_{AD}(1-\sin^{2}\theta\cos^{2}\phi), \\
H_{yz} &= \eta_{AD}\sin^{2}\theta\sin\phi\cos\phi.
\end{align}
\end{subequations}
To eliminate the $\phi$ dependence we now make use of invariant eigenvalues again with a rotation matrix
\begin{equation}
    \boldsymbol{Q} = \begin{pmatrix}
        1 & 0 & 0 \\
        0 & \cos(\phi) & \sin(\phi) \\
        0 & \sin(\phi) & -\cos(\phi)
    \end{pmatrix}.
\end{equation}
This gives the rotated matrix $\boldsymbol{H}_{\tilde{F},\text{rot}}$ of Section~\ref{sec:wellposed}.
\section*{Appendix~C - Details of the HLLE-Mod derivations}
\subsection*{C.1 - Linearised Operators}
For the ideal and Hall terms (Section~\ref{sec:whist}), the linearised operator acting on the perturbation $\delta\vec{W}$ defined there is
\begin{equation}\label{eq:M-ideal-Hall}
\boldsymbol{M}^{(\text{I},\eta_{H})} = \sqrt{-1}k_{x}\begin{pmatrix}
0 & -\rho_{0} & 0 & 0 & 0 & 0 & 0\\
0 & 0 & 0 & 0 & -\frac{\gamma-1}{\rho_{0}} & -\frac{B_{0}b_{y}}{\rho_{0}} & -\frac{B_{0}b_{z}}{\rho_{0}}\\
0 & 0 & 0 & 0 & 0 & \frac{B_{0}b_{x}}{\rho_{0}} & 0\\
0 & 0 & 0 & 0 & 0 & 0 & \frac{B_{0}b_{x}}{\rho_{0}}\\
0 & -\gamma\varepsilon_{0} & 0 & 0 & 0 & 0 & 0\\
0 & -B_{0}b_{y} & B_{0}b_{x} & 0 & 0 & 0 & \sqrt{-1}\eta_{H}k_{x}b_{x}\\
0 & -B_{0}b_{z} & 0 & B_{0}b_{x} & 0 & -\sqrt{-1}\eta_{H}k_{x}b_{x} & 0
\end{pmatrix}.
\end{equation}
For the ideal and Peltier/Seebeck terms (Section~\ref{sec:hlle_ps}) it is
\begin{equation}
\renewcommand{\arraystretch}{1.3}
\boldsymbol{M}^{(\text{I},J)} = \sqrt{-1}k_{x}\begin{pmatrix}
0 & -\rho_{0} & 0 & 0 & 0 & 0 & 0\\
0 & 0 & 0 & 0 & -\frac{\gamma-1}{\rho_{0}} & -\frac{B_{0}b_{y}}{\rho_{0}} & -\frac{B_{0}b_{z}}{\rho_{0}}\\
0 & 0 & 0 & 0 & 0 & \frac{B_{0}b_{x}}{\rho_{0}} & 0\\
0 & 0 & 0 & 0 & 0 & 0 & \frac{B_{0}b_{x}}{\rho_{0}}\\
0 & -\gamma\varepsilon_{0} & 0 & 0 & 0 & -\sqrt{-1}Kb_{z} & \sqrt{-1}Kb_{y}\\
-\frac{\sqrt{-1}Kb_{z}}{\rho_{0}} & -B_{0}b_{y} & B_{0}b_{x} & 0 & \frac{\sqrt{-1}Kb_{z}}{\varepsilon_{0}} & 0 & 0\\[3pt]
\frac{\sqrt{-1}Kb_{y}}{\rho_{0}} & -B_{0}b_{z} & 0 & B_{0}b_{x} & -\frac{\sqrt{-1}Kb_{y}}{\varepsilon_{0}} & 0 & 0
\end{pmatrix},
\end{equation}
where $K \equiv \Delta\kappa_{J}k_{x}b_{x}$.
For the ideal and gyro-viscous terms (Section~\ref{sec:hlle_gv}), with $g_{1}$ and $g_{2}$ as defined there, it is
\begin{equation}
\renewcommand{\arraystretch}{1.3}
\boldsymbol{M}^{(\text{I},\nu_{\wedge})} = \sqrt{-1}k_{x}\begin{pmatrix}
0 & -\rho_{0} & 0 & 0 & 0 & 0 & 0\\
0 & 0 & \sqrt{-1}k_{x}g_{1}b_{z} & -\sqrt{-1}k_{x}g_{1}b_{y} & -\frac{\gamma-1}{\rho_{0}} & -\frac{B_{0}b_{y}}{\rho_{0}} & -\frac{B_{0}b_{z}}{\rho_{0}}\\
0 & -\sqrt{-1}k_{x}g_{1}b_{z} & 0 & \sqrt{-1}k_{x}g_{2}b_{x} & 0 & \frac{B_{0}b_{x}}{\rho_{0}} & 0\\
0 & \sqrt{-1}k_{x}g_{1}b_{y} & -\sqrt{-1}k_{x}g_{2}b_{x} & 0 & 0 & 0 & \frac{B_{0}b_{x}}{\rho_{0}}\\
0 & -\gamma\varepsilon_{0} & 0 & 0 & 0 & 0 & 0\\
0 & -B_{0}b_{y} & B_{0}b_{x} & 0 & 0 & 0 & 0\\
0 & -B_{0}b_{z} & 0 & B_{0}b_{x} & 0 & 0 & 0
\end{pmatrix},
\end{equation}

\subsection*{C.2 - Combined Hall and Peltier/Seebeck Bound}
We show that no root of $F(\Omega,u)$ exceeds
\begin{equation}
c_{\text{TEw}}^{2} = \Psi^{2} + \frac{1}{4}D + \frac{1}{2}\sqrt{DE}, \qquad \Psi = \max\left(c_{f},c_{w}\right),
\end{equation}
for any $u\in[0,1]$, either sign of $s$, any $\gamma>1$ and any $D,E\geq0$.

The polynomials in the decomposition $F = G - u(1-u)\mathcal{T}$ of Eq.~(\ref{eq:decomp}) are
\begin{subequations}
\begin{align}
R(\Omega,u) &= \Omega^{2} - c_{f}^{2}\Omega + c_{a}^{2}c_{s}^{2}u,\\
G(\Omega,u) &= \left(\Omega - c_{a}^{2}u\right)R(\Omega,u) - Eu\Omega\left(\Omega - c_{s}^{2}\right),\\
\mathcal{T}(\Omega,u) &= D\left[\Omega^{2} - \left(c_{a}^{2}+c_{\text{iso}}^{2}\right)\Omega + c_{a}^{2}c_{\text{iso}}^{2}u\right] - 2sc_{a}c_{\text{iso}}\sqrt{(\gamma-1)DE}\,\Omega,
\end{align}
\end{subequations}
where $G$ is the ideal + Hall polynomial of Eq.~(\ref{eq:F-Hall}), $c_{f}^{2} = c_{a}^{2}+c_{s}^{2}$ and $c_{s}^{2} = \gamma c_{\text{iso}}^{2}$.

We use
\begin{equation}
\label{eq:psi}
\delta_{f} = \Psi^{2} - c_{f}^{2}, \qquad
\delta_{E} = \Psi^{2} - c_{a}^{2} - E, \qquad
\delta_{w} = \left(\Psi^{2} - c_{a}^{2}\right)^{2} - E\Psi^{2},
\end{equation}
which are all positive. The first is because $\Psi\geq c_{f}$ by definition. For the other two, $c_{w}^{2} = c_{a}^{2} + \frac{1}{2}E + \frac{1}{2}\sqrt{E(E+4c_{a}^{2})}$ is the larger root of the quadratic $(\Omega-c_{a}^{2})^{2} - E\Omega$ (Eq.~\ref{eq:Hall-bound}), and $\Psi^{2}\geq c_{w}^{2}$. Since $\sqrt{E(E+4c_{a}^{2})}\geq E$, $c_{w}^{2}\geq c_{a}^{2}+E$, so $\delta_{E}\geq0$. The quadratic is positive above its larger root, so $\delta_{w}\geq0$. We define
\begin{equation}
\Psi^{2} - c_{a}^{2} = \delta_{f} + c_{s}^{2}, \qquad \Psi^{2} - c_{s}^{2} = \delta_{f} + c_{a}^{2}, \qquad \Psi^{2} - 2c_{a}c_{s} = \delta_{f} + \left(c_{a}-c_{s}\right)^{2},
\end{equation}
which are also positive.

The coefficient $A_{2}$ of $\Omega^{2}$ in $F$, plus $2c_{\text{TEw}}^{2}$, is a sum of positive terms:
\begin{subequations}
\begin{align}
A_{2} &= -\left[c_{f}^{2} + \left(c_{a}^{2}+E\right)u + Du(1-u)\right],\\
2c_{\text{TEw}}^{2} + A_{2} &= \delta_{f} + \delta_{E} + \left(c_{a}^{2}+E\right)(1-u) + \frac{1}{4}D\left[1 + (1-2u)^{2}\right] + \sqrt{DE} \geq 0.
\end{align}
\end{subequations}
By the test of Section~\ref{sec:hlle_mod}, it is therefore enough to show that $F(c_{\text{TEw}}^{2},u)\geq0$.

The sign $s$ and $\gamma$ enter $F$ only through $\mathcal{T}$, via $c_{\text{iso}}^{2}$ and $sc_{\text{iso}}\sqrt{\gamma-1} = sc_{s}\sqrt{1-1/\gamma}$, which lies between $-c_{s}$ and $c_{s}$. Let $\hat{F}$ be $F$ at the extreme $s=-1$, $\gamma\to\infty$ at fixed $c_{s}$, where $c_{\text{iso}}\to0$ and $c_{\text{iso}}\sqrt{\gamma-1}\to c_{s}$:
\begin{equation}
\hat{F}(\Omega,u) = G(\Omega,u) - u(1-u)\Omega\left[D\left(\Omega-c_{a}^{2}\right) + 2c_{a}c_{s}\sqrt{DE}\right].
\end{equation}
For any $s$ and $\gamma$,
\begin{equation}
F(\Omega,u) - \hat{F}(\Omega,u) = u(1-u)\left\{Dc_{\text{iso}}^{2}\left(\Omega - c_{a}^{2}u\right) + 2c_{a}\sqrt{DE}\,\Omega\left[c_{s} + sc_{\text{iso}}\sqrt{\gamma-1}\right]\right\},
\end{equation}
and both terms are positive when $\Omega\geq c_{a}^{2}u$. This holds at $\Omega = c_{\text{TEw}}^{2}$, since $c_{\text{TEw}}^{2}\geq\Psi^{2}\geq c_{a}^{2}u$, so $F(c_{\text{TEw}}^{2},u)\geq\hat{F}(c_{\text{TEw}}^{2},u)$, and it is enough to show that $\hat{F}(c_{\text{TEw}}^{2},u)\geq0$.

Since $G$ is a monic cubic in $\Omega$, its Taylor series about $\Psi^{2}$ ends at $h^{3}$, where $h=\frac{1}{4}D + \frac{1}{2}\sqrt{DE}$. 
$G(\Psi^{2}+h,u) = g_{0} + g_{1}h + g_{2}h^{2} + h^{3}$, where $g_{n} = \partial_{\Omega}^{n}G(\Psi^{2},u)/n!$:
\begin{subequations}
\begin{align}
g_{0} &= G(\Psi^{2},u),\\
g_{1} &= 3\Psi^{4} - 2\left[c_{f}^{2} + \left(c_{a}^{2}+E\right)u\right]\Psi^{2} + \left[c_{a}^{2}\left(c_{f}^{2}+c_{s}^{2}\right) + Ec_{s}^{2}\right]u,\\
g_{2} &= 3\Psi^{2} - c_{f}^{2} - \left(c_{a}^{2}+E\right)u.
\end{align}
\end{subequations}
Substituting into $\hat{F}$ and collecting powers of $\sqrt{D}$ gives
\begin{equation}
\hat{F}(c_{\text{TEw}}^{2},u) = \sum_{k=0}^{6} a_{k}D^{k/2},
\end{equation}
with
\begin{subequations}
\begin{align}
a_{0} &= g_{0},\\
a_{1} &= \frac{1}{2}\sqrt{E}\left[g_{1} - 4c_{a}c_{s}\Psi^{2}u(1-u)\right],\\
a_{2} &= \frac{1}{4}\left[g_{1} + Eg_{2} - 4u(1-u)\left(\Psi^{2}\left(\Psi^{2}-c_{a}^{2}\right) + c_{a}c_{s}E\right)\right],\\
a_{3} &= \frac{1}{8}\sqrt{E}\left[2g_{2} + E - 4u(1-u)\left(2\Psi^{2} - c_{a}^{2} + c_{a}c_{s}\right)\right],\\
a_{4} &= \frac{1}{16}\left[g_{2} + 3E - 4u(1-u)\left(2\Psi^{2} - c_{a}^{2} + E\right)\right],\\
a_{5} &= \frac{1}{32}\sqrt{E}\left[3 - 8u(1-u)\right],\\
a_{6} &= \frac{1}{64}(1-2u)^{2}.
\end{align}
\end{subequations}
As $D\geq0$, it is enough to show that every $a_{k}\geq0$.

The $u^{2}$ coefficient of $G$, $-c_{a}^{4}c_{s}^{2}$, does not depend on $\Omega$, so $g_{1}$ and $g_{2}$ are linear in $u$. The end point values are
\begin{equation}
\begin{aligned}
g_{1}\big|_{u=0} &= \Psi^{4} + 2\Psi^{2}\delta_{f},\\
g_{1}\big|_{u=1} &= c_{a}^{2}c_{s}^{2} + \delta_{w} + c_{a}^{2}\left(\delta_{f}+\delta_{E}\right) + \delta_{f}\left(c_{s}^{2}+\delta_{f}+\delta_{E}\right),\\
\left(g_{1}+Eg_{2}\right)\big|_{u=0} &= \Psi^{2}\left(\Psi^{2}+2E\right) + \delta_{f}\left(2\Psi^{2}+E\right),\\
\left(g_{1}+Eg_{2}\right)\big|_{u=1} &= \Psi^{2}\left(\Psi^{2}-c_{a}^{2}\right) + 2\delta_{f}\left(\Psi^{2}-c_{a}^{2}\right) + \left(c_{a}^{2}+E\right)\delta_{E},\\
\left(2g_{2}+E\right)\big|_{u=0} &= 4\Psi^{2} + 2\delta_{f} + E,\\
\left(2g_{2}+E\right)\big|_{u=1} &= 2\Psi^{2} + 3\delta_{f} + \delta_{E} + c_{s}^{2}.
\end{aligned}
\end{equation}
Every term after the first is positive, so dropping them gives, for $u\in[0,1]$,
\begin{subequations}
\label{eq:g-bounds}
\begin{align}
g_{1} &\geq (1-u)\Psi^{4} + uc_{a}^{2}c_{s}^{2},\\
g_{1} + Eg_{2} &\geq (1-u)\Psi^{2}\left(\Psi^{2}+2E\right) + u\Psi^{2}\left(\Psi^{2}-c_{a}^{2}\right),\\
2g_{2} + E &\geq 4(1-u)\Psi^{2} + 2u\Psi^{2}.
\end{align}
\end{subequations}
Each $a_{k}$ now reduces to a sum of positive terms. For $a_{0}$ we evaluate Eq.~(\ref{eq:chord}) at $\Omega=\Psi^{2}$, using $G(\Omega,0) = \Omega^{2}(\Omega-c_{f}^{2})$ and $G(\Omega,1) = (\Omega-c_{s}^{2})[(\Omega-c_{a}^{2})^{2}-E\Omega]$. For $a_{1}$, $a_{2}$ and $a_{3}$ we insert the bounds of Eq.~(\ref{eq:g-bounds}), together with $c_{a}c_{s}\leq\frac{1}{2}\Psi^{2}$ for $a_{3}$. For $a_{4}$ we use $4u(1-u) = 1-(1-2u)^{2}$. The results are
\begin{subequations}
\begin{align}
a_{0} &= (1-u)\Psi^{4}\delta_{f} + u\left(\Psi^{2}-c_{s}^{2}\right)\delta_{w} + c_{a}^{4}c_{s}^{2}u(1-u),\\
a_{1} &\geq \frac{1}{2}\sqrt{E}\left[(1-u)\Psi^{4} + uc_{a}^{2}c_{s}^{2} - 4c_{a}c_{s}\Psi^{2}u(1-u)\right] \nonumber\\
      &= \frac{1}{2}\sqrt{E}\left(\left[(1-u)\Psi^{2} - uc_{a}c_{s}\right]^{2} + u(1-u)\left(\Psi^{2}-c_{a}c_{s}\right)^{2}\right),\\
4a_{2} &\geq (1-u)\Psi^{2}\left(\Psi^{2}+2E\right) + u\Psi^{2}\left(\Psi^{2}-c_{a}^{2}\right) - 4u(1-u)\left[\Psi^{2}\left(\Psi^{2}-c_{a}^{2}\right) + c_{a}c_{s}E\right] \nonumber\\
      &= (1-2u)^{2}\Psi^{2}\left(\Psi^{2}-c_{a}^{2}\right) + (1-u)^{2}\Psi^{2}\left(c_{a}^{2}+2E\right) \nonumber\\
      &\quad + u(1-u)\left[c_{a}^{2}\Psi^{2} + 2E\left(\Psi^{2}-2c_{a}c_{s}\right)\right],\\
a_{3} &\geq \frac{1}{8}\sqrt{E}\left[4(1-u)\Psi^{2} + 2u\Psi^{2} - 10u(1-u)\Psi^{2}\right] \nonumber\\
      &= \frac{1}{4}\sqrt{E}\,\Psi^{2}\left[(1-2u)^{2} + (1-u)^{2}\right],\\
16a_{4} &= \delta_{f} + c_{a}^{2}(1-u) + E(2-u) + (1-2u)^{2}\left(\Psi^{2} + \delta_{f} + c_{s}^{2} + E\right),\\
a_{5} &= \frac{1}{32}\sqrt{E}\left[1 + 2(1-2u)^{2}\right],\\
a_{6} &= \frac{1}{64}(1-2u)^{2}.
\end{align}
\end{subequations}
Every term on the right is a product of positive factors, so every $a_{k}\geq0$.

Hence $F(c_{\text{TEw}}^{2},u)\geq\hat{F}(c_{\text{TEw}}^{2},u)\geq0$ for all $u\in[0,1]$, either sign of $s$ and all $\gamma>1$, so no root of $F$ exceeds $c_{\text{TEw}}^{2}$ for any $D,E\geq0$.
\section*{Appendix~D - HLLE-Mod Noise Tests}
\subsection*{D.1 - Noise Initial Conditions}
Each noise field is generated over the whole mesh in two steps. First, every cell is given an independent Gaussian deviate of unit variance by the Box--Muller transform \cite{box_1958},
\begin{equation}
    g = \sqrt{-2\ln u_{1}}\cos(2\pi u_{2}),
\end{equation}
where $u_{1}$ and $u_{2}$ are uniform deviates on $(0,1)$ from Athena++'s \texttt{ran2} generator \cite{press_1992}. Each field and component receives a different seed. We then take the discrete Fourier transform \cite{frigo_2005}, multiply each mode by a filter $W(\vec{k})$, and transform back. The filter removes the mean, $W(0)=0$, and is otherwise $W=1$ for the temperature and for each velocity component. 
The magnetic noise is the discrete curl of the magnetic vector potential $\vec{A}$, whose components sit on the cell edges. On the $x$-faces
\begin{equation}
\begin{split}
    \delta B_{x}(i-\frac{1}{2},j,k) ={}& \frac{A_{z}(i-\frac{1}{2},j+\frac{1}{2},k) - A_{z}(i-\frac{1}{2},j-\frac{1}{2},k)}{\Delta y_{i,j,k}} \\
    &- \frac{A_{y}(i-\frac{1}{2},j,k+\frac{1}{2}) - A_{y}(i-\frac{1}{2},j,k-\frac{1}{2})}{\Delta z_{i,j,k}}.
\end{split}
\end{equation}
The filter for the magnetic field is
\begin{equation}
    W(\vec{k}) = \frac{1}{|\tilde{\vec{k}}|}, \qquad |\tilde{\vec{k}}|^{2} = \sum_{d=1}^{3}\left(\frac{2}{\Delta x_{d}}\sin\left(\frac{k_{d}\Delta x_{d}}{2}\right)\right)^{2},
\end{equation}
for each component of the magnetic vector potential $\vec{A}$, with $k_{d}=2\pi m_{d}/L_{d}$ for $m_{d}=0,\dots,N_{d}-1$.

Finally, each noise field is multiplied by a constant so that its RMS over the cells is the amplitude,
\begin{equation}
    \langle\delta T^{2}\rangle^{1/2} = \sigma_{T}, \qquad \langle|\vec{v}|^{2}\rangle^{1/2} = \sigma_{v}, \qquad \langle|\delta\vec{B}_{\text{cc}}|^{2}\rangle^{1/2} = \sigma_{B},
\end{equation}
where $\delta\vec{B}_{\text{cc}}$ is the cell-centred field (the mean of the two face values). The initial state is then $\rho=\rho_{0}$, $P=\rho_{0}(T_{0}+\delta T)$, $\vec{v}$, and $\vec{B}=\vec{B}_{0}+\delta\vec{B}$, with the temperature noise at uniform density.

\subsection*{D.2 - Noise Slices for Individual Effects}
\begin{figure}[H]
    \centering
\begin{subfigure}{\linewidth}
    \centering
        \includegraphics[scale=0.95]{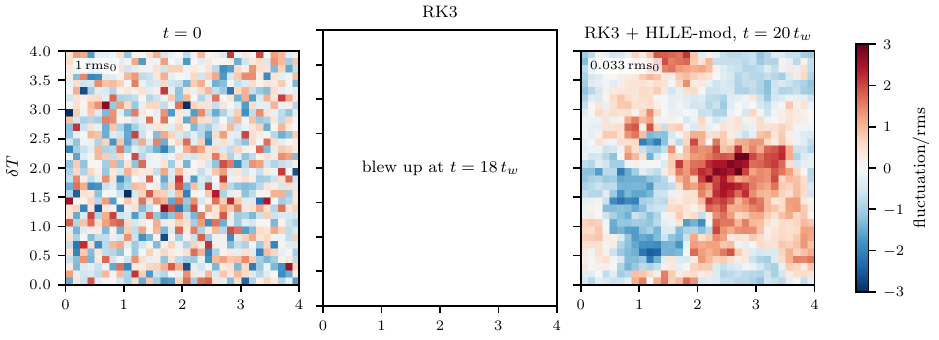}
        \caption{Righi--Leduc.}
    \label{fig:RL-Noise-slice}
\end{subfigure}
\begin{subfigure}{\linewidth}
    \centering
        \includegraphics[scale=0.95]{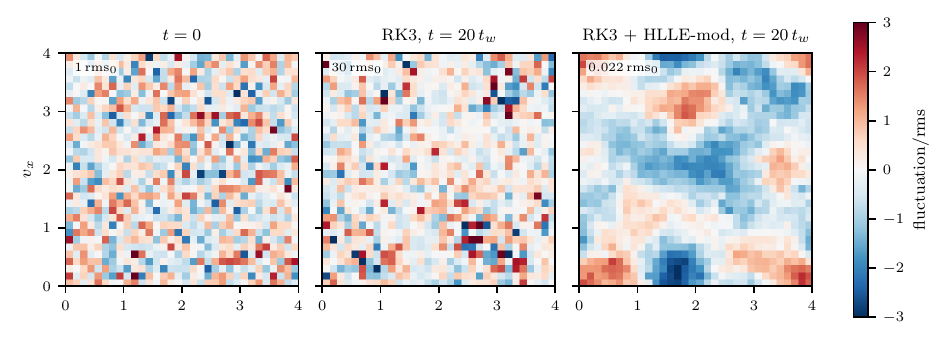}
        \caption{Gyroviscosity.}
    \label{fig:GV-Noise-slice}
\end{subfigure}
\begin{subfigure}{\linewidth}
    \centering
        \includegraphics[scale=0.95]{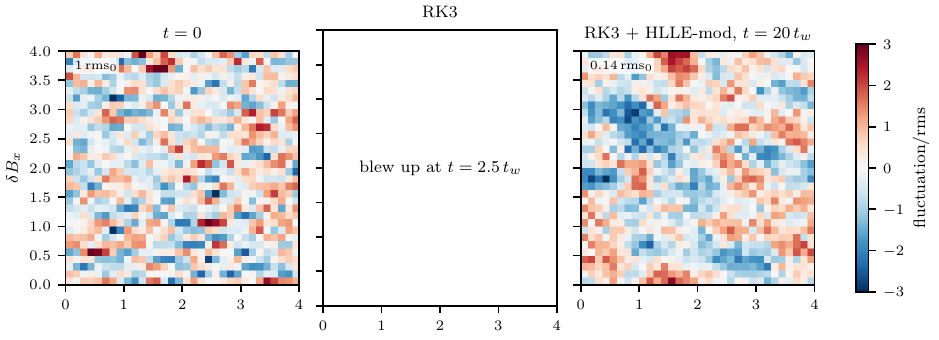}
        \caption{Hall Effect.}
    \label{fig:Hall-Noise-slice}
\end{subfigure}
    \caption{Slices through the noise fields at $t=0$ and $t=20t_{w}$ (or the time of blow-up). Each field is shown normalised by its own RMS.}
    \label{fig:Noise-RMS-slice}
\end{figure}
 
\begin{figure}[H]
    \ContinuedFloat
    \centering
\begin{subfigure}{\linewidth}
    \centering
        \includegraphics[scale=0.95]{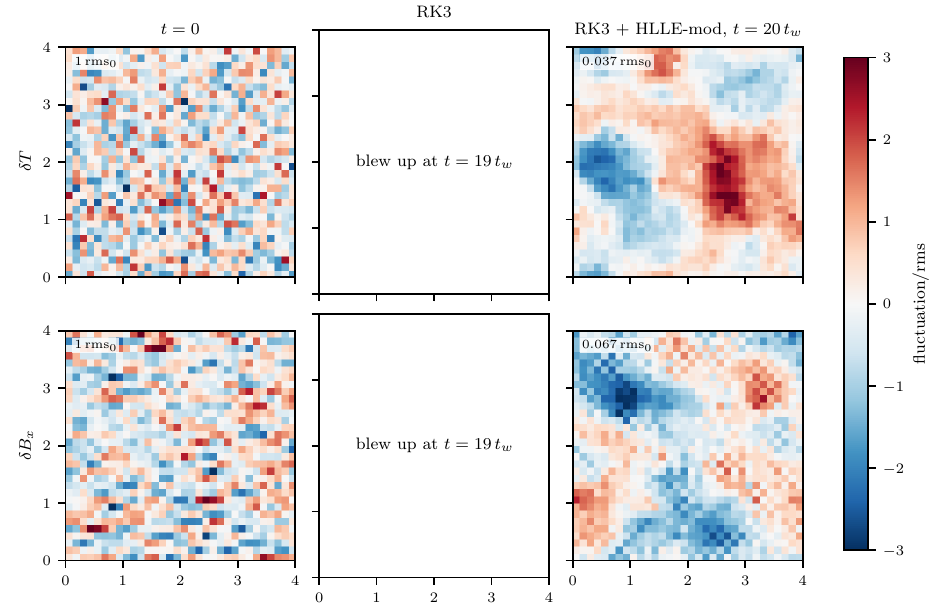}
        \caption{Peltier/Seebeck.}
    \label{fig:PS-Noise-slice}
\end{subfigure}
\begin{subfigure}{\linewidth}
    \centering
        \includegraphics[scale=0.95]{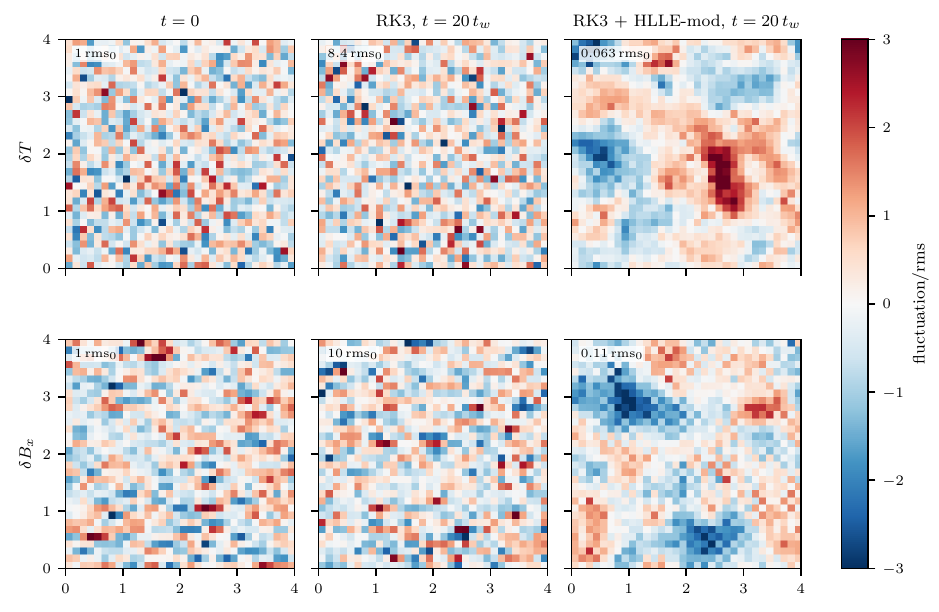}
        \caption{Peltier/Seebeck + Hall.}
    \label{fig:HPS-Noise-slice}
\end{subfigure}
    \caption[]{Slices through the noise fields at $t=0$ and $t=20t_{w}$ (or the time of blow-up). Each field is shown normalised by its own RMS (continued).}
\end{figure}
Figure \ref{fig:Noise-RMS-slice} shows the same slices as Fig.~\ref{fig:All-Noise-slice} for each effect on its own. With the standard solver the Righi--Leduc, Hall, and Peltier/Seebeck runs blow up at $18t_{w}$, $2.5t_{w}$, and $19t_{w}$, respectively. The gyroviscous and Peltier/Seebeck + Hall runs reach $20t_{w}$, but with the noise amplified to around 30 ($v_{x}$) and 8--10 ($\delta T$, $\delta B_{x}$) times its initial RMS. It remains at the grid scale, and for gyroviscosity it is concentrated in patches.

With HLLE-Mod the cell-to-cell noise is removed in every case, leaving larger-scale structure at around 2--14 per cent of the initial RMS. The Hall field, seeded with $\sigma_{B}=1$ rather than $10^{-6}$, retains the most small-scale structure and the largest fraction ($\sim14$ per cent) of its initial RMS.
\section*{Appendix~E - Bounded Timestep and Well-Posed Numerical Checks}
\subsection*{E.1 - Physical Diffusion Coefficients}\label{app:coeffs}
For neutral gas, the Parker conductivity of neutral hydrogen of \cite{parker_1953} is used
\begin{equation}
\label{parker-con}
    \kappa_{P} = 2.5\times10^{5}\sqrt{\nicefrac{T}{10^{4}}}~\text{erg}~\text{s}^{-1}~\text{cm}^{-1}~\text{K}^{-1}.
\end{equation}
For ionised gas the Spitzer conductivity for ionised hydrogen is used \cite{spitzer_2013,El-Badry_2019}
\begin{equation}
    \label{spitzer-con}
   \kappa_{s} = \frac{1.70\times10^{11}(\nicefrac{T}{10^{7}})^{5/2}}{1+0.029\ln\left(\frac{T}{10^{7}}\left(\frac{n_{e}}{10^{-2}}\right)^{-1/2}\right)}~\text{erg}~\text{s}^{-1}~\text{cm}^{-1}~\text{K}^{-1}.
\end{equation}
We take $n_{e}=\max(1.2x_{\text{ion}},10^{-6})n_{H}$, where $n_{e}$ is the electron number density and $n_{H}$ the number density of hydrogen nuclei, appropriate for fully ionised gas of solar composition. The floor prevents the ambipolar diffusion coefficient from diverging in neutral gas.  The number densities of hydrogen ions and neutral hydrogen are $n_{i}=x_{\text{ion}}n_{H}$ and $n_{n}=(1-x_{\text{ion}})n_{H}$. Each conductivity $\kappa_{\text{phys}}$ is converted to the diffusivity of Eq.~(\ref{eqn:conduction}) as $\kappa_{\text{code}}=\kappa_{\text{phys}}\tilde{m}/(\rho k_{B})$, where $\tilde{m} = \mu(T)m_{P}$ is the mean particle mass, $\mu(T)$ is the mean molecular weight, and $m_{P}$ is the mass of a proton. \\ \\
The anisotropic thermal conduction is constructed as \cite{Braginskii_1965}
\begin{subequations}
\begin{equation}
\kappa_{n} = \kappa_{P}, \qquad \kappa_{\parallel} = \kappa_{s},
\end{equation}
\begin{equation}
\kappa_{\perp} = \frac{\kappa_{s}}{3.1616}\frac{4.664\beta_{e}^{2}+11.92}{\beta_{e}^{4}+14.79\beta_{e}^{2}+3.7703},
\end{equation}
\begin{equation}
\kappa_{\wedge} = \frac{\kappa_{s}\beta_{e}}{3.1616}\frac{\nicefrac{5}{2}\beta_{e}^{2}+21.67}{\beta_{e}^{4}+14.79\beta_{e}^{2}+3.7703},
\end{equation}
\end{subequations}
where $\beta_{e}$ is the electron Hall parameter
\begin{equation}
    \beta_{e} = \frac{\omega_{ce}}{\nu_{ei}+\nu_{en}},
\end{equation}
and $\omega_{ce}$ is the angular electron gyro-frequency
\begin{equation}
    \omega_{ce} = \frac{e|\vec{B}|}{m_{e}c_{\text{light}}},
\end{equation}
with $|\vec{B}|$ the field in Gaussian units, $\sqrt{4\pi}$ times the code field. $\nu_{ei}$ and $\nu_{en}$ are the electron--ion and electron--neutral collision rates \cite{pinto_2008,richardson_2019}
\begin{subequations}
    \begin{equation}
        \nu_{ei} = 2.91\times10^{-6}n_{e}\ln\Lambda T_{\text{eV}}^{-3/2}~\text{s}^{-1},
    \end{equation}
    \begin{equation}
        \nu_{en} = 10^{-9}n_{n}T^{1/2}\left(2.841 + 0.093\log_{10}(T) + 0.245\log_{10}^{2}(T) - 0.089\log_{10}^{3}(T)\right)~\text{s}^{-1},
    \end{equation}
\end{subequations}
where $T_{\text{eV}}=k_{B}T/(1~\text{eV})=8.617\times10^{-5}T$. The fit for $\nu_{en}$ becomes negative above $T\approx3.4\times10^{4}$~K, so above $10^{4}$~K we hold the rate coefficient at it's value at $10^{4}$~K. \\ \\
We take the bulk viscosity to come entirely from H$_{2}$. As we do not track non-equilibrium chemistry we weight it by the tabulated H$_{2}$ mass fraction $x_{\text{H}_{2}}$. It is the sum of the rotational and vibrational contributions of H$_{2}$ \cite{cramer_2012}
\begin{equation}
 \zeta = x_{\text{H}_{2}}\left(\zeta_{\text{rot}} + \zeta_{\text{vib}}\right),
\end{equation}
with
\begin{subequations}
    \begin{equation}
     \zeta_{\text{rot}} = (\gamma_{\text{H}_{2}}-1)^{2} \frac{c_{\text{rot}}}{R}P\tau_{\text{rot}},
    \end{equation}
    \begin{equation}
     \zeta_{\text{vib}} = (\gamma_{\text{H}_{2}}-1)^{2} \frac{c_{\text{vib}}}{R}P\tau_{\text{vib}}.
    \end{equation}
\end{subequations}
Here $\gamma_{\text{H}_{2}}(T)$ is the ratio of specific heats of H$_{2}$ \cite{decampli_1978}
\begin{equation}
\gamma_{\text{H}_{2}}-1 = \frac{1}{3/2 + c_{\text{rot}}/R + c_{\text{vib}}/R}.
\end{equation}
$c_{\text{rot}}$ and $c_{\text{vib}}$ are the rotational and vibrational heat capacities respectively \cite{mcquarrie_2000,popovas_2016}
\begin{subequations}
    \begin{equation}
      c_{\text{rot}}/R =  \sum_{s\in\{p,o\}} x_{s} \frac{\langle E^{2}\rangle_{s} - \langle E\rangle_{s}^{2}}{(k_{B}T)^{2}},
    \end{equation}
    \begin{equation}
     c_{\text{vib}}/R = (\theta_{\text{vib}}/T)^{2} \exp(-\theta_{\text{vib}}/T)/(1-\exp(-\theta_{\text{vib}}/T))^{2},
    \end{equation}
\end{subequations}
and $R$ is the gas constant. The moments of the level energies are
\begin{equation}
    \langle E^{n}\rangle_{s} =
    \frac{\displaystyle\sum_{J\in s}(2J+1)\varepsilon_{J}^{n}
          \exp(-\varepsilon_{J}/k_{B}T)}
         {\displaystyle\sum_{J\in s}(2J+1)\exp(-\varepsilon_{J}/k_{B}T)},
\end{equation}
where
\begin{equation}
 \varepsilon_{J} = hc_{\text{light}}B_{\text{rot}}J(J+1),
\end{equation}
are the rigid-rotor levels and $h$ is Planck's constant.
We sum over the two nuclear-spin species (para, $s=p$ for even $J$; and
ortho, $s=o$ for odd $J$) with frozen normal-hydrogen ratio
$x_{p}=1/4$, $x_{o}=3/4$, with $B_{\text{rot}}=59.32~\text{cm}^{-1}$ and $\theta_{\text{vib}} = 5984.0$~K \cite[p.~250]{huber_1979}. \\ \\
We then take our relaxation times as \cite{Winter_1967,audibert_1974,dove_1974}
\begin{subequations}
\begin{equation}
    P\tau_{\text{rot}} = \left(1.822\times 10^{-4} + 4.64\times 10^{-6}T\right)~\text{Pa s},
\end{equation}
\begin{equation}
    P\tau_{\text{vib}} = 10^{34.74T^{-1/3} - 8.686}~\text{atm s},
\end{equation}
\end{subequations}
where $1~\text{Pa s}=10~\text{dyn s cm}^{-2}$ and $1~\text{atm s}=1.01325\times10^{6}~\text{dyn s cm}^{-2}$. \\ \\
For neutral gas we use the Sutherland viscosity for hydrogen \cite{Sutherland_1893}
\begin{equation}
\label{neutral-visc}
   \nu_{\text{neu}} = \frac{1}{\rho}8.76\times 10^{-5}\left( \frac{T}{300} \right)^{3/2}\frac{372}{T+72}~\text{cm}^{2}~\text{s}^{-1}.
\end{equation}
For ionised plasma the Spitzer model is used \cite[Eq.~5-54]{spitzer_2013}
\begin{equation}
    \label{ion-visc}
    \nu_{\text{ion}} = \frac{1}{\rho}2.21\times10^{-15}\frac{T^{5/2}}{\ln\Lambda}~\text{cm}^{2}~\text{s}^{-1},
\end{equation}
which is Braginskii's $\eta_{0}=0.96n_{i}k_{B}T\tau_{i}/\rho = \nu_{\parallel}$. The anisotropic viscosity is constructed as \cite{Braginskii_1965}
\begin{subequations}
\begin{equation}
\nu_{\parallel} = \nu_{\text{ion}},
\end{equation}
\begin{equation}
\nu_{\perp1} = \frac{\nu_{\text{ion}}}{0.96}\frac{4.8\beta_{i}^{2}+2.23}{16\beta_{i}^{4}+16.12\beta_{i}^{2}+2.33},
\qquad
\nu_{\perp2} = \frac{\nu_{\text{ion}}}{0.96}\frac{1.2\beta_{i}^{2}+2.23}{4\left(\beta_{i}^{4}+4.03\beta_{i}^{2}+2.33\right)},
\end{equation}
\begin{equation}
\nu_{\wedge1} = \frac{\nu_{\text{ion}}}{0.96}\frac{\beta_{i}\left(8\beta_{i}^{2}+4.76\right)}{16\beta_{i}^{4}+16.12\beta_{i}^{2}+2.33},
\qquad
\nu_{\wedge2} = \frac{\nu_{\text{ion}}}{0.96}\frac{\beta_{i}\left(\beta_{i}^{2}+2.38\right)}{2\left(\beta_{i}^{4}+4.03\beta_{i}^{2}+2.33\right)}.
\end{equation}
\end{subequations}
These are Braginskii's $\eta_{1}$, $\eta_{2}/4$, $\eta_{3}$ and $\eta_{4}/2$ divided by $\rho$, where $\eta_{1}$ and $\eta_{3}$ are $\eta_{2}$ and $\eta_{4}$ evaluated at $2\beta_{i}$. The factors of 4 and 2 are those absorbed into $\boldsymbol{\Pi}^{(2)}$ and $\boldsymbol{\Pi}^{(4)}$, so $\nu_{\perp1}=\nu_{\perp2}$ and $\nu_{\wedge1}=\nu_{\wedge2}$ only in the strongly magnetised limit. Here $\beta_{i}$ is the ion Hall parameter
\begin{equation}
    \beta_i=\frac{\omega_{ci}}{\nu_{ii}+\nu_{in}},
\end{equation}
and $\omega_{ci}$ is the ion angular gyro-frequency
\begin{equation}
\omega_{ci}=\frac{e|\vec{B}|}{m_{p} c_{\text{light}}}.
\end{equation}
$\nu_{ii}$ and $\nu_{in}$ are the ion--ion and ion--neutral collision rates \cite{pinto_2008,richardson_2019}
\begin{subequations}
    \begin{equation}
        \nu_{ii} = 4.80\times 10^{-8}n_{i}\ln\Lambda T_{\text{eV}}^{-3/2}~\text{s}^{-1},
    \end{equation}
    \begin{equation}
        \nu_{in} = \frac{m_{n}}{m_{i}+m_{n}}0.649\times10^{-9}n_{n}T^{0.375}~\text{s}^{-1},
    \end{equation}
\end{subequations}
where $m_{n}/(m_{i}+m_{n})=1/2$ for H$^{+}$--H converts the rate coefficient of \cite{pinto_2008} to a momentum-transfer rate. $\ln\Lambda$ is the Coulomb logarithm \cite{Braginskii_1965}
\begin{equation}
    \label{coul-log}
   \ln\Lambda = \begin{cases}
        23.4 - 1.15\log_{10}n_{e} + 3.45\log_{10}T_{\text{eV}}, & T_{\text{eV}} < 50,\\
        25.3 - 1.15\log_{10}n_{e} + 2.30\log_{10}T_{\text{eV}}, & T_{\text{eV}} \geq 50.
    \end{cases}
\end{equation}
Our magnetic diffusivities are
\begin{equation}
    \label{mag-diffusion}
    \eta = \frac{c_{\text{light}}^{2}m_{e}(\nu_{ei}+\nu_{en})}{4\pi e^{2}n_{e}},\quad \eta_{AD}=\frac{|\vec{B}|^{2}\rho_{n}^{2}}{4\pi\rho^{2}\rho_{i}\nu_{in}},\quad \eta_{H} = \frac{c_{\text{light}}|\vec{B}|}{4\pi e n_{e}}.
\end{equation}
Here, $\rho_{n}$ and $\rho_{i}$ are the mass densities of neutral and ionised hydrogen respectively. These are diffusivities, so they take the same values in code units.\\ \\
Our thermoelectric coefficients are \cite{Braginskii_1965}
\begin{subequations}
\begin{equation}
\kappa_{J,\parallel} = 0.7110\frac{c_{\text{light}}k_{B}T}{\sqrt{4 \pi}e},
\end{equation}
\begin{equation}
\kappa_{J,\perp} = \frac{c_{\text{light}}k_{B}T}{\sqrt{4 \pi}e} \frac{5.101\beta_{e}^{2}+2.681}{\beta_{e}^{4}+14.79\beta_{e}^{2}+3.7703},
\end{equation}
\begin{equation}
\kappa_{J,\wedge} = \frac{c_{\text{light}}k_{B}T}{\sqrt{4 \pi}e}\frac{\beta_{e}\left(1.5\beta_{e}^{2}+3.053\right)}{\beta_{e}^{4}+14.79\beta_{e}^{2}+3.7703}.
\end{equation}
\end{subequations}
We retrieve our neutral/ionised mass fractions and thermodynamic variables from a linearly interpolated lookup table \cite{Sutherland_2010}. \\ \\
For all plots we exclude the domain where single fluid non-relativistic MHD breaks down (these are the grey areas). These regimes are either due to the single fluid model for ambipolar diffusion breaking down at
\begin{equation}
    \eta_{AD} >= c_{A}\Delta x,
\end{equation}
where $c_{A}$  is the Alfv\'en speed. Or where relativistic corrections become necessary when either the electron thermal velocity or the Alfv\'en speed become relativistic (Alfv\'en speeds can become superluminal with no relativistic correction)
\begin{subequations}
    \begin{equation}
        \frac{k_{B}T}{m_{e}c_{\text{light}}^{2}} \geq 0.03,
    \end{equation}
        \begin{equation}
        c_{A} \geq 0.3c_{\text{light}}.
    \end{equation}
\end{subequations}
\subsection*{E.2 - Magnetic system}
\begin{figure}[H]
    \centering
    \includegraphics[width=0.8\linewidth]{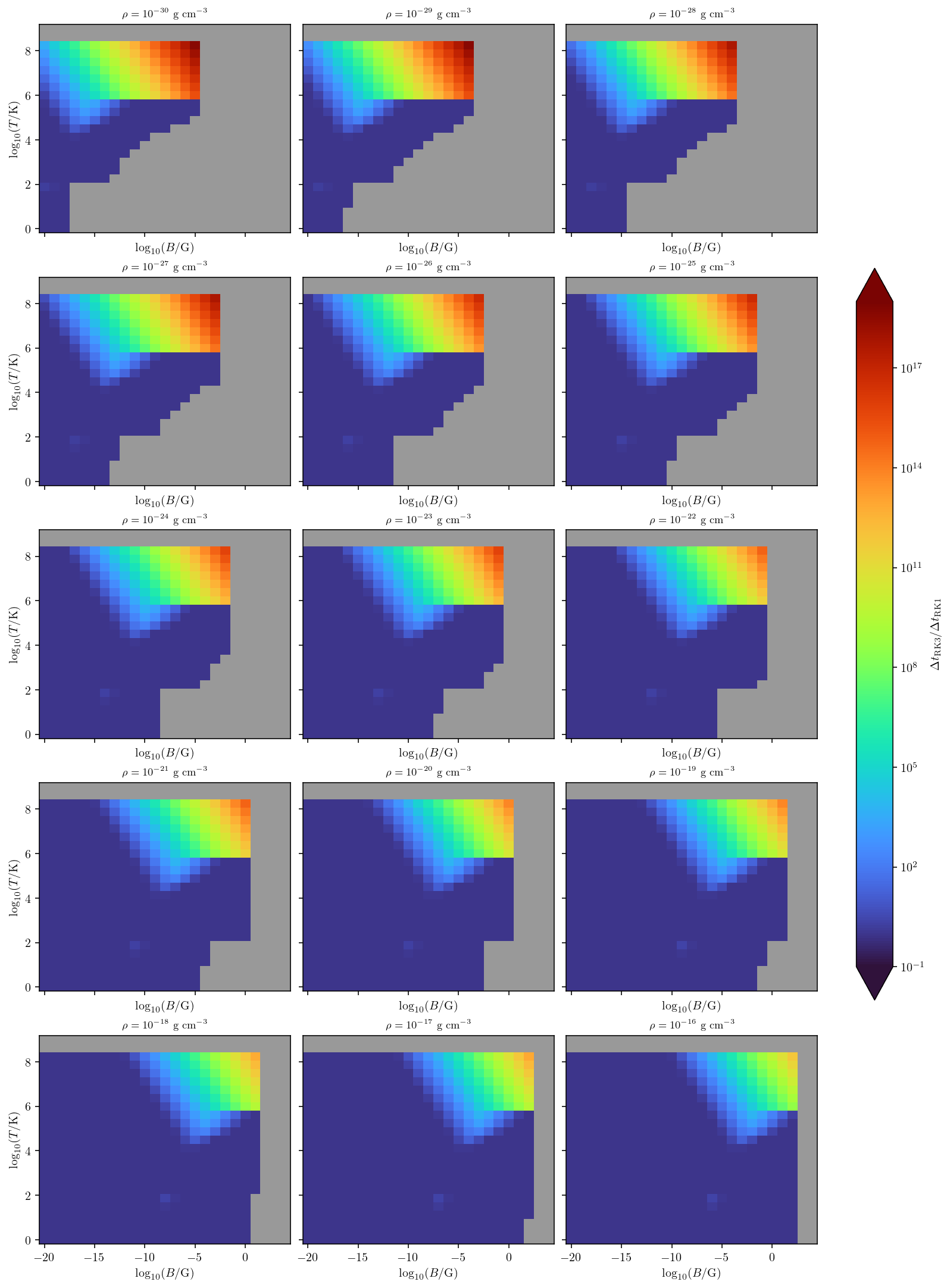}
    \caption{Ratio of RK3 and Forward Euler magnetic timesteps. From left to right, top to bottom the mass density ranges from $10^{-30}$ g cm$^{-3}$ to $10^{-16}$ g cm$^{-3}$ in increments of one decade per panel. $\psi=\pi/2$, $\theta = \pi/2$, $\phi=0$. Grey cells are where the regime is no longer physically valid.}
    \label{fig:mag_rk1_vs_rk3}
\end{figure}
Figure \ref{fig:mag_rk1_vs_rk3} shows the ratio $\Delta t_{\text{RK3}}/\Delta t_{\text{RK1}}$ for the unsplit magnetic timestep. We find that the least restrictive ratio is $\approx0.87$. This is in regimes where all three coefficients (Ohmic, ambipolar, and Hall) are at their smallest. The ratio of timesteps approaches the ratio of RK1 to RK3 stability factors $2/2.51\approx0.80$. Our most restrictive ratio is $\sim 10^{18}$. This is the low density, high temperature, high magnetic field regime where the Hall effect dominates. The timestep for RK1 scales as $\Delta t_{\eta,RK1} \propto 1/\eta_{H}^{2}$, where as RK3 scales as $\Delta t_{\eta,RK3} \propto  1/(\eta +\eta_{AD}+\eta_{H})$. The eigenvalues of the system are almost purely imaginary, and forward Euler stabilisation comes from the extremely small real part of the eigenvalue that scales $\propto 1/(\eta+\eta_{AD})$.
\subsection*{E.3 - Thermo-Magnetic System}
Here we present our numerical verification of our our bounded timestep and well-posedness criterion for the coupled thermo-magnetic system. We take $\Delta x =1$ cm and $\Delta \perp = 1/2$ cm.
\begin{figure}[H]
    \centering
    \includegraphics[width=0.8\linewidth]{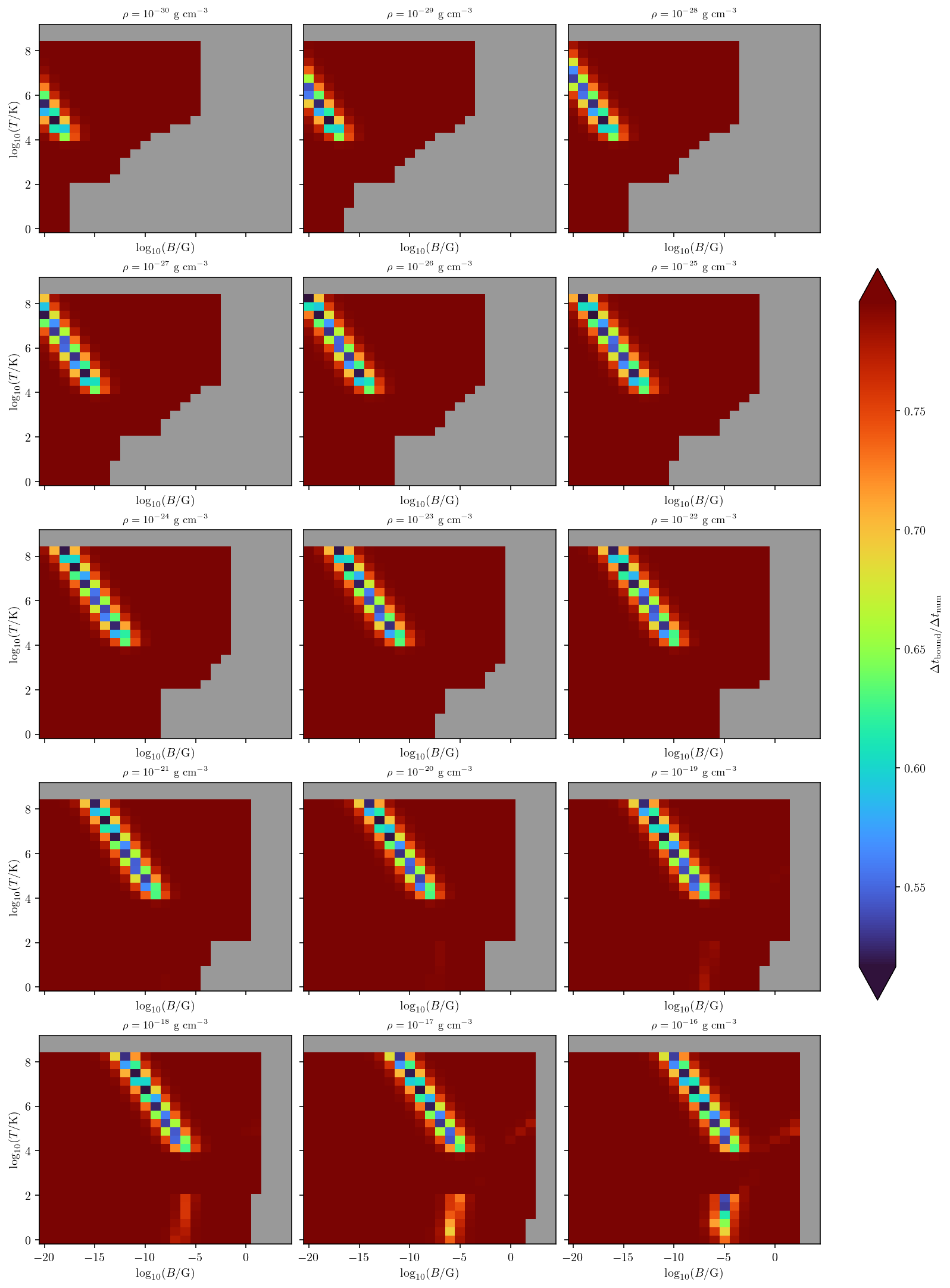}
    \caption{Thermo-magnetic timestep bound for RK3 using physical diffusion coefficients setting $b_{y}=b_{z}=0$, and $b_{y}=x$ - the parallel regime. From left to right, top to bottom the mass density ranges from $10^{-30}$ g cm$^{-3}$ to $10^{-16}$ g cm$^{-3}$ in increments of one decade per panel. $\psi=\pi/2$, $\theta = 0$, $\phi=\pi/2$. Grey cells are where the regime is no longer physically valid.}
    \label{fig:dt_bound_tmte_para}
\end{figure}
\noindent Figure \ref{fig:dt_bound_tmte_para} shows the timestep ratio for the parallel $b_{x}=1$ case. Over most of the parameter space a single dissipative term sets the timestep, and the ratio of timesteps approaches the ratio of RK1 to RK3 stability factors $2/2.51\approx0.80$. Along the diagonal band in each panel, where $\beta_{e}\approx0.5$, the Righi--Leduc term is comparable to parallel conduction and the ratio falls to 0.52. Conduction is fastest at the grid scale ($\psi=\pi/2$) and the Righi--Leduc term at $\psi=\pi/4$, so no single mode carries both at full strength. In cold, dense, magnetised gas the Hall effect is comparable to ambipolar diffusion and the ratio falls to 0.53.
\begin{figure}[H]
    \centering
    \includegraphics[width=0.8\linewidth]{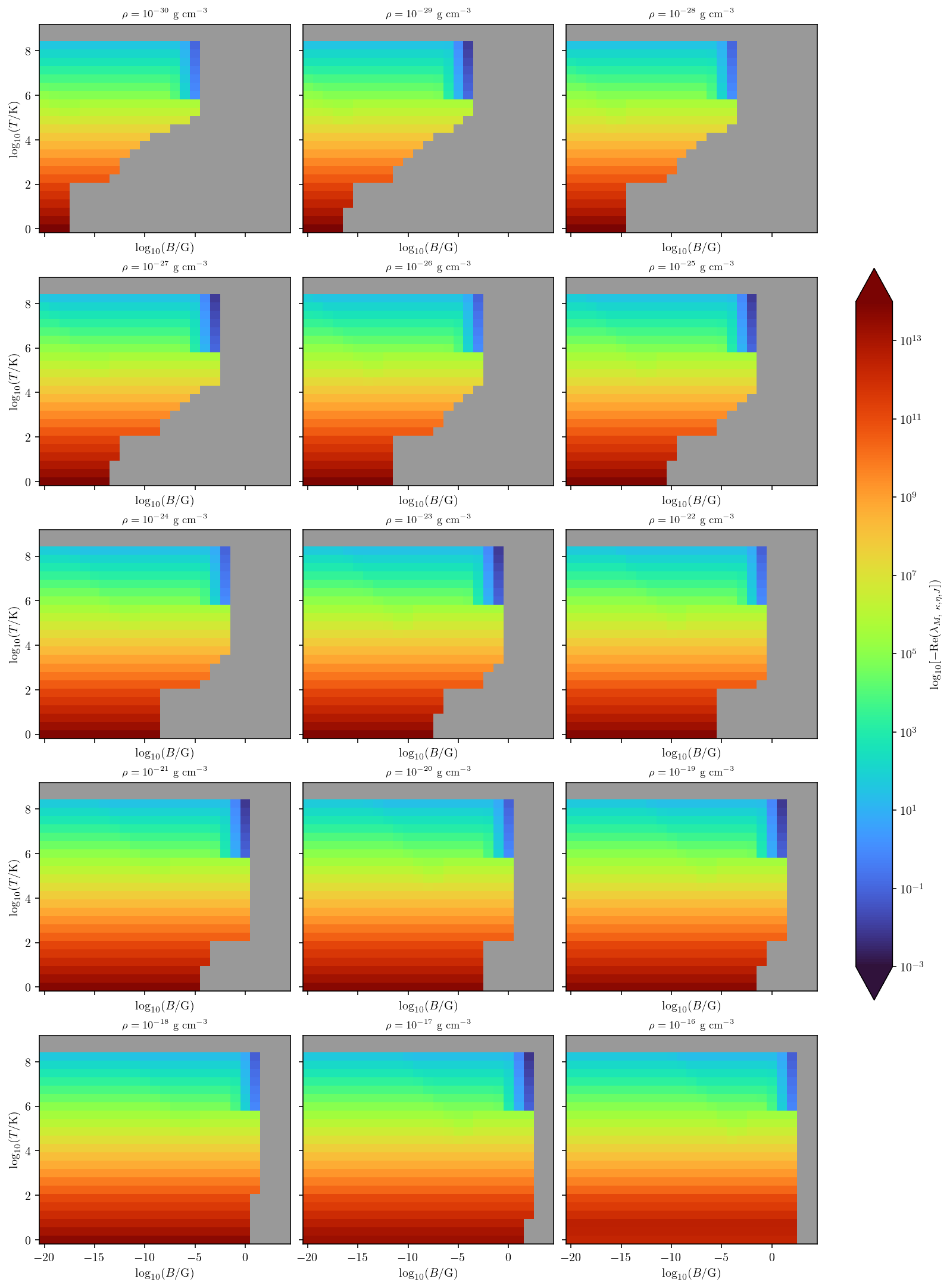}
    \caption{Numerically determined largest magnitude eigenvalue for the thermo-magnetic system using physical diffusion coefficients. $b_{x}=b_{z}=0$, and $b_{y}=1$ - the perpendicular regime. From left to right, top to bottom the mass density ranges from $10^{-30}$ g cm$^{-3}$ to $10^{-16}$ g cm$^{-3}$ in increments of one decade per panel. $\psi=\pi/2$, $\theta = \pi/2$, $\phi=0$. Grey cells are where the regime is no longer physically valid.}
    \label{fig:well-posed-numerical}
\end{figure}
Figure \ref{fig:well-posed-numerical} shows our numerically determined largest real component eigenvalue for the thermo-magnetic system (we take the negative for ease of logarithmic plotting). Taking $b_{y}=1$ maximises the potential for ill-posedness. In this regime, only Ohmic resistivity and neutral conduction (when the gas is not fully ionized) are able to suppress growing modes. Across the entire domain, the eigenvalue remains negative, which ensures the system is always well-posed.
\begin{figure}[H]
    \centering
    \includegraphics[width=0.8\linewidth]{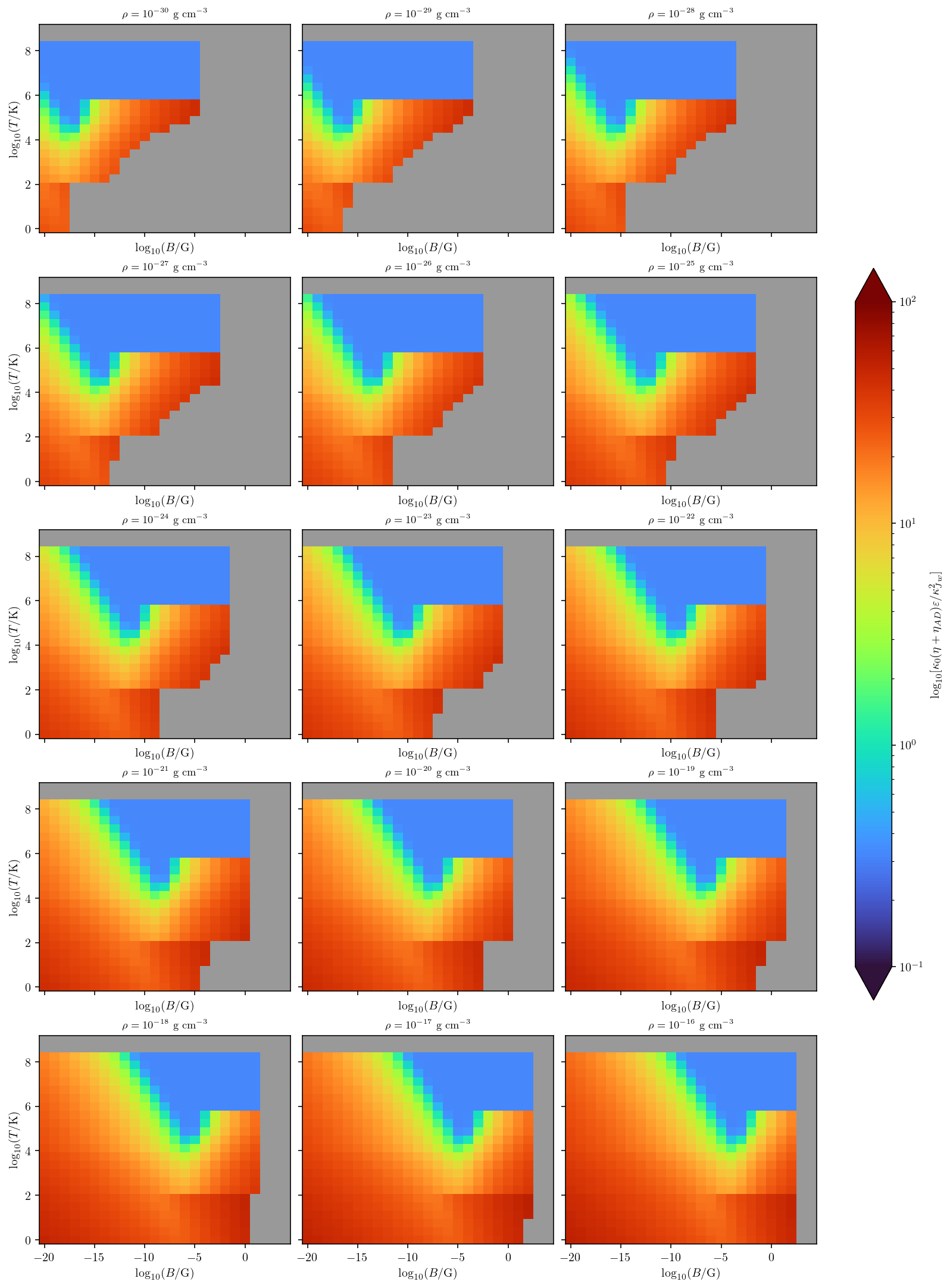}
    \caption{Analytically determined logarithm of the well-posedness criterion for the thermo-magnetic system using physical diffusion coefficients. $b_{x}=b_{z}=0$, and $b_{y}=1$ - the perpendicular regime. From left to right, top to bottom the mass density ranges from $10^{-30}$ g cm$^{-3}$ to $10^{-16}$ g cm$^{-3}$ in increments of one decade per panel. $\psi=\pi/2$, $\theta = \pi/2$, $\phi=0$. Grey cells are where the regime is no longer physically valid.}
    \label{fig:well-posed-analytical}
\end{figure}
\noindent Figure \ref{fig:well-posed-analytical} shows the logarithm of the well-posedness criterion expressed as a ratio. Our worst-case scenario gives an actual ratio of $\sim2.02$. This is in the low density, low temperature, high magnetic field regime where Ohmic resistivity and neutral conduction are at their lowest.

\subsection*{E.4 - Viscosity}
\begin{figure}[H]
    \centering
    \includegraphics[width=0.75\linewidth]{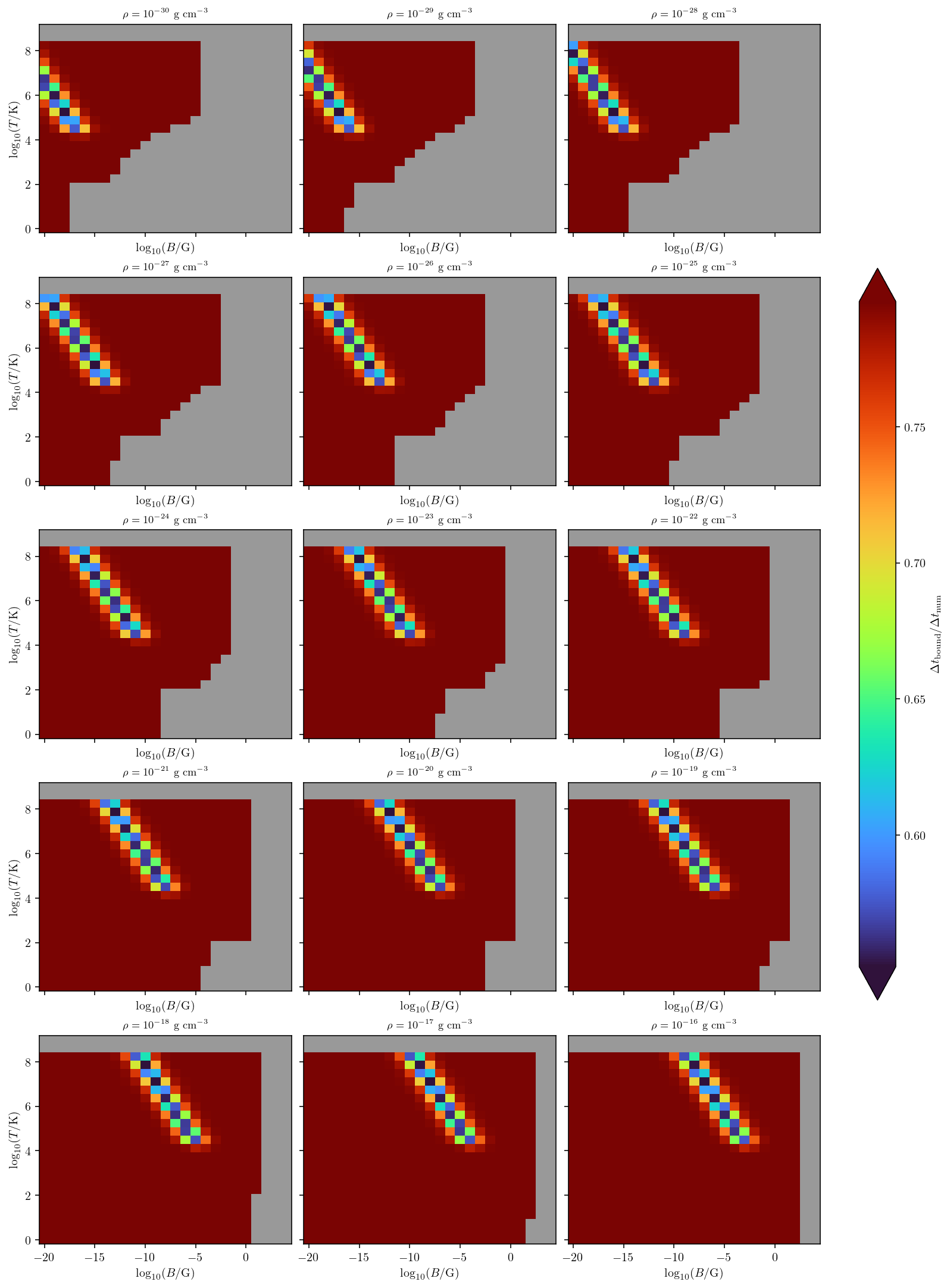}
    \caption{Ratio of numerically determined unsplit viscosity timestep and inverse sum bound for $b_{x}=1$ - the parallel regime. From left to right, top to bottom the mass density ranges from $10^{-30}$ g cm$^{-3}$ to $10^{-16}$ g cm$^{-3}$ in increments of one decade per panel. $\psi=\pi/2$, $\theta = 0$, $\phi=0$. Grey cells are where the regime is no longer physically valid.}
    \label{fig:dt_visc_bound_perp}
\end{figure}
\noindent Figure \ref{fig:dt_visc_bound_perp} shows the ratio between the numerically determined and inverse sum bounded timestep for the unsplit viscosity system in the parallel case. Our least conservative ratios are in regimes when a single dissipative term dominates $2/2.51\approx0.80$, while our most conservative ratio is $\approx 0.55$. The latter is only in a band around $\omega_{\rm ci}\tau_{\rm i}\sim1$ (below $0.7$ for $0.08\lesssim\omega_{\rm ci}\tau_{\rm i}\lesssim6$), where gyroviscosity is strongest relative to dissipation. Unlike the Hall effect, in physical regimes, gyroviscosity does not lead to extremely restrictive timesteps relative to the dissipative viscosity with unsplit integration at RK1.
\bibliographystyle{elsarticle-num-names}
\bibliography{references}

\end{document}